\documentclass{aa}  

\usepackage{graphicx}
\usepackage[colorlinks=true,citecolor=blue]{hyperref}
\usepackage{txfonts}
\usepackage{color}
\usepackage{xcolor}
\usepackage{threeparttable}
\usepackage{multirow}
\usepackage{graphicx}
\usepackage{subcaption}
\usepackage{orcidlink}
\usepackage{placeins}

\usepackage{soul}

\begin{document} 

\title{Flashing with colours: Photometric indicators of an early interaction in type II supernovae}

   \author{A. Mendez-Llorca\inst{1,2}$^,$\thanks{Corresponding author: mendez.ll.axel@gmail.com}\orcidlink{0009-0000-0130-8435},
          G. Folatelli\inst{1,2}\orcidlink{0000-0001-5247-1486},
          C. P. Guti\'errez\inst{3,4}\orcidlink{0000-0003-2375-2064},
          K. Ertini\inst{1,2}\orcidlink{0000-0001-7251-8368},
          C. R. Burns\inst{5}\orcidlink{0000-0003-4625-6629},
          N. Morrell\inst{6},\\
          M. D. Stritzinger\inst{7}\orcidlink{0000-0002-5571-1833},
          M. Bersten\inst{1,2}\orcidlink{0000-0002-6991-0550},
          J. M. DerKacy\inst{8}\orcidlink{0000-0002-7566-6080},
          L. Ferrari\inst{1,2}\orcidlink{0009-0000-6303-4169},
          L. Galbany\inst{3,4}\orcidlink{0000-0002-1296-6887},
          J. T. Hinkle\inst{9,10}$^,$\thanks{NHFP Einstein Fellow}\orcidlink{0000-0001-9668-2920}\\
          W. B. Hoogendam\inst{11}\orcidlink{0000-0003-3953-9532},
          M. A. Tucker\inst{12,13},
          C. Ashall\inst{11}\orcidlink{0000-0002-5221-7557},
          K. Auchettl\inst{14,15}\orcidlink{0000-0002-4449-9152},
          E. Y. Hsiao\inst{16}\orcidlink{0000-0003-1039-2928},
          K. Medler\inst{11}\orcidlink{0000-0001-7186-105X},
          C. Pfeffer\inst{11} \and
          H. Xiao\inst{16}\orcidlink{0009-0006-9436-7197}
          }

   \institute{Instituto de Astrofísica de La Plata, Paseo del Bosque s/n, 1900 Buenos Aires, Argentina
         \and
         Facultad de Ciencias Astronómicas y Geofísicas, Universidad Nacional de La Plata, Paseo del Bosque s/n B1900FWA, La Plata, Argentina
        \and
        Institut d'Estudis Espacials de Catalunya (IEEC), 08860 Castelldefels (Barcelona), Spain 
        \and
        Institute of Space Sciences (ICE, CSIC), Campus UAB, Carrer de Can Magrans, s/n, E-08193 Barcelona, Spain
        \and
        Observatories of the Carnegie Institution for Science, 813 Santa Barbara St., Pasadena, CA 91101, USA
        \and
        Carnegie Observatories, Las Campanas Observatory, Casilla 601, La Serena, Chile
        \and 
        Department of Physics and Astronomy, Aarhus University, Ny Munkegade 120, DK-8000 Aarhus C, Denmark
        \and
        Space Telescope Science Institute, 3700 San Martin Drive, Baltimore, MD 21218-2410, USA
        \and
        Department of Astronomy, University of Illinois Urbana-Champaign, 1002 West Green Street, Urbana, IL 61801, USA
        \and
        NSF-Simons AI Institute for the Sky (SkAI), 172 E. Chestnut St., Chicago, IL 60611, USA
        \and 
        Institute for Astronomy, University of Hawai'i, 2680 Woodlawn Drive, Honolulu, HI 96822, USA
        \and
        Center for Cosmology and Astroparticle Physics, The Ohio State University, Columbus, OH, USA
        \and 
        Department of Astronomy, The Ohio State University, Columbus, OH, USA
        \and
        OzGrav, School of Physics, The University of Melbourne, Parkville, VIC, Australia
        \and
        Department of Astronomy and Astrophysics, University of California, Santa Cruz, CA, USA
        \and
        Department of Physics, Florida State University, 77 Chieftan Way, Tallahassee, FL 32306, USA
        }

   \date{}
 
\abstract
{We present early-time optical photometry and spectroscopy for a sample 27 previously unpublished type II supernovae (SNe~II) observed by the Precision Observations of Infant Supernova Explosions (POISE) collaboration, combined with a literature sample of 59 events to form a total sample of 86 objects. The resulting light curves comprise near-daily cadence, percent-level precision \textit{uBgVri} photometry. From spectra acquired within four days of the estimated explosion time, transient narrow emission lines of highly ionised elements, referred to as `flash features', can be identified. Based on the presence or absence of these lines, the SNe were categorised. 
We estimate key photometric properties, including rise times, peak absolute magnitude, and colour evolution.
Our tests indicate that flash SNe are, on average, more luminous at the peak than their non-flash counterparts, which is consistent with previous studies. Interestingly, statistically significant differences were found in the rise times in the $B$ band, with flash SNe exhibiting a tail towards longer rise times. Flash SNe are generally bluer than non-flash events, with distinct colour distributions at 20 days post explosion, likely reflecting intrinsic differences rather than host-galaxy extinction. Colour-curve breaks within 30 days of explosion provide a clear separation between flash and non-flash SNe. The change in slopes across the break, combined with the absolute peak magnitude, yield a  strong diagnostic of the flash-spectroscopic properties. This indicates that photometric observations alone may be used to identify SNe experiencing an interaction with dense circumstellar material, even in the absence of early-time spectroscopy. }

\keywords{supernovae: general -- circumstellar matter -- stars: massive -- techniques: photometric}

\authorrunning{Mendez-Llorca et al.}
\titlerunning{Photometric indicators of early interaction in SNe~II}

\maketitle

\section{Introduction}

Massive stars ($\mathrm{M_{ZAMS}}\,\gtrsim 8\,\mathrm{M}_{\odot}$) are believed to end their lives in powerful explosions triggered by the gravitational collapse of their cores (see \citealp{Smartt.2009.intro} for a review), known as core-collapse supernovae (CC SNe). Among CC SNe, type II supernovae (SNe~II) are characterised by strong hydrogen features in their spectra \citep{Minkowski1941}. This has been interpreted as SNe~II arising from progenitors that retain a significant fraction of their hydrogen-rich envelopes at the time of the explosion. 
Current evidence indicates that normal SNe~II (i.e. type II plateau and linear) arise from massive red supergiant (RSG) stars, supported by multiple direct progenitor detections (\citealp{VanDyk2003,Smartt2009,VanDyk.Handbook.2017}). Nonetheless, several open questions remain about the origin of SNe~II progenitors, most importantly, the range of initial masses, their mass-loss histories, and the influence of a binary interaction (\citealp{Smith2014,Smartt2015,Martinez2022-II}).

Although it is well established that RSG stars undergo mass loss ($\dot{M}\sim 10^{-6}\, \mathrm{M}_\odot\,\mathrm{yr}^{-1}$, \citealp{Beasor2018}), the underlying physical mechanisms are not fully understood. Several scenarios have been proposed, including stellar winds, which are thought to be driven by a combination of pulsations and radiation pressure on dust that formed in the outer envelopes (e.g. \citealp{Heger1997,vanLoon2005,Yoon2010}; \citealp{Fuller2024,Bronner2025,Laplace2026}), a binary interaction, and eruptive mass-loss events \citep[e.g.][]{Smith2014,ercolino2024,Matsuoka2024}. Moreover, enhanced or eruptive mass-loss episodes occurring during the final years prior to core collapse have been suggested \citep{Puls2008,Moriya2011,Vink2012,Shivvers2015,Dessart2016,Fuller2017}, which could increase the amount of material released into the surrounding medium. These processes are expected to produce circumstellar material (CSM) structures in close proximity to the progenitor.  

Over the past decade, a growing body of observational evidence has revealed signatures of nearby CSM around some SNe~II. One of the most distinctive is the appearance in early-time spectra of narrow, high-ionisation emission lines with electron-scattered wings, including He \textsc{ii}, N \textsc{iii}, and C \textsc{iv}, commonly referred to as `flash features' (e.g. \citealp{Niemela1985,Garnavich1994,Gal-Yam2014}; \citealp{Khazov2016}; \citealp{Yaron2017}). These features are generally interpreted as recombination emission arising from dense, compact CSM, which is initially ionised by the shock-breakout radiation field. In some cases, early ejecta-CSM interaction may also contribute during the first hours to days after the explosion (e.g. \citealp{Yaron2017,Dessart2017}). Their rapid disappearance, typically within hours to about ten days, implies that the emitting material is located close to the progenitor star \citep{Yaron2017}. However, flash features are not observed in all SNe~II \citep{Bruch2021,JG2024}, indicating diversity in the density, extent, or geometry of the immediate circumstellar environment. Producing detectable flash emission requires sufficiently dense and compact material ($\rho \gtrsim 10^{-14}\, \mathrm{g}\,\mathrm{cm}^{-3}$ at $r \approx 10^{14} -10^{15}$ cm), which, under standard steady-wind assumptions, corresponds to mass-loss rates of the order of $\dot{M}\sim 10^{-3}-10^{-2}\, \mathrm{M}_\odot\,\mathrm{yr}^{-1}$ (see \citealp{Dessart2017}). Such conditions, however, may also arise from extended or structured stellar atmospheres, and they do not necessarily imply extreme steady winds. 

Additional indicators of an interaction include early-time light- and colour-curve behaviour, such as long rise times (\citealp{GonzalezGaitan2015,Moriya2018,Forster2018}), high ultraviolet and optical luminosities (\citealp{DessartHiller2022, JG2024}), radio and X-ray emission (\citealp{ChevalierFranson2017,Matsuoka2019,Moriya2021}), and high-velocity absorption features (\citealp{Chugai2007,gutierrez2017,DessartHiller2022}). These observations provide efficient tools for constraining the physical properties of the CSM (e.g. \citealp{Grefenstette2023,Martinez2024,Zimmerman2024,Andrews2024_22jox,Irani2024,Ertini2025}). 
In particular, the early-time morphology of the light curves has been closely related to the structure of the outer layers of the progenitor and to the distribution of CSM (e.g. \citealp{Dessart2017,Moriya2018}). 
Interestingly, by comparing early-time observations with radiation hydrodynamic models, \cite{Forster2018} argue that the majority of SNe~II interact with a nearby CSM ($r \approx 10^{14}$ cm). 

Early-time observations of SNe~II, obtained within hours to a few days after their explosion, have become increasingly common \citep{Rehemtulla2025}, enabling more detailed investigations of this stage. The advent and improvement of wide-field transient surveys, such as the All Sky Automated Survey for SuperNovae (ASAS-SN, \citealp{Shappee.2014.asassn}), the Zwicky Transient Factory (ZTF, \citealp{Bellm19, Graham19, Masci.2019.ZTF}), the Asteroid Terrestrial-impact Last Alert System (ATLAS, \citealp{Tonry2018,Smith20}), the Gravitational-wave Optical Transient Observer (GOTO, \citealp{GOTO.2022}), and, most recently, the Vera-Rubin Legacy Survey of Space and Time (LSST, \citealp{Ivezi2019-LSST}) and the La Silla Schmidt Southern Survey (LS4, \citealp{LS4.2025}), have dramatically increased both the discovery rate and the early follow-up of SNe~II. The growing availability of early-time spectra allowed diverse studies aimed at disentangling the relative contributions of the CSM interaction and shock-cooling emission, and at identifying correlations and differences among SN~II subgroups using spectroscopic and photometric properties \citep{Bruch2021,Bruch2023,JG2024,JG2025,Ertini.2026}. In particular, \cite{Bruch2021,Bruch2023} estimated that a substantial fraction of SNe~II ($> 40 \%$) with spectra obtained within two days from explosion exhibit flash features. 

Studies comparing the observational properties of SNe~II distinguished by the presence or absence of flash-ionisation features have reported differing results. \cite{Bruch2023} analyse 17 SNe~II and report that there are no significant differences in the distributions of $g$- and $r$-band peak absolute magnitudes and rise times, concluding that CSM interaction does not contribute additional energy to the light curves.
In contrast, \cite{JG2024}, using a sample of 74 SNe~II, find that events displaying flash features are generally brighter in the optical and pseudo-bolometric peak luminosities.
Moreover, they infer higher progenitor mass-loss rates for these events, suggesting a boost in luminosity and thus an enhancement of pre-explosion mass loss relative to SNe~II without flash features. Similarly, \cite{Ertini.2026} find significant differences in the peak luminosities and the post-maximum decline rates from 68 SNe~II in the ATLAS $o$ filter, though not in rise times. They identify that the absorption-to-emission ratio of the $\mathrm{H}\alpha$ P-Cygni profile is systematically lower in SNe~II with flash features and correlates with peak luminosity, possibly indicating suppression of the absorption caused by the CSM interaction. 

Colour curves of SNe~II provide key insights into the nature of these events. 
\cite{dejaeger2018} identify multiple correlations between colour-curve parameters and spectral features within the first 100 days, showing that redder SNe~II tend to be less luminous at maximum light. They attribute this trend to differences in progenitor and explosion properties and possibly to the presence of CSM rather than to extinction by dust. \cite{JG2024} further report that most SNe~II exhibiting strong flash features are bluer at early phases, within the first 40 days from explosion, than their non-flash counterparts. From a theoretical perspective, \cite{Moriya2018} show that the early evolution of colour curves is strongly influenced by the presence of CSM, leading to distinct behaviours until the interaction ceases. More recently \cite{Shrestha2024_24ggi} and \cite{Andrews2025} suggest that the observed minimum in colour evolution may serve as an indicator of the duration of flash features.

In this paper, we present photometric and spectroscopic observations for a large sample of SNe~II with early-time coverage, classified according to the presence of flash-ionisation features. Our sample includes 27 unpublished SNe followed up by the Precision Observations of Infant Supernova Explosions\footnote{\url{https://poise.obs.carnegiescience.edu/}} (POISE, \citealp{Poise}) collaboration. Motivated by the photometric quality of the POISE data and the relative scarcity of early-time colour-curve analyses of SNe~II, we explore possible photometric indicators of CSM interaction. 

The paper is organised as follows. Section~\ref{sec:data} presents the observations, a description of the sample, and the methods used to estimate distances, first-light epochs, and extinction. We outline the flash-spectroscopic classification scheme in Sect.~\ref{sec:classification}. Section~\ref{sec:stats.methods} describes the statistical tests used in the analysis and the comparison cases. In Sect.~\ref{sec:light_curves}, we analyse the multi-band light curves, while Sect.~\ref{sec:colour.curves} focuses on the definition and analysis of colour-curve parameters and their correlations with photometric and spectroscopic properties. In Sect.~\ref{sec:discussion}, we discuss our results in the context of previous observational and theoretical studies. We summarise our main conclusions in Sect.~\ref{sec:conclusions}. Section~\ref{sec:data_availability} provides access to the photometric and spectroscopic data. 

\section{Data sample}
\label{sec:data}

The POISE collaboration \citep{Poise} has been conducting two follow-up campaigns per year since 2021, with the exception of 2024, when only one campaign was carried out in the second semester. Each campaign typically spans 60-70 days. The main focus is on obtaining early-time observations of all SN types. The observations comprise high-precision, almost daily multi-band photometry, plus optical and near-infrared spectroscopy. In this work, we analyse a sample of 27 SNe~II, all observed across five campaigns from February 2021 to April 2023. Table \ref{t:POISE.sample} presents our POISE SN~II sample and its key information.

Follow-up observations of 24 SNe were initiated within three days of their discovery. The day of first light was estimated for all of the SNe, as described in Sect.~\ref{sec:first-light}. The majority of them (18 out of 27 SNe) had their first photometric observation within three days of the inferred first-light epoch. Light-curve coverage for each SN lasted between $\approx$~20 and 70 days, with a mean of 45 days. Photometric observations were obtained with an average cadence of $\approx$~1.5 days. The redshift range spans from 0.0049 to 0.059, with the majority of SNe (24 out of 27 SNe) at $z \leq 0.03$. Given the low redshift of this sample, no K-corrections were applied to the photometry.

\subsection{Photometry}\label{sec:photometry}

The POISE collaboration obtained optical photometry in \textit{uBgVri} bands using the 1.0 m Swope Telescope at Las Campanas Observatory (LCO) in Chile. Image reduction was carried out using the POISE photometric pipeline, following the methodology established by the Carnegie Supernova Project (CSP), as described in \cite{Contreras2010} and \cite{Krisciunas2017}. Bias and flat-field corrections were applied to the raw data. 
Host-galaxy subtraction was performed using reference images from public surveys, such as Pan-STARRS \citep{Kaiser2002} or SkyMapper \citep{Wolf2018}, or templates obtained during later campaigns, once the SN had faded below detection thresholds. Photometric calibration in \textit{BV} was done with \cite{Landolt1992} standards, while calibration in \textit{ugri} was done using \cite{Smith2002} standards. Colour terms required to convert instrumental magnitudes to the standard magnitudes were determined using the standards for calibration \cite[see][for more details]{Krisciunas2017}. Atmospheric extinction corrections were applied following \cite{Krisciunas2017}. The colour terms were applied in reverse to convert \cite{Landolt1992} and \cite{Smith2002} standard magnitudes to Swope natural-system magnitudes. Then, the local sequence stars in the field of each SN were calibrated in the natural system. Relative photometric calibration of each SN was done using the calibrated natural magnitudes of the local sequence stars. Finally, the resulting POISE light curves are given in the CSP natural system for the Swope telescope. 

To convert the CSP natural-system photometry to the AB system, we used the Eq.~47 and the offsets listed in Table 16 of \cite{Krisciunas2017}. While this approach may introduce systematic uncertainties, these are not expected to affect the results of this work. The optical photometry of the POISE SN sample used in this work has an average precision of 0.025 mag. The photometry tables and the corresponding light curves can be found in Sect.~\ref{sec:data_availability} and on the POISE webpage.

\subsection{Spectroscopy}

All SNe from POISE have at least one optical spectroscopic observation. Since our goal is to identify and classify flash-spectroscopic features (see Sect.~\ref{sec:classification}), only spectra obtained within four days of first light (see Sect.~\ref{sec:first-light}) are considered. For seven of these SNe, the spectra were acquired through POISE and its collaborators. For the remaining SNe, we used publicly available spectra from the Transient Name Server\footnote{\url{https://www.wis-tns.org/}} (TNS) and \cite{JG2024}. These observations are summarised in Table~\ref{t:POISE.spectra}.

Three of the POISE spectra (SN 2021bmw, SN 2021cgu, and SN 2023dtd) were taken with the 6.5 m Baade Magellan telescope at LCO, using the Inamori-Magellan Areal Camera and Spectrograph (IMACS, \citealp{IMACS2011}). The IMACS data were processed through standard \textsc{IRAF}\footnote{IRAF was distributed by the National Optical Astronomy Observatory, which is operated by the Association of Universities for Research in Astronomy, Inc., under cooperative agreement with the National Science Foundation.} procedures, including bias subtraction, flat-fielding, one-dimensional extraction, and wavelength calibration with a HeNeAr lamp exposure obtained right after the SN observation. Flux calibration was performed using spectra of spectrophotometric standards obtained during the same observing nights. We also used specific standard stars for telluric feature removal, although residuals of some strongly saturated telluric bands often remain. Of the remaining four POISE spectra, three (SN 2021adlr, SN 2022fae, and SN 2022aacn) were obtained with the Supernova Integral Field Spectrograph (SNIFS, \citealt{Lantz2004}) on the University of Hawaii 2.2 m Telescope (UH2.2m) at Mauna Kea as part of the Spectroscopic Classification of Astronomical Transients (SCAT, \citealp{SCAT.2022}) project; and one (SN 2022ffg) was acquired with the Low Resolution Imaging Spectrometer (LRIS, \citealp{Oke1995}) on the Keck I telescope at Mauna Kea. The SNIFS spectra were reduced using the SCAT pipeline as described in \cite{SCAT.2022}, and the reduction of the LRIS spectrum was performed using \textsc{PypeIt} \citep{PypeIt2020}, following standard procedures.

Figures~\ref{f:c1+2_spectra}-\ref{f:nf_spectra} present the early-time optical spectra of all SNe~II in this work, divided by flash-spectroscopic classification (see Sect.~\ref{sec:classification}). The corresponding POISE spectra can be found in Sect.~\ref{sec:data_availability}. 

\subsection{Distances}\label{sec:distances}

For SNe located in galaxies with recession velocities relative to the Cosmic Microwave Background (CMB) greater than $2\,000$ km s$^{-1}$ ($z_{\mathrm{CMB}} \gtrsim 0.0067$), we computed distances using a flat $\Lambda$CDM cosmology. Specifically, we adopted the luminosity distance formula from Eq. (5) of \cite{2010Folatelli} with the following parameters: $H_0 = 71.76\, \pm\, 0.58\, \mathrm{(stat) \pm1.19 \,\mathrm{(sys)}} $  km s$^{-1}$ Mpc$^{-1}$ \citep{Uddin2024},  $\Omega_M$ = 0.27 and $\Omega_{\Lambda}$ = 0.73 \citep{Burns2018}. Redshift values were obtained from the NASA/IPAC Extragalactic Database\footnote{\url{https://ned.ipac.caltech.edu/}} (NED). Distance uncertainties were estimated by propagating the uncertainties through this formulation, assuming a redshift uncertainty of 0.001, corresponding to peculiar velocities of $\approx$ 300 km s$^{-1}$ relative to the Hubble flow. The majority of the objects in the sample (24 out of 27 SNe) met this recession-velocity criterion. These values are given in Table \ref{t:POISE.sample}.
For the remaining three SNe (SN 2022aagp, SN 2022acko, and SN 2023bvj), we used redshift-independent distances also obtained from NED or other sources, primarily from Cepheids when available. The corresponding values and references are given in Table \ref{t:POISE.sample}. 

\subsection{First-light epochs}
\label{sec:first-light}

The present analysis requires a characterisation of the time when the SN explosion occurred. From an observational viewpoint, we can define the time when optical radiation from the SN is first released, which we call the `time of first light'. For all SNe, the first-light epochs were estimated as the midpoint between the last non-detection (LND) and the first detection. The definition of these epochs is described below. The associated uncertainties were taken as half of the time interval between these two epochs (see Table~\ref{t:POISE.sample}). The estimates were obtained using forced photometry in flux from the ATLAS forced-photometry server\footnote{\url{https://fallingstar-data.com/forcedphot/}} \citep{Shingles21} and the ZTF forced-photometry service\footnote{\url{ht tps://irsa.ipac.calt ech.edu/data/ZTF/docs/forcedphot .pdf}} \citep{masci2023}. Since the measurements are based on image differences, we define a non-detection as a flux consistent with zero within 1$\sigma$ of the uncertainty. The first detection is defined as the first observation with a flux measurement exceeding 3$\sigma$ above zero. Its error bar does not overlap with that of the LND, and which is followed by subsequent observations showing increasing flux, indicative of the rise of the light curve. The LND was adopted as the observation immediately before the first detection. The first-light epochs for our sample are listed in Table~\ref{t:POISE.sample}, with a mean uncertainty on these measurements of 1.3 days. Hereafter all temporal parameters presented in this work are referenced to the time of first light and expressed in the rest frame. 

\subsection{Extinction}
\label{sec:extinction}

All Milky Way (MW) extinction values were obtained from NED, based on the dust maps of \cite{Schlafly2011} and assuming a \cite{Fitzpatrick1999} extinction law with $R_V = 3.1$. In this work, host-galaxy extinction is neglected, following the arguments presented by \cite{Anderson2014} and \cite{dejaeger2018} in their analyses of SNe~II. \cite{Anderson2014} cautioned against using the equivalent width (EW) of Na \textsc{i} D absorption lines as a reliable proxy for extinction, especially when these lines are saturated or exhibit large EW values. Furthermore, low-resolution spectra can significantly affect EW measurements (see \citealt{Poznanski2011, Phillips2013}); in particular, broad SN features may contaminate the narrow Na \textsc{i} D lines, leading to artificial EW evolution \citep{Gonzalez-gaitan2024} and thus introducing additional uncertainty. 

\cite{dejaeger2018} find that colour variations among SNe~II cannot be fully explained by host-galaxy extinction. Consequently, the colour-excess method based on the (\textit{$V-I$}) at the end of the plateau \citep{Olivares2010} becomes unreliable, as it assumes uniform intrinsic colours across all SNe~II. Recently \cite{Hinds2025} also report an intrinsic colour dispersion among SNe~II, estimating from a large SNe~II sample that only $10 \%$ of them are affected by host-galaxy extinction. 

Although we did not correct our colours and absolute magnitudes for host-galaxy extinction, we used the colour excesses derived from the EW of the Na \textsc{i} D line to identify SNe that may be strongly extincted. For this purpose, we employed the relation given by \citet{Stritzinger2018} to derive $E(B-V)_\mathrm{host}$ for our sample (the values are listed in Table~\ref{t:POISE.sample}). SNe with $E(B-V)_\mathrm{host}> 0.3$ mag are considered highly extincted. These are SN~2021bmw, SN~2023bvj, SN~2023cps, and SN~2023dtd (four out of 27 SNe, i.e.\ $\approx15\%$ of the sample). As shown in Sect.~\ref{sec:literature.sample}, this fraction is similar in the larger literature sample. Therefore, we can assume that our results will not be significantly affected by the lack of host-galaxy extinction correction.

\begin{figure}
\centering
\resizebox{0.95\hsize}{!}
{\includegraphics{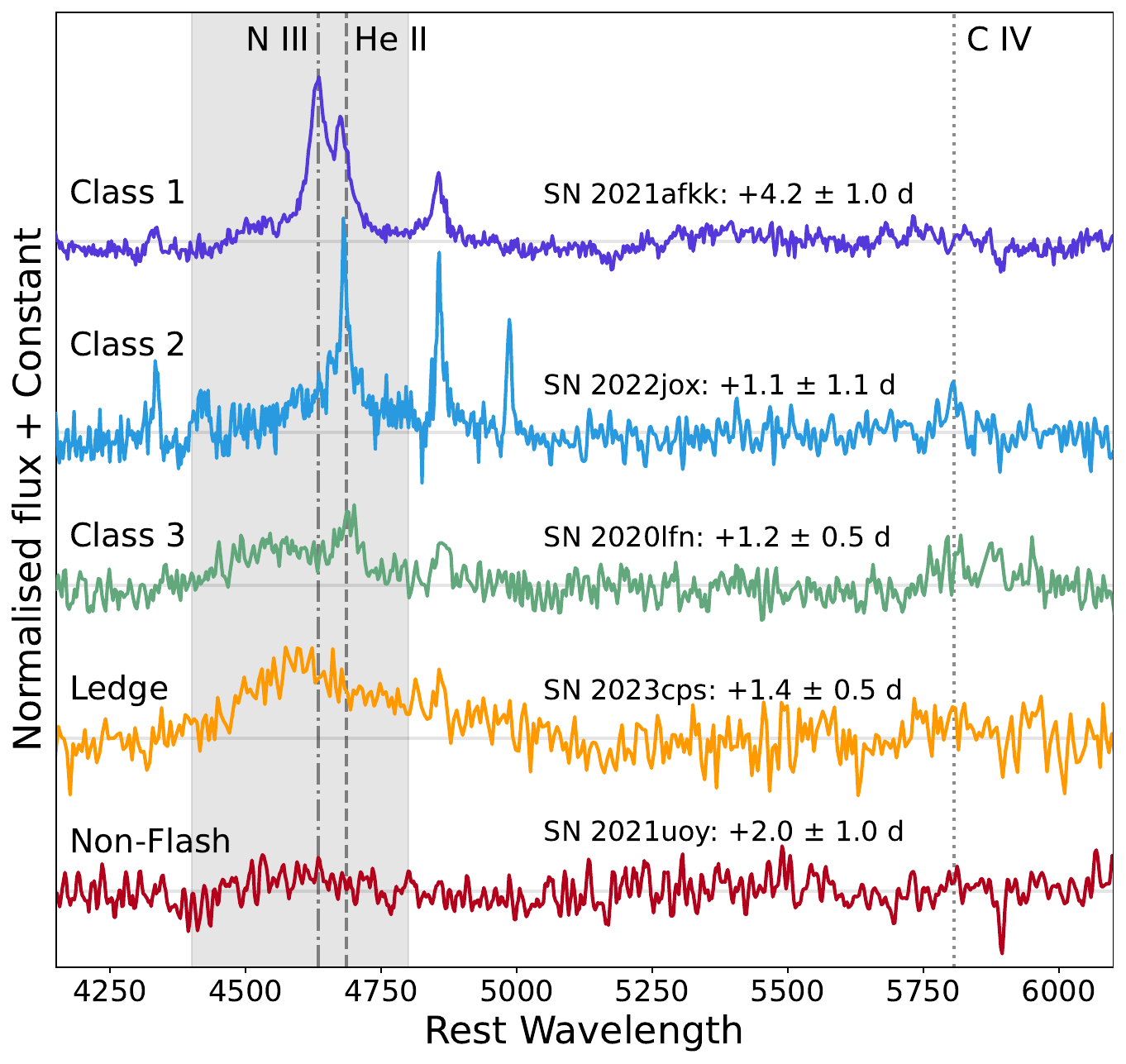}}
\caption{Early-time spectra of all flash-spectroscopic classes (see Sect.~\ref{sec:classification}) in the region where flash features appear. Class 1 (blue) shows intense He \textsc{ii} and N \textsc{iii} emission; class 2 (light blue) displays only intense He \textsc{ii} but lacks N \textsc{iii}; class 3 (green) presents weak He \textsc{ii} emission; ledge (orange) exhibits a broad, asymmetric emission feature in the 4400-4800 \r{A} range (shaded area); and non-flash (red) does not show any of these features. All spectra are normalised by fitting the continuum between 4000 and 6200 \r{A} with a fourth-degree Legendre polynomial, excluding the 4400-4900 \r{A} region to avoid the spectral features. Vertical lines mark the position of the most prominent flash-ionisation features. Each spectrum is labelled with the SN name, phase relative to the first-light epoch, and first-light uncertainty.}
\label{f:spectral.class}
\end{figure}

\subsection{Literature sample}\label{sec:literature.sample}

To increase the statistical significance of our analysis, we supplemented our dataset by incorporating SNe~II from the literature, hereafter referred to as the literature sample. Our primary reference was the compilation presented by \cite{JG2024}, from which we selected objects that satisfy the following criteria: (i) photometric data available in at least two optical bands overlapping with the POISE system, allowing the construction of at least one of the colours analysed in this work (i.e. \textit{$g-r$}, \textit{$g-i$}, \textit{$u-g$}, or \textit{$B-V$}); (ii) well-sampled early-time light curves, covering at least 40 days after first light; and (iii) at least one early-phase optical spectrum obtained within four days from first light, if the SN did not exhibit flash-ionisation features.  

A total of 55 SNe~II were selected from the \cite{JG2024} dataset. Owing to the similarity in the effective wavelength of the \textit{u} and \textit{U} filters, we treated them as equivalent to increase the sample size in that band. However, photometry obtained with the \textit{Swift}-UVOT filters was excluded from the analysis due to the lack of such data for the POISE SNe. 

From \cite{Anderson2024}, we included two additional SNe~II that satisfied the above criteria, SN~2012aw and SN~2013hj. We also added SN~2023ixf (photometry from \citealp{Hiramatsu2023_23ixf} and spectroscopy from \citealp{Bostroem.23ixf.spectrum.2023}) and SN~2024ggi (photometry from \citealp{Shrestha2024_24ggi} and spectroscopy from \citealp{24ggi.spectrum.2024}), two of the most outstanding SNe~II of the last decade, both showing flash-ionisation signatures (first spectroscopic reports: \citealp{Perley2023.tns.ixf} and \citealp{Hoogendam2024.tns.ggi}, respectively). In addition, due to the limited temporal coverage of SN~2022acko ($\approx20$ days from first light) in POISE, we incorporated the photometry from \cite{Bostroem.acko.2023} to enlarge its coverage (up to $\approx$ 40 days) and fill large gaps. Altogether, the final combined dataset comprises 86 SNe~II.  
 
For all SNe in the literature sample, we determined their distances (as described in Sect. \ref{sec:distances}), Galactic extinctions (see Sect. \ref{sec:extinction}), and first-light epochs (see Sect. \ref{sec:first-light}). In the latter case, when survey data were unavailable, first-light estimates were taken from TNS reports or the literature. As in Sect.~\ref{sec:extinction}, we identified SNe with $E(B-V)_\mathrm{host}> 0.3$ mag as likely being highly extincted. For this we employed the colour excesses based on the EW of the Na \textsc{i} D line provided by \cite{JG2024}, and we added our own measurements from public spectra of SN~2012aw, SN~2013hj, SN~2016iyz, SN~2019nyk, SN~2019ofc, SN~2023ixf, and SN~2024ggi. Based on this criterion, we find that nine out 59 SNe in the literature sample (around 15\%, namely SN~2014G, SN~2017gmr, SN~2018cuf, SN~2019ust, SN~2020fqv, SN~2021aek, SN~2021jtt, SN~2022jzc, and SN~2022pgf) may be highly extincted. Details of the literature sample are summarised in Table \ref{t:Lit.sample}.

\begin{figure*}
\centering
\resizebox{0.93\hsize}{!}
{\includegraphics{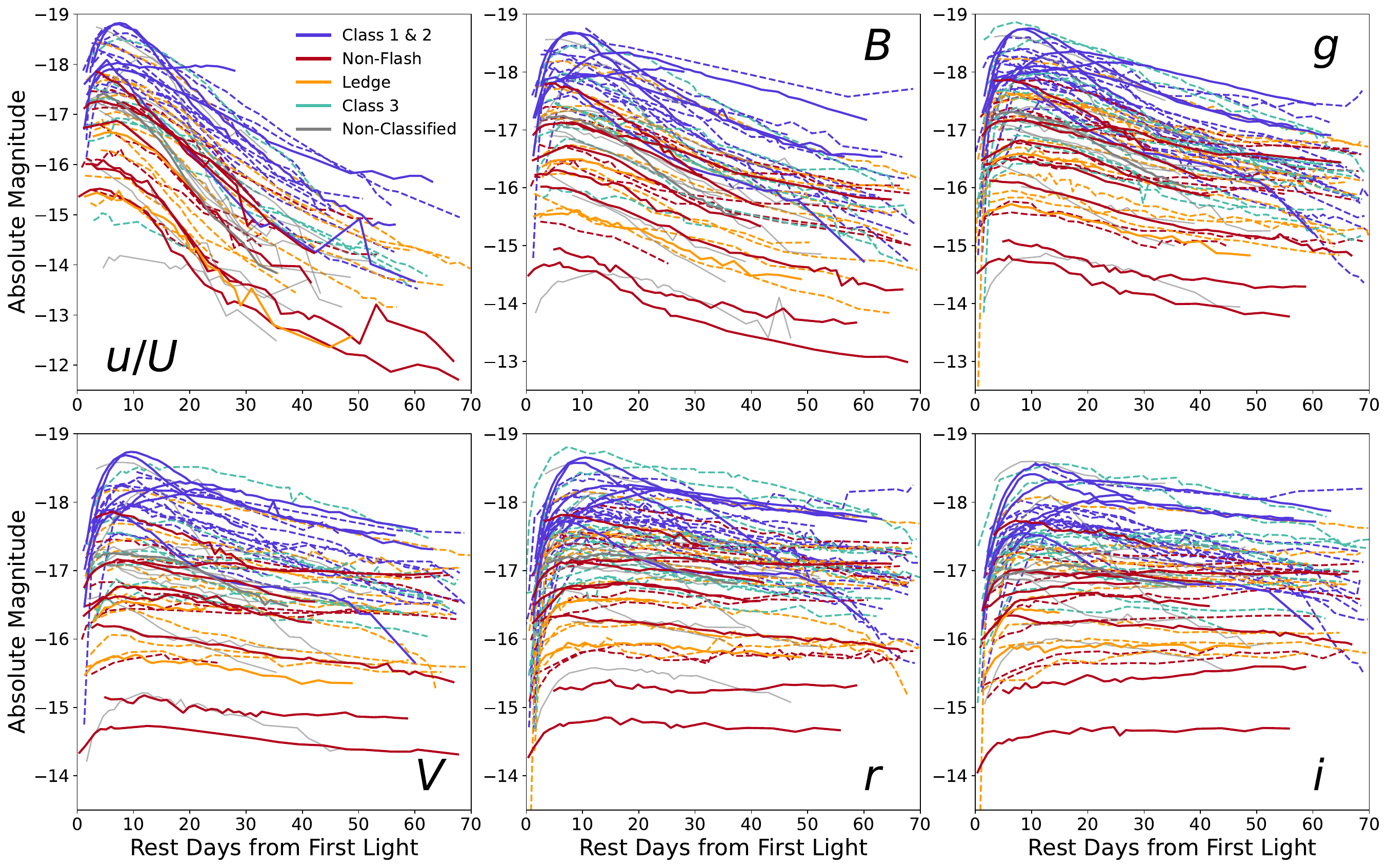}}
\caption{Left to right, top to bottom: Light curves in $u/U$, $B$, $g$, $V$, $r$, and $i$ bands, corrected only for MW extinction. Line colours represent the SNe flash-spectroscopic classification of the SNe (see Sect. \ref{sec:classification}): `class 1+2' (blue), `class 3' (green), `ledge' (orange), and `non-flash' (red). Solid and dashed lines show SNe from the POISE and literature samples, respectively. `Non-classified' SNe (grey) are displayed for completeness. Class 1+2 SNe are consistently brighter across all bands. In contrast, SNe~II showing only a ledge-like feature have luminosities comparable to those SNe~II without flash features.
}
\label{f:LC_class}
\end{figure*}

\section{Flash-spectroscopic classification}
\label{sec:classification}

A fundamental aspect of our analysis is to explore the possible connection between the photometric properties of SNe~II and the presence or absence of flash features. These narrow, high-ionisation lines are indicative of a dense CSM ($\rho \gtrsim 10^{-14}\, \mathrm{g}\,\mathrm{cm}^{-3}$ at $r \approx 10^{14} -10^{15}$ cm, \citealp{Dessart2017}), thereby providing direct evidence of early circumstellar interaction (CSI). However, the absence of such lines does not necessarily indicate the absence of CSM around the progenitor. 

Our flash-spectroscopic classification is partially based on the criteria established by \cite{JG2024} and defined according to the presence or absence and relative intensity of specific spectral lines. The classification was performed via visual inspection of optical spectra obtained within four days of first light. Additionally, we required that the sum of the spectral phase (days since first light) and the uncertainty in the first-light epoch is $\leq$ 5 days (e.g. SN 2022ewj does not fulfil this criterion). Figure \ref{f:spectral.class} presents a representative spectrum for each category, while Figs.~\ref{f:c1+2_spectra}-\ref{f:nf_spectra} show the spectra used to classify each SN, organised by class. SNe exhibiting high-ionisation emission lines with electron scattered broad wings, such as He~\textsc{ii} $\lambda$4686, N~\textsc{iii} $\lambda$4641, and C~\textsc{iv} $\lambda$$\lambda$5801,5812, were classified as `flash' SNe and further divided into three subclasses:
    `Class 1' shows intense emission lines of He \textsc{ii} and N \textsc{iii}, comprising 19 SNe (four from POISE and 15 from the literature); 
    `Class 2' exhibits intense emission of He \textsc{ii} but no N \textsc{iii}, with nine SNe (two from POISE and seven from the literature);
    and `Class 3' presents only weak He \textsc{ii} emission, which is neither blueshifted nor Doppler-broadened, including 12 SNe (all from the literature).

To improve statistical robustness and comparisons, we combined classes 1 and 2 into a single subsample, hereafter `class 1+2', while keeping class 3 separate. 
This is because class 3 objects exhibit only weak He \textsc{ii} emission and thus may be subject to misclassification, particularly in low S/N spectra. For instance, we reclassified four SNe from the \cite{JG2024} dataset (SN~2016X, SN~2017gmr, SN~2020fqv, and SN~2021gvm) as class 3 after our inspection, whereas they were originally labelled as `comparison' SNe (i.e. lacking narrow, high-ionisation lines).

The requirement of having at least one spectrum within four days of first light is set to minimise misclassification of SNe with very short flash phases. However, if an SN showed clear flash-ionisation features in later spectra, we included it in one of these classes. This was the case for SN~2019qch (7.5 days after first light), SN~2021zj (4.9 days), SN~2021tyw (8.3 days), SN~2021afkk (4.2 days), SN~2022dml (4.5 days), SN~2022prv (10 days), and SN~2023bvh (5.6 days). 

Beyond the classification scheme by \cite{JG2024}, we classified the remaining SNe with the following criteria. SNe displaying only a broad, asymmetric emission feature around 4400-4800 \r{A} (shaded region in Fig.~\ref{f:spectral.class}), denoted a `ledge-like feature' (e.g. \citealp{Bullivant2018,Andrews2019-17gmr,Andrews2024_22jox,Soumagnac2020,Hosseinzadeh2022_21yja}; \citealp{Chugai2023_ledge}), and lacking any narrow flash-ionisation lines, were categorised as `ledge' SNe. 
Events showing both flash features and the ledge feature, either simultaneously or at different phases, were instead assigned to the previously defined flash classes. This group consists of 14 SNe (three from POISE and 11 from the literature). 

Finally, SNe lacking flash or ledge features were classified as `non-flash', comprising 22 SNe (eight from POISE and 14 from the literature). We note that some of the SNe classified as non-flash exhibit an emission feature to the blue of the He \textsc{ii} $\lambda$4686 line (see e.g.\ SN~2019edo, SN~2021gmj, and SN~2023aagp in Fig.~\ref{f:nf_spectra}). This emission is, however, narrower and more symmetric than that identified as the ledge feature, and thus these SNe are classified as non-flash. An additional ten SNe from the POISE sample lacked sufficiently early spectra to determine a reliable flash classification; these are labelled as `non-classified', but they are included in the measurements and plots for completeness. 

We note that the short-lived nature of the narrow flash-ionisation features may lead to misclassification of some non-flash and ledge SNe that may not have been observed early enough to detect the emission. Consequently, our comparative analyses are likely conservative, as potential contamination by misidentified flash SNe in these samples would tend to reduce the statistical significance of the observed differences, rather than enhance them.

\section{Statistical methods}
\label{sec:stats.methods}

Since one of our main goals is to identify photometric parameters that can distinguish the different flash-spectroscopic classes (see Sect.~\ref{sec:classification}), we performed two-sample statistical tests: the Kolmogorov-Smirnov (K-S) test, using the function \textsc{ks$\_$2samp} \citep{Virtanen2020.KS.test} from \textsc{scipy}, and the Anderson-Darling (A-D, \citealp{A-D.Pettitt1976}) test, via \textsc{anderson$\_$ksamp} from \textsc{scipy}. Both tests are complementary because K-S tests are suited to detecting differences around their central values, whereas A-D tests are more effective in finding differences at the tails of the distributions. Since the A-D test yields an approximate p value, statistical significance was inferred by comparing the observed statistic with the tabulated critical values provided by the algorithm, which allows us to constrain the p values to a given significance range. The null hypothesis, that both samples originate from the same parent distribution, is typically reject at p $<0.05$. However, given the number of parameters tested simultaneously, we adopted a more conservative threshold of p $<0.005$ to mitigate the risk of false positives arising from multiple comparisons. For the A-D tests, the tabulated critical values only extend down to p $ < 0.01$, meaning that a more precise significance threshold cannot be resolved by this method alone; results falling below this limit are therefore interpreted as indicative of statistically significant differences between the samples and are used to complement the conclusions drawn from the K-S tests. 

The comparison cases considered for the statistical tests are (a) class 1+2 vs. non-flash and (b) flash (class 1+2+3) vs. non-flash + ledge. The first case contrasts events with clear and intense narrow flash-ionisation emission features against those without any emission feature, while the second compares all flash classes against SNe lacking narrow emission lines. In all comparisons, parameters with upper or lower limits were excluded from the tests, ensuring that only well-constrained parameters were used.

\begin{figure*}
\centering
\includegraphics[width=.95\textwidth]{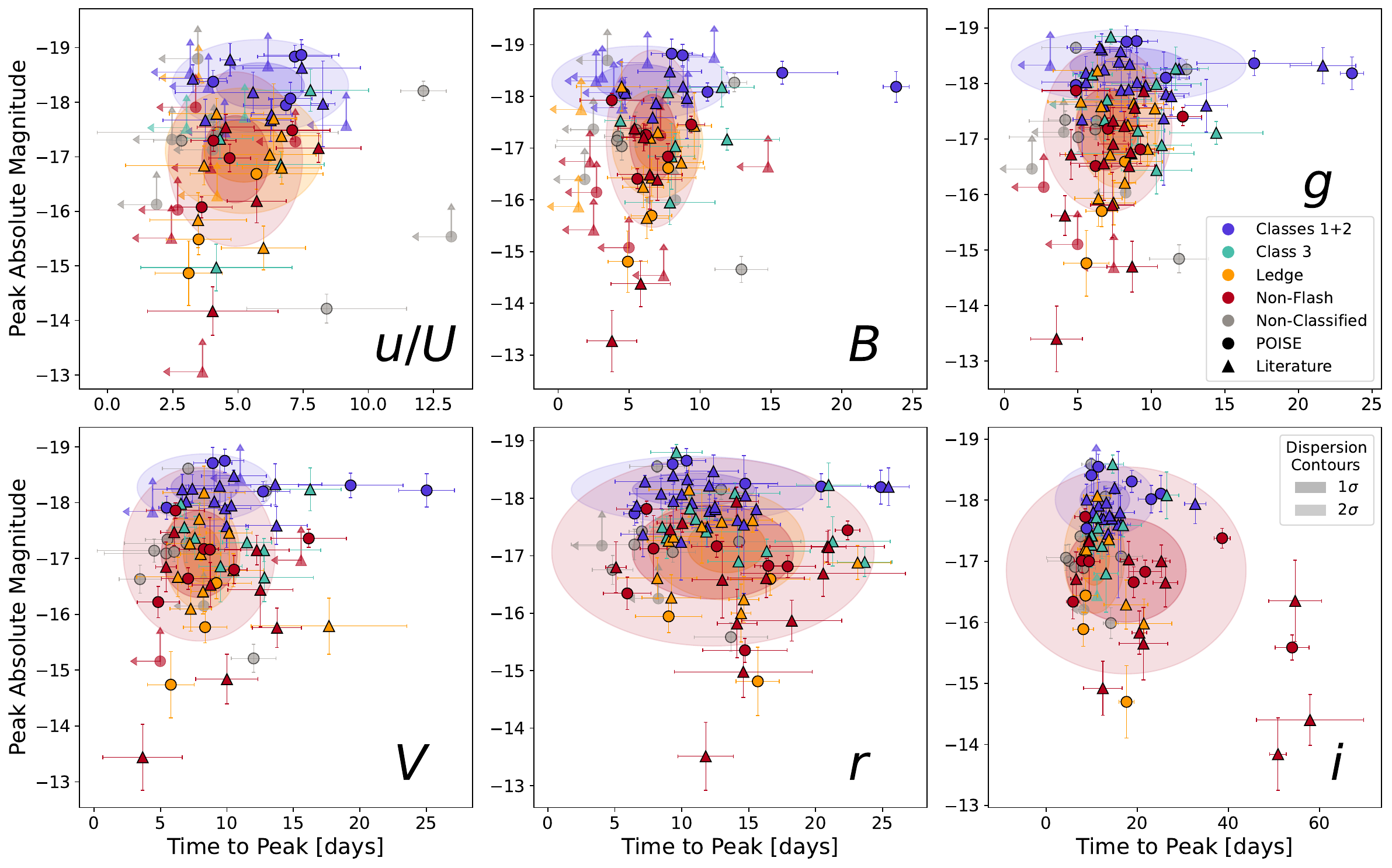}
\caption{Peak absolute magnitude vs. time to peak in $u/U$, $B$, $g$, $V$, $r$, and $i$ bands (left to right, top to bottom). Class 1+2 (blue), class 3 (green), ledge (orange), and non-flash SNe (red) are shown for the POISE (circles) and literature (triangles) samples. Fainter markers without black edges indicate upper and lower limits for time to peak and absolute magnitude, respectively. Non-classified SNe (grey) are included for completeness. The shaded ellipses are centred on the weighted mean of each parameter for class 1+2 (blue), ledge (orange), and non-flash SNe (red), with widths corresponding to 1$\sigma$ and 2$\sigma$ weighted standard deviation (from darker to lighter tones). Class 1+2 SNe exhibit brighter peak magnitudes than non-flash and ledge SNe, while class 3 SNe are mixed with the other classes. In the bluer bands (i.e. \textit{uBgV}), class 1+2 SNe present slightly longer rise times on average than non-flash SNe, whereas in the redder bands (i.e. $r$ and $i$) this trend reverses.}
\label{f:rise_times_vs_peak_mag}
\end{figure*}

\section{Light-curve properties} 
\label{sec:light_curves}

We analyse the light-curve properties of the complete sample, grouping the SNe based on the flash-spectroscopic classification outlined in Sect. \ref{sec:classification}. Hereafter all magnitudes are reported in the AB system for consistency across the sample; the conversion for the POISE photometry is described in Sect.~\ref{sec:photometry}. Figure \ref{f:LC_class} shows the resulting absolute-magnitude light curves, corrected for Galactic extinction (see Sect. \ref{sec:extinction}) and using the distances and first-light epochs described in Sects. \ref{sec:distances} and \ref{sec:first-light}, respectively. 

\subsection{Light-curve parameters}\label{sec:light.curve.parameters}

The phase of maximum light and peak absolute magnitudes were measured for all the SNe in all available bands. For this purpose, low-degree Legendre polynomials (degree 2--4) were fitted to the photometric data using the \textsc{legendre}\footnote{\url{https://numpy.org/doc/stable/reference/routines.polynomials.legendre.html}} module from the \textsc{numpy} \citep{harris2020numpy} library in \textsc{python}. The initial guess for the time and magnitude at maximum was set to the data point with the highest brightness. In cases where the first photometric data point of the light curve already corresponded to the maximum brightness and the light curve did not display a transition from rising to declining, the corresponding point was treated as an upper limit for the time to peak and, consequently, as a lower limit for the peak apparent brightness. 

The fits were performed via Monte Carlo simulations with $2\,000$\footnote{Throughout this work, different numbers of realisations were used to ensure the convergence of the parameter uncertainties in the Monte Carlo simulations. A larger number of iterations was required in cases with a greater number of fitted parameters.} realisations, in which the magnitudes were randomly perturbed according to a normal distribution with a width equal to their uncertainties. The fitting range was allowed to vary uniformly from 2 to 5 days before and 5 to 8 days after the initial guess. The median and median absolute deviation (MAD) of all realisations were adopted as the final estimates and uncertainties for the time and apparent magnitude at maximum light. The uncertainties\footnote{All uncertainties reported throughout this work are at 1$\sigma$, unless otherwise stated.} in the absolute peak magnitude include the contribution from the distance modulus added in quadrature with the photometric uncertainty of the measurement. The photometric uncertainties of individual measurements are below 0.2 mag, with a mean of 0.03 mag (below 0.5 mag with an average of 0.05 mag in the $u$ band). In turn, distance-modulus uncertainties range from 0.02 to 0.81 mag with a mean of 0.2 mag. Therefore, the uncertainty in the peak magnitude is dominated by the distance error. Tables \ref{t:POISE.peak.magnitude} and \ref{t:lit.peak.magnitude} present the resulting absolute peak magnitudes for the POISE and literature samples, respectively. 

Based on these fits, we computed the time to peak (or rise time) in each band as the rest-frame interval between first light (see Sect.~\ref{sec:first-light}) and the date of maximum. Uncertainties in the time to peak were also calculated by summing the uncertainties in the epochs of first and maximum light in quadrature. The times to peak for all bands of the POISE and literature samples are presented in Tables \ref{t:POISE.rise.times} and \ref{t:lit.rise.times}, respectively.

\begin{figure*}
\centering
\includegraphics[width=.47\textwidth]{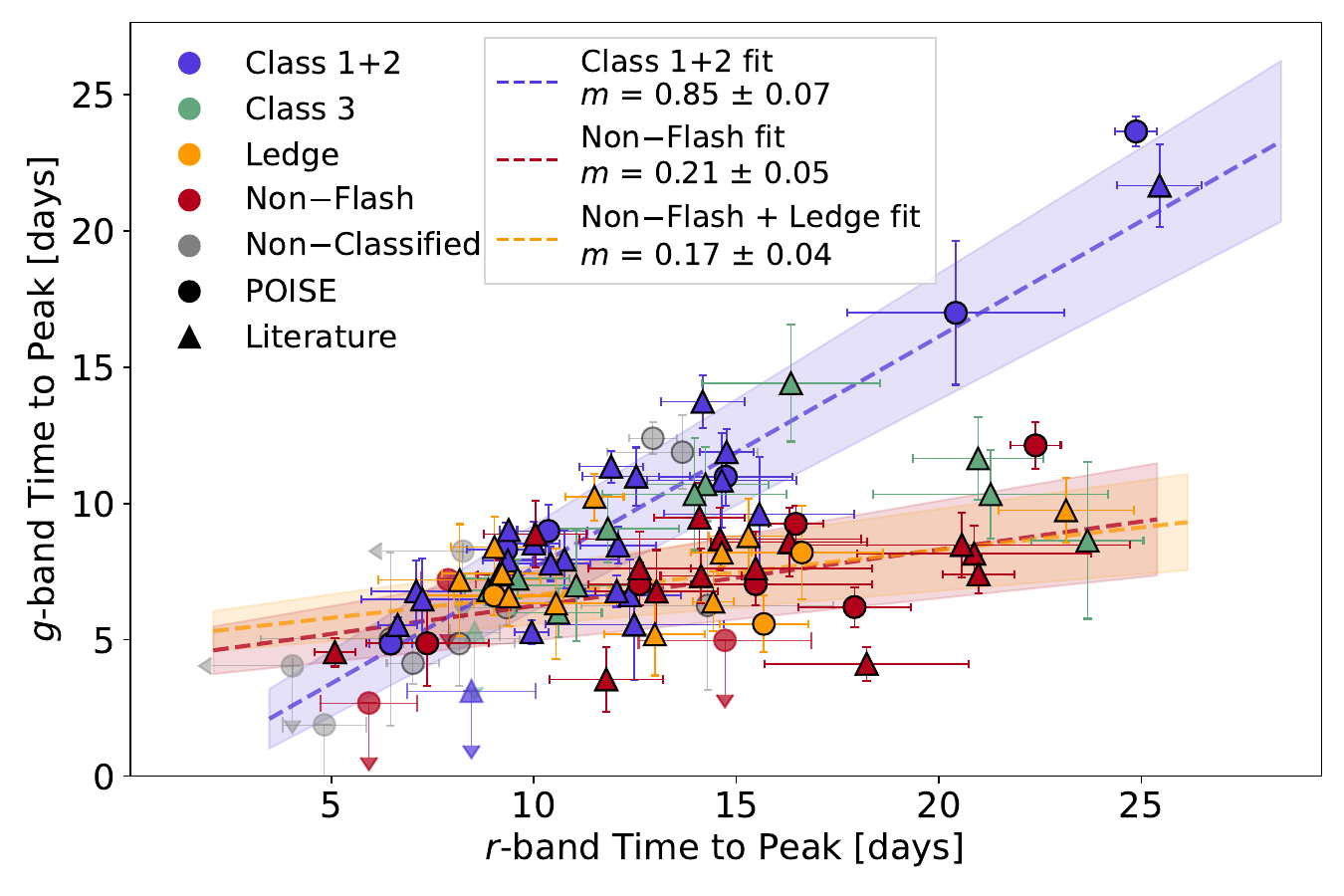}
\includegraphics[width=.47\textwidth]{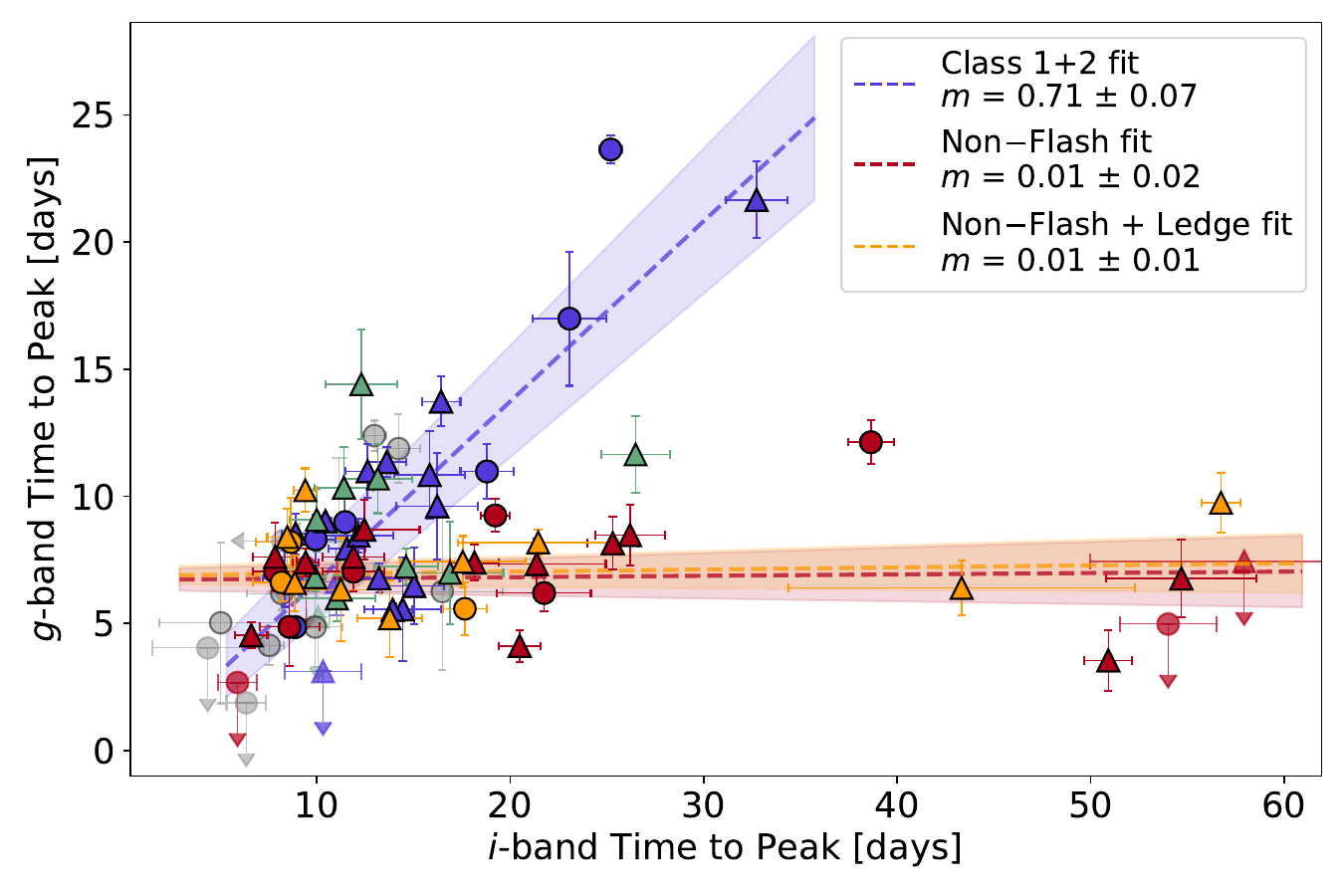}
\caption{Comparison of epoch of maximum light (time to peak) in $g$ band vs. $r$ band (left panel) and $i$ band (right panel). Class 1+2 (blue), class 3 (green), ledge (orange), and non-flash SNe (red) from the POISE (circles) and literature (triangles) samples. Fainter markers without black edges represent SNe with at least one upper limit in either bands. Non-classified SNe (grey) are plotted for completeness. Dashed lines represent separate linear fits obtained via Monte Carlo simulations for class 1+2 (blue), non-flash (red), and non-flash + ledge SNe (orange), with the shaded regions showing the 1$\sigma$ range of the fits. The fitted parameter values are given in each panel.} 
\label{f:delta_t_max}
\end{figure*}

\begin{table*}
\centering
\caption{Weighted mean values and weighted standard deviation of the peak absolute magnitude ($M_\mathrm{{peak}}$) for SNe~II, grouped by the flash-spectroscopic classification and photometric band.}
\resizebox{\textwidth}{!}{
\begin{tabular}{ccccccc}
\hline\hline
\begin{tabular} [c]{@{}c@{}} SNe \\ Class \end{tabular} & \begin{tabular}[c]{@{}c@{}}$\langle M_\mathrm{peak}^u \rangle$ \\ {[}mag{]}\end{tabular} & \begin{tabular}[c]{@{}c@{}}$\langle M_\mathrm{peak}^B \rangle$ \\ {[}mag{]}\end{tabular} & \begin{tabular}[c]{@{}c@{}}$\langle M_\mathrm{peak}^g \rangle$ \\ {[}mag{]}\end{tabular} & \begin{tabular}[c]{@{}c@{}}$\langle M_\mathrm{peak}^V \rangle$ \\ {[}mag{]}\end{tabular} & \begin{tabular}[c]{@{}c@{}}$\langle M_\mathrm{peak}^r \rangle$ \\ {[}mag{]}\end{tabular} & \begin{tabular}[c]{@{}c@{}}$\langle M_\mathrm{peak}^i \rangle$ \\ {[}mag{]}\end{tabular}  \\ \hline
Class 1+2  & $-18.30 \pm 0.42\, (13)$ & $-18.27 \pm 0.35 \, (12)$ & $-18.31 \pm 0.33 \, (26)$ & $-18.28 \pm 0.30 \, (17)$ & $-18.15 \pm 0.29 \, (27)$ & $-18.01 \pm 0.30 \, (21)$ \\
Class 3    & $-16.92 \pm 1.09\, (4 )$ & $-17.39 \pm 0.64 \, (7 )$ & $-17.87 \pm 0.77 \, (11)$ & $-17.46 \pm 0.45 \, (8 )$ & $-17.86 \pm 0.67 \, (12)$ & $-17.79 \pm 0.56 \, (10)$ \\
Ledge      & $-17.11 \pm 0.58\, (11)$ & $-17.09 \pm 0.52 \, (11)$ & $-17.30 \pm 0.58 \, (14)$ & $-17.07 \pm 0.49 \, (13)$ & $-17.18 \pm 0.47 \, (14)$ & $-17.07 \pm 0.46 \, (10)$ \\
Non-flash  & $-16.96 \pm 0.80\, (8)$  & $-17.18 \pm 0.86 \, (11)$ & $-17.18 \pm 0.74 \, (18)$ & $-17.08 \pm 0.78 \, (15)$ & $-17.06 \pm 0.82 \, (21)$ & $-16.85 \pm 0.85 \, (19)$ \\
POISE      & $-17.38 \pm 1.14\, (16)$ & $-17.49 \pm 0.95 \, (20)$ & $-17.67 \pm 0.93 \, (21)$ & $-17.59 \pm 0.89 \, (25)$ & $-17.53 \pm 0.92 \, (25)$ & $-17.47 \pm 0.87 \, (26)$ \\
All        & $-17.34 \pm 0.90\, (40)$ & $-17.36 \pm 0.81 \, (47)$ & $-17.73 \pm 0.80 \, (76)$ & $-17.53 \pm 0.80 \, (62)$ & $-17.62 \pm 0.75 \, (82)$ & $-17.47 \pm 0.76 \, (69)$\\\hline

\end{tabular}
}
\tablefoot{Numbers in brackets indicate the number of SNe in each group. `All' also includes Non-classified SNe.}
\label{t:mean_peak_magnitude}
\end{table*}

\begin{table*}
\centering
\caption{Weighted mean values and weighted standard deviation of the time to peak ($t_{\mathrm{peak}}$) for SNe~II, grouped by the flash-spectroscopic classification and photometric band.}
\begin{tabular}{ccccccc}
\hline\hline
\begin{tabular}[c]{@{}c@{}} SNe \\ Class \end{tabular} & \begin{tabular}[c]{@{}c@{}}$\langle t_\mathrm{peak}^u \rangle$ \\ {[}days{]}\end{tabular} & \begin{tabular}[c]{@{}c@{}}$\langle t_\mathrm{peak}^B \rangle$ \\ {[}days{]}\end{tabular} & \begin{tabular}[c]{@{}c@{}}$\langle t_\mathrm{peak}^g \rangle$ \\ {[}days{]}\end{tabular} & \begin{tabular}[c]{@{}c@{}}$\langle t_\mathrm{peak}^V \rangle$ \\ {[}days{]}\end{tabular} & \begin{tabular}[c]{@{}c@{}}$\langle t_\mathrm{peak}^r \rangle$ \\ {[}days{]}\end{tabular} & \begin{tabular}[c]{@{}c@{}}$\langle t_\mathrm{peak}^i \rangle$ \\ {[}days{]}\end{tabular}  \\ \hline
Class 1+2  & $5.86 \pm 1.69 \, (13)$ & $5.81 \pm 3.13 \, (12)$ & $8.44 \pm 3.99 \,(26)$ & $8.35 \pm 2.55 \,(17)$ & $10.88 \pm 4.57\, (27)$ & $12.90 \pm 5.44 \,(21)$ \\
Class 3    & $5.46 \pm 1.63 \, (4) $ & $7.13 \pm 2.48 \, (7) $ & $7.85 \pm 1.99 \,(11)$ & $8.17 \pm 2.39 \,(8) $ & $11.14 \pm 4.54\, (12)$ & $11.15 \pm 3.41 \, (10)$ \\
Ledge      & $5.23 \pm 1.50 \, (11)$ & $7.00 \pm 1.60 \, (11)$ & $7.72 \pm 1.41 \,(14)$ & $8.15 \pm 1.41 \,(13)$ & $13.34 \pm 2.92\, (14)$ & $10.57 \pm 3.33 \, (10)$ \\
Non-flash  & $4.90 \pm 1.29 \, (8) $ & $6.66 \pm 1.64 \, (11)$ & $7.02 \pm 2.19 \,(18)$ & $7.92 \pm 2.85 \,(15)$ & $12.33 \pm 6.02\, (21)$ & $17.58 \pm 13.15\, (19)$ \\
POISE      & $5.46 \pm 2.64 \, (16)$ & $8.35 \pm 5.41 \, (20)$ & $9.03 \pm 5.76 \,(21)$ & $8.38 \pm 3.91 \,(25)$ & $13.20 \pm 6.77\, (25)$ & $13.78 \pm 8.16 \, (26)$ \\
All        & $5.75 \pm 2.02 \, (40)$ & $6.10 \pm 2.98 \, (47)$ & $8.06 \pm 3.39 \,(76)$ & $8.28 \pm 2.58 \,(62)$ & $11.38 \pm 4.68\, (82)$ & $13.01 \pm 7.35\, (69)$\\\hline
\end{tabular}
\tablefoot{Numbers in brackets indicate the number of SNe in each group. `All' also includes Non-classified SNe.}
\label{t:mean_peak_time}
\end{table*}

\subsection{Rise-time and peak-magnitude distributions}\label{sec:rise.distributions}

Figure \ref{f:rise_times_vs_peak_mag} presents the peak absolute magnitude vs. rise time for all available bands, distinguishing SNe~II by their flash-spectroscopic classification defined in Sect.~\ref{sec:classification}. The corresponding weighted mean values of both quantities for each class are shown using ellipses in the figure and are summarised in Tables \ref{t:mean_peak_magnitude} and \ref{t:mean_peak_time}, respectively.  

A clear difference in peak absolute magnitude is observed among the flash-spectroscopic classes across all bands: class 1+2 SNe are, on average, more luminous than the other groups (see Table~\ref{t:mean_peak_magnitude}), with peak absolute magnitudes typically brighter than $\approx$ $-17.5$ mag. In contrast, ledge and non-flash objects are generally fainter, exhibiting a broader luminosity range spanning from $-14.5$ to $-18$ mag. Class 3 shows no pronounced trend and lies among the other classes. \cite{JG2024} report similar patters, although in their work, class 3 SNe resemble class 1+2 more closely than SNe without flash features. The faintest events, fainter than $-15$ mag, correspond to SN~2016aqf (in the \textit{i} band, non-flash), SN~2019enr (non-flash), SN~2020fqv (in the \textit{u} band, class 3), SN~2021adxd (in the \textit{uBgV} bands, non-classified), SN~2022jzc (in the \textit{gVri} bands, non-flash), and SN~2022acko (ledge).

Figure \ref{f:CDFs.LC.all.bands} (second and fourth columns) illustrates the cumulative distribution functions (CDFs) of the peak absolute magnitude in all bands. Each panel shows the corresponding p values from the K-S and A-D tests for the comparison cases (a) class 1+2 vs. non-flash and (b) flash vs. non-flash + ledge (see Sect.~\ref{sec:stats.methods} for details). These p values are also listed in Table~\ref{t:KS_AD_peak_m_t}. In all bands and for both comparison cases, the tests reveal statistically significant differences, with p$^\mathrm{KS}<10^{-4}$ and p$^\mathrm{AD}<0.01$. The most significant difference is found in the $i$ band for both comparison cases. 

We believe that the distinction in peak absolute magnitudes found here cannot be explained by differences in (uncorrected for) dust extinction. On the one hand, dust extinction is stronger in the bluer bands, and so it would produce a larger effect there, which is not seen. On the other hand, such a systematic difference in dust extinction would be at odds with the similar distributions of Na I D equivalent widths that \cite{JG2024} report for SNe with and without flash features. In our case, for example, the distribution of flash-spectroscopic classes among the SNe highly reddened by their host galaxy (see Sects.~\ref{sec:extinction} and \ref{sec:literature.sample}) is quite balanced, with four class 1+2 (SN~2014G, SN~2019ust, SN~2021aek, and SN~2022pgf), three class 3 (SN~2017gmr, SN~2020fqv, and SN~2021jtt), one ledge (SN~2023cps), and five non-flash SNe (SN~2018cuf, SN~2021bmw, SN~2022jzc, SN~2023bvj, and SN~2023dtd). Therefore, the observed differences are most likely driven by intrinsic properties correlated with the presence of flash features rather than by host-galaxy extinction. 

Regarding the rise times, the variation among classes is less pronounced than that observed for the peak magnitudes. In the bluer bands (i.e. \textit{uBgV}), class 1+2 SNe show, on average, slightly longer rise times than non-flash SNe, whereas in the redder bands (i.e. $r$ and $i$) this trend reverses (see Table~\ref{t:mean_peak_time}). Across all bands, class 3 SNe span the full range of values, and in the $i$ band, they closely resemble the behaviour of the class 1+2 SNe. In the $B$, $g$, $V$, and $r$ bands, we identify three class 1+2 SNe (SN~2022ffg, SN~2022prv, and SN~2023bvh) that exhibit the longest times to peak, exceeding $\approx15$ days. A possible interpretation of this behaviour is discussed in Sect. \ref{sec:disc.models}. Nonetheless, in the $V$ band there are three SNe from the other classes that also exceed 15 days in rise time: SN~2019pjs (ledge), SN~2020nif (class 3), and SN~2021adlr (non-flash). Conversely, non-flash SNe show a narrower range of rise times, typically shorter than 10 days in \textit{uBg} and 15 days in $V$, while in the redder bands the range broadens up to approximately 25 days. Furthermore, four non-flash SNe (SN~2016aqf, SN~2020mjm, SN~2022jzc, and SN~2023bvj) exhibit extremely long times to peak in the $i$ band, exceeding 45 days. 

Figure \ref{f:CDFs.LC.all.bands} (first and third columns) presents the CDFs for time to peak in all bands. The resulting p values of the comparison cases a and b are shown in each panel and summarised in Table~\ref{t:KS_AD_peak_m_t}. We find statistically significant differences in the distributions only for the $B$ band in case (b). The low p value indicates that this distinction is unlikely to arise by chance, as it remains below the significance threshold after correcting for multiple testing. 
To our knowledge, this behaviour has not been reported in previous analyses of SNe~II. 

Interestingly while ledge SNe resemble non-flash SNe more closely in the bluer bands, in the redder bands their behaviour becomes more similar to that of class 1+2 SNe, as can be seen in Fig.~\ref{f:rise_times_vs_peak_mag} and in the corresponding CDFs (see Fig.~\ref{f:CDFs.LC.all.bands}). Moreover, the p values are higher in the redder bands when ledge SNe are grouped together with the non-flash SNe. Owing to the limited size of the class 3 sample, we cannot derive definitive conclusions for this subclass. 

\begin{table}
\centering
\caption{K-S and A-D tests' p values for the comparison cases a and b of the time to peak ($t_\mathrm{peak}$) and peak absolute magnitude ($M_\mathrm{peak}$). }
\resizebox{\columnwidth}{!}{
\begin{tabular}{cccccc}
\hline\hline
Parameter & Band & p$_a^\mathrm{KS}$ & p$_b^\mathrm{KS}$ & p$_a^\mathrm{AD}$ & p$_b^\mathrm{AD}$ \\ \hline
\multirow{6}{*}{$M_\mathrm{peak}$} & $u$ & $\mathbf{9.8\,\times\,10^{-6}}$ & $\mathbf{6.8\,\times\,10^{-5}}$    &  $\mathbf{< 0.01}$     &  $\mathbf{< 0.01}$  \\
                                   & $B$ & $\mathbf{3.4\,\times\,10^{-5}}$ & $\mathbf{3\,\times\,10^{-5}}$      &  $\mathbf{< 0.01}$     &  $\mathbf{< 0.01}$  \\
                                   & $g$ & $\mathbf{1.5\,\times\,10^{-7}}$ & $\mathbf{4.4\,\times\,10^{-7}}$    &  $\mathbf{< 0.01}$     &  $\mathbf{< 0.01}$  \\
                                   & $V$ & $\mathbf{1.1\,\times\,10^{-7}}$ & $\mathbf{1.6\,\times\,10^{-6}}$    &  $\mathbf{< 0.01}$     &  $\mathbf{< 0.01}$  \\
                                   & $r$ & $\mathbf{1.1\,\times\,10^{-7}}$ & $\mathbf{5.9\,\times\,10^{-7}}$    &  $\mathbf{< 0.01}$     &  $\mathbf{< 0.01}$  \\
                                   & $i$ & $\mathbf{6.1\,\times\,10^{-10}}$& $\mathbf{5.1\,\times\,10^{-11}}$   &  $\mathbf{< 0.01}$     &  $\mathbf{< 0.01}$  \\\hline
\multirow{6}{*}{$t_\mathrm{peak}$} & $u$ & $0.32$                          & $0.23$                             &  $\geq 0.25$  &  $\geq 0.25$  \\
                                   & $B$ & $0.02$                          & $\mathbf{2\,\times\,10^{-3}}$      &  $< 0.025$    &  $\mathbf{< 0.01}$ \\
                                   & $g$ & $0.16$                          & $0.02$                             &  $\geq 0.05$  &  $< 0.025$  \\
                                   & $V$ & $0.18$                          & $0.05$                             &  $\geq 0.25$  &  $\geq 0.05$  \\
                                   & $r$ & $0.06$                          & $0.19$                             &  $\geq 0.25$   &  $\geq 0.25$  \\
                                   & $i$ & $0.03$                          & $0.03$                             &  $< 0.025$    &  $< 0.025$  \\\hline
\end{tabular}}
\tablefoot{Most significant distinctions (p$^\mathrm{KS}< 5\times10^{-3}$ and p$^\mathrm{AD}< 0.01$) are highlighted in bold.}
\label{t:KS_AD_peak_m_t}
\end{table}

\begin{figure*}
\centering
\includegraphics[width=.9\textwidth]{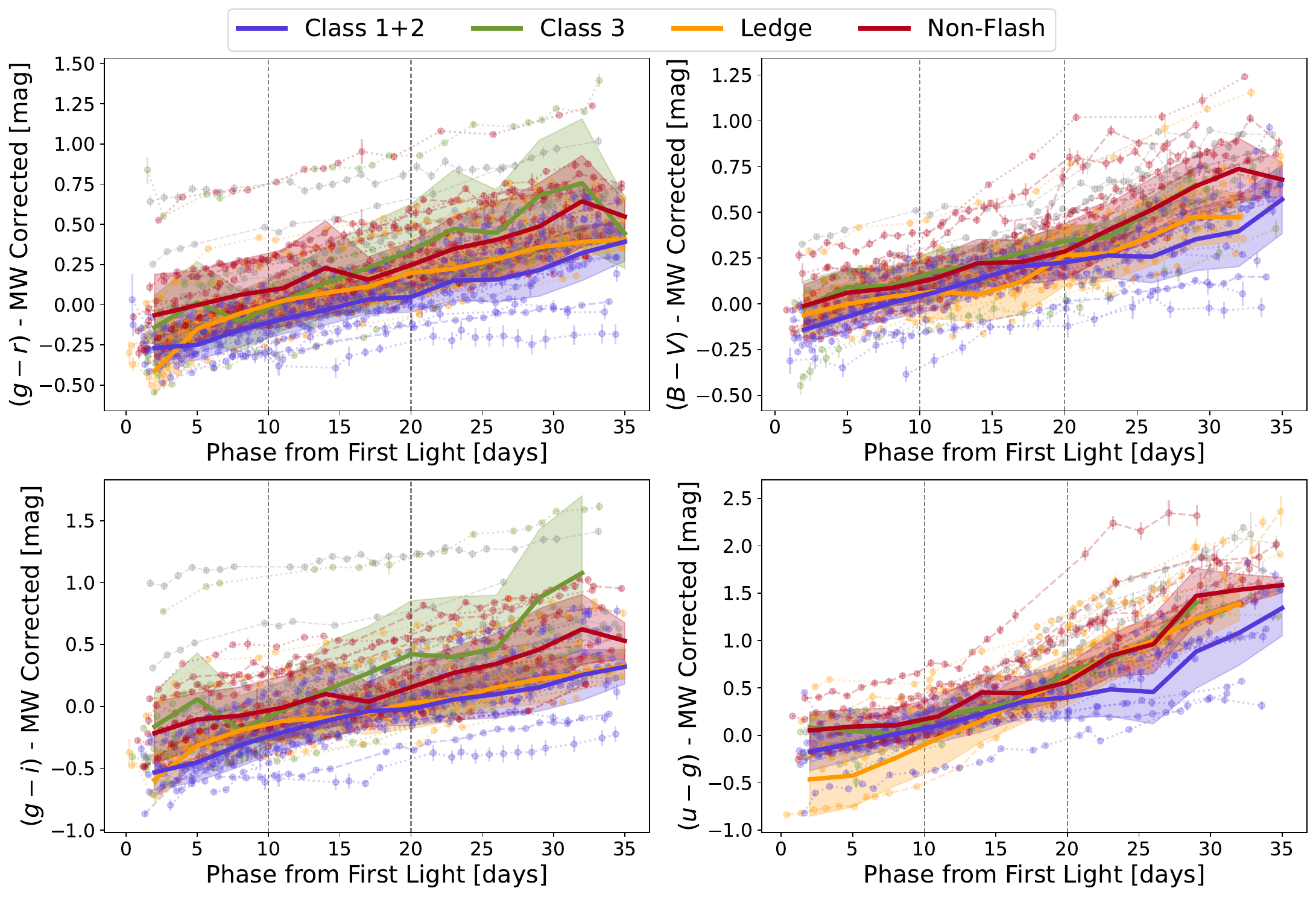}
\caption{From left to right, top to bottom: Early-time, MW-extinction-corrected (\textit{$g-r$}), (\textit{$B-V$}), (\textit{$g-i$}), and (\textit{$u-g$}) colour curves for class 1+2 (blue), class 3 (green), ledge (orange), and non-flash SNe (red). Non-classified SNe (grey) are included for completeness. The data points for each SN are connected by dashed lines for the POISE sample and by dotted lines for the literature sample. The solid lines show the weighted mean colour curve for each group, with the shaded regions indicating the 1$\sigma$ uncertainty, accounting for both the dispersion and measurement uncertainties. Vertical lines mark the phases of 10 and 20 days. Class 1+2 SNe are, on average, bluer than non-flash SNe in all colours, while class 3 and ledge are spread between the two groups.}
\label{f:colour_curves}
\end{figure*}

Figure \ref{f:delta_t_max} compares the time to peak between the $g$ and $r$ bands (left panel) and between the $g$ and $i$ bands (right panel) for all SNe. We see a clear separation between the class 1+2 and non-flash SNe, as well as between class 1+2 and non-flash + ledge SNe. The class 1+2 exhibit an almost linear correlation between rise times in redder and bluer bands. This indicates that the time to peak increases nearly proportionally across wavelengths, implying a weak dependence on the effective wavelength. In contrast, non-flash and ledge SNe display a much flatter relation. Their rise times in the bluer bands vary slightly, whereas those in the redder bands extend substantially. This delay of the maximum at longer wavelengths is suggestive of a post-shock cooling phase dominating the light-curve evolution of non-flash and ledge SNe. The behaviour of class 1+2 objects may be interpreted as resulting from a stronger influence of CSI (see Sect.~\ref{sec:disc.models} for further discussion). Finally, class 3 SNe do not show a consistent trend and are distributed among all the other classes. 

\section{Colour-curve properties} 
\label{sec:colour.curves}

In this section, we analyse the (\textit{$g-r$}), (\textit{$B-V$}), (\textit{$g-i$}), and (\textit{$u-g$}) colour curves. Colours were computed by subtracting the MW-extinction-corrected magnitudes in the two bands whenever the observations were separated from each other by less than 0.5 days. This was the case for most of the POISE sample, except for the (\textit{$u-g$}) colour, given the more limited coverage in the $u$ band. Whenever one of the bands lacked data to directly compute a colour, interpolations were performed based on Legendre-polynomial fits to the light curves (mostly in (\textit{$u-g$}) and for the literature sample). For this purpose, we employed the \texttt{legendre} module from \texttt{numpy}. We further required a maximum gap of 7 days around the interpolation epoch. The fits were performed using data points in the range of $\pm6$ days around the interpolation epoch, under the requirement of having a minimum of two data points on each side. To determine the associated uncertainties, Monte Carlo simulations were employed, with $1\,000$ realisations where the polynomial degree was randomly set to 2, 3, or 4\footnote{This choice spans the main polynomial degrees that adequately describe the colour curves while avoiding overfitting.}, and the magnitudes were varied within their uncertainties, adopting a normal distribution. 

Figure \ref{f:colour_curves} presents the early colour curves of the SNe~II in (\textit{$g-r$}), (\textit{$B-V$}), (\textit{$g-i$}), and (\textit{$u-g$}), corrected for time dilation and MW reddening. As also noted by \cite{JG2024} in (\textit{$g-r$}), class 1+2 SNe are, in general, bluer than non-flash objects at early phases. We observe the same trend across the remaining colours. As discussed in Sect.~\ref{sec:rise.distributions}, host-galaxy extinction is unlikely to be the primary driver of the observed differences in magnitudes and, consequently, in colours. This interpretation is further supported by the fact that, as we show in Sect.~\ref{sec:slopes.correlations}, the difference in colours between SNe of different classes is not constant in time, as would be expected from a difference in extinction. The colour differences are therefore most likely intrinsic and governed by physical conditions related to the appearance of flash features. 

Furthermore, the colour curves of class 3 and ledge SNe span a wide range of values and cannot be directly associated with either the class 1+2 or the non-flash groups. In Sect.~\ref{sec:colour.phases}, we further examine the colours evaluated at specific phases. The overall colour evolution is analysed in detail in Sect.~\ref{sec:slopes.correlations}. 

\begin{figure}
\centering
\includegraphics[width=.47\textwidth]{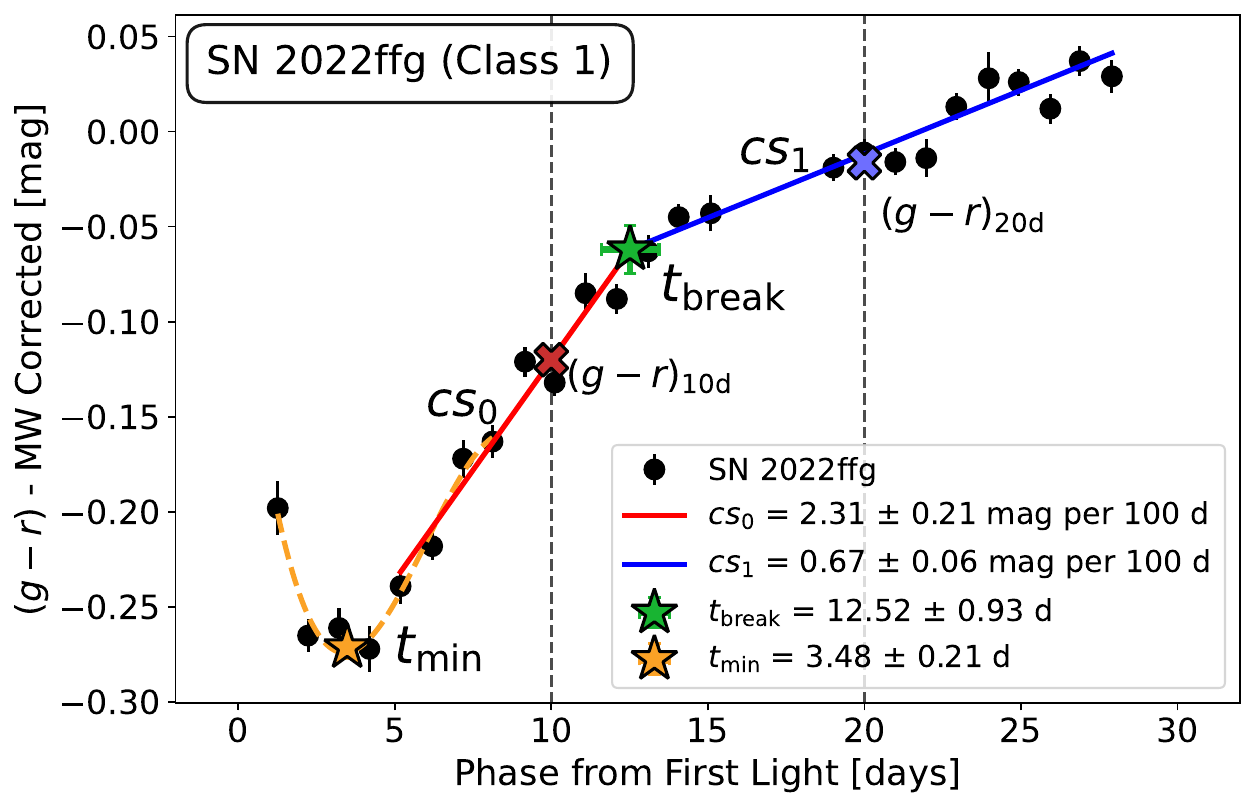}
\caption{Example of colour-curve parameters measured in (\textit{g-r}) for SN~2022ffg (class 1, i.e. showing intense emission lines of He \textsc{ii} and N \textsc{iii} in early spectroscopy). The colour-curve minimum ($t_\mathrm{min}$, yellow star) is shown together with an example of the Legendre-polynomial fit (dashed yellow line). The (\textit{$g-r$}) colours at 10 days (red cross) and 20 days (blue cross) are also shown. The median colour-curve slopes $cs_0$ and $cs_1$ from the Monte Carlo realisations are represented in red and blue, respectively. The green star corresponds to the phase and colour of the break ($t_\mathrm{break}$) with its uncertainties. The $cs_0$ fit starts after the colour minimum. This SN exhibits a flattening after the break, a behaviour typical of class 1+2 SNe. 
}
\label{f:slopes_example}
\end{figure}

At the earliest phases, some SNe~II (primarily the class 1+2 SNe, along with a few class 3 and ledge objects) exhibit an initial evolution towards bluer colours for several days, reaching a minimum in their colour curves before transitioning towards redder colours. This characteristic has previously been reported in radiation-hydrodynamic models of interacting SNe~II \citep[e.g.][]{Moriya2018}. In Sect.~\ref{sec:disc.models}, we further discuss the interpretation and possible physical scenarios underlying this early behaviour. In contrast, non-flash-SNe and most ledge SNe evolve directly towards redder colours without showing such a turn at early phases, even among the earliest observed SNe. Nonetheless, within the first $\approx5$ days, we also identify an additional break in the (\textit{$g-r$}), (\textit{$B-V$}), and (\textit{$g-i$}) colour curves for $\approx$ 30 SNe, predominantly from the ledge class, with a few events from the class 3 and non-flash groups. After this break, the colour evolution becomes shallower for all these events. However, photometric coverage within a few days after first light is scarce, preventing us from performing a full analysis of the colour behaviour at this stage. 

Beyond approximately five days after first light, all SNe~II follow the same general trend, evolving continuously towards redder colours in a nearly linear fashion. For many objects, however, we notice a break in the colour evolution occurring at some time between five and 30 days. This is similar to what \cite{dejaeger2018} report at epochs later than $\approx$ 30 days. We observe that the curves can become either steeper or shallower after the break, whereas \cite{dejaeger2018} find that they always become shallower post-break at $\gtrsim$30 days (see Sect.~\ref{sec:disc.colours} for further comparison). The change in slopes can be rather subtle in some cases. Therefore, we characterise these changes using piecewise linear fits as described in Sect. \ref{sec:colour.curves.parameters}.

\subsection{Colour-curve parameters}
\label{sec:colour.curves.parameters}

Figure \ref{f:slopes_example} shows the (\textit{$g-r$}) colour curve of SN 2022ffg, together with the colour-curve parameters adopted in this work. To characterise the behaviour of the colour curves exhibiting breaks at early phases (see Sect.~\ref{sec:colour.curves}), we describe them using two linear segments, with the break defined as the point connecting both. Since for the majority of the SNe the break occurs between 10 and 20 days, the colour at these phases was also measured (as explained below). Consequently, the following parameters were defined to quantify the colour-curve behaviours:

\begin{itemize}
\item (colour)$_\mathrm{10d}$: the colour corrected for MW extinction at 10 days after first light.

\item (colour)$_\mathrm{20d}$: the colour corrected for MW extinction at 20 days after first light.

\item $t_\mathrm{min}$: the phase of the minimum in the colour curve in days, indicating the turn from bluer to redder evolution. 
    
\item $cs_{0,\,\mathrm{colour}}$: the colour-curve slope measured from either the colour minimum or 5 days after first light (whichever occurs later) until the break. 

\item $cs_{1,\,\mathrm{colour}}$: the post-break colour-curve slope, measured from the break to either the last observation or 30 days after first light (whichever comes first).

\item $\Delta cs_\mathrm{colour}$: the colour-curve slope difference between $cs_0$ and $cs_1$, i.e. $\Delta cs_\mathrm{colour} = cs_{0,\,\mathrm{colour}} - cs_{1\,,\mathrm{colour}}$.

\item $t_{\mathrm{break}}$: the phase of the break in days, indicating the transition between $cs_{0,\,\mathrm{colour}}$ and $cs_{1,\,\mathrm{colour}}$ in the colour curve. 
\end{itemize}

\noindent All colour-curve slopes and slope differences are expressed in magnitudes per 100 days.

The colour values at different phases and $t_\mathrm{min}$ were estimated from the colour curves using the same fitting technique as that applied to obtain light-curve peak times and magnitudes (see Sect.~\ref{sec:light.curve.parameters}). Uncertainties were estimated performing $2\,000$ Monte Carlo simulations. For the colours at 10 and 20 days, the fitting range in each realisation was varied uniformly between 3 and 5 days before and after the corresponding phase, while the polynomial degree was varied between 2 and 4. The resulting (\textit{$g-r$}) values for the POISE and literature samples are presented in Tables~\ref{t:POISE.slopes_gr} and \ref{t:lit.slopes_gr}, respectively. The values for the remaining colours are available in machine-readable form (see Sect.~\ref{sec:data_availability}).  

In the case of the $t_\mathrm{min}$ measurement, the bluest data point was adopted as the initial guess and the fitting interval was restricted to the range from the first available observation up to 6-8 days (varying uniformly in each realisation) after this initial estimate. Since the minimum in the colour curves is not always detected, primarily due to the lack of sufficiently early-time observations, the measurement of $t_\mathrm{min}$ was performed only for the (\textit{$g-r$}) colour, which provides the largest number of suitable cases. In our sample, this parameter was estimated for 12 SNe (11 from class 1+2 and one from class 3), with the resulting values listed in Table~\ref{t:tIIn.tmin}. 

We note that even if in some cases the colour curves could be equally well described by a single linear function between 5 and 30 days, with no clear break, we still used the two-segment fits to derive the parameters defined above. In these cases, the break times are not well defined and therefore they have large associated uncertainties. However, the slopes are not affected. 

To estimate these parameters, we fitted piecewise linear functions with one break using the \textsc{Python} library \texttt{pwlf}\footnote{\url{https://github.com/cjekel/piecewise_linear_fit_py/blob/master/pwlf/pwlf.py}}. To avoid unreliable measurements, the following minimum data requirements were imposed for the colour curves: (i) at least seven data points covering the fitted range, from 5 days (or the first data point fitted) until 30 days from first light; (ii) gaps shorter than 8 days between observations; (iii) at least three data points between the first fitted epoch and 15 days from first light; and (iv) a minimum of three data points per segment (i.e. for each slope).

The estimation of uncertainties was carried out by performing $5\,000$ Monte Carlo realisations, varying the colour values randomly following a normal distribution within their observational uncertainties, and varying the phases uniformly based on the uncertainties of the first-light epoch. This variation led to differences in the selected phase range and number of data points per realisation, particularly for SNe with larger uncertainties in their first-light epoch. Finally, for each parameter, we adopted the median and MAD of all realisations as the characteristic values and their uncertainty, respectively.

To ensure robust slope measurements, a `gold' sample was defined according to the quality of the colour curves and their fitting. Based on the characteristics of the POISE photometry, the following more stringent criteria were adopted: (i) at least ten (eight for (\textit{$u-g$})) data points in the entire fitted range; (ii) gaps between observations shorter than 5 days ($\leq$ 6 days for (\textit{$u-g$})); and (iii) each segment with four or more data points (not applied to (\textit{$u-g$})). In addition, to be included in the gold sample, the fits were required to have a root mean square (RMS) deviation $\leq 0.04$ mag ($\leq 0.06$ mag for (\textit{$u-g$})). Fits with RMS $\geq 0.055$ mag ($\geq 0.075$ mag for (\textit{$u-g$})) were discarded from the analysis. The remaining fits constitute the `complete' sample, of which the gold sample is a subset. These thresholds were derived from the typical RMS values of the POISE sample. We note that this division is performed for each colour separately. Therefore, the same SN may belong to the gold sample in one colour and to the complete sample (or even be discarded) in another. Tables~\ref{t:POISE.slopes_gr} and \ref{t:lit.slopes_gr} list the resulting (\textit{$g-r$}) colour-curve parameters for the POISE and literature samples, respectively, and the tables for the remaining colours are in machine-readable form (see Sect.~\ref{sec:data_availability}).
\begin{figure}
\centering
\includegraphics[width=.47\textwidth]{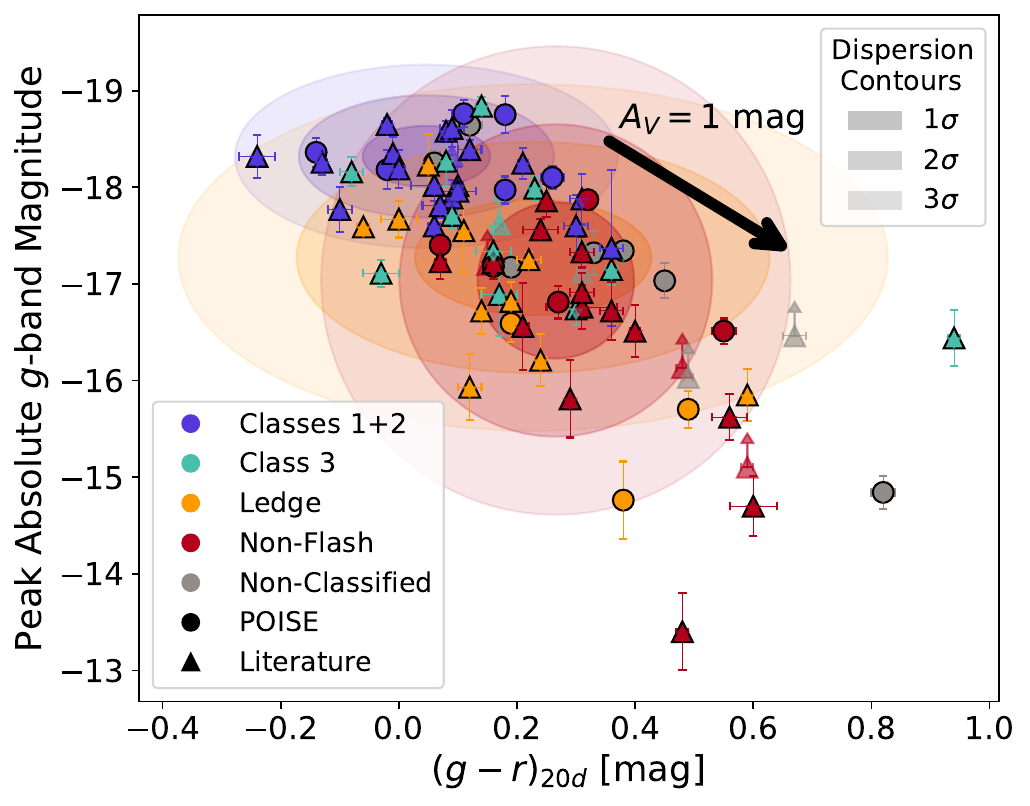}
\caption{(\textit{$g-r$}) colour at 20 days after first light vs. peak absolute $g$-band magnitude. Class 1+2 (blue), class 3 (green), ledge (orange), and non-flash SNe (red) are shown for the POISE (circles) and the literature (triangles) samples. Fainter markers without black edges represent SNe with upper limits on their peak absolute magnitudes. Non-classified SNe (grey) are plotted for completeness. The shaded ellipses indicate the weighted mean values of both parameters for each group and their 1$\sigma$, 2$\sigma$, and 3$\sigma$ regions (from darker to lighter tones) for class 1+2 (blue), ledge (orange), and non-flash SNe (red). Class 1+2 SNe are, on average, more luminous and bluer than non-flash SNe, while class 3 and ledge SNe lie between the two groups in both parameters. The black arrow shows the reddening vector for $A_V=1$ mag, adopting the extinction law of \cite{Schlafly2011} with $R_V =3.1$.
}
\label{f:colour.phases}
\end{figure}

\begin{table}
\centering
\caption{K-S and A-D tests' p values for the comparison cases a and b of the colours at 10 and 20 days from the first light.}
\resizebox{\columnwidth}{!}{%
\begin{tabular}{cccccc}
\hline\hline
Colour & Phase & p$_a^\mathrm{KS}$ & p$_b^\mathrm{KS}$ & p$_a^\mathrm{AD}$ & p$_b^\mathrm{AD}$ \\ \hline
\multirow{2}{*}{(\textit{$g-r$})}  & 10 d & $\mathbf{1.9\,\times\,10^{-4}}$ & $\mathbf{1.8\,\times\,10^{-3}}$  &  $\mathbf{< 0.01}$     &  $\mathbf{< 0.01}$  \\
                                   & 20 d & $\mathbf{2.3\,\times\,10^{-5}}$ & $\mathbf{5.5\,\times\,10^{-4}}$  &  $\mathbf{< 0.01}$     &  $\mathbf{< 0.01}$ \\
\multirow{2}{*}{(\textit{$B-V$})}  & 10 d & $0.13$                          & $0.13$                           &  $\geq 0.1$            &  $\geq 0.1$  \\
                                   & 20 d & $\mathbf{1.9\,\times\,10^{-3}}$ & $\mathbf{1\,\times\,10^{-4}}$    &  $\geq 0.05$           &  $\geq 0.05$  \\
\multirow{2}{*}{(\textit{$g-i$})}  & 10 d & $9.6\,\times\,10^{-3}$          & $0.12$                           &  $\mathbf{< 0.01}$     &  $\mathbf{< 0.01}$  \\
                                   & 20 d & $0.05$                          & $0.12$                           &  $< 0.05$              &  $\geq 0.05$  \\
\multirow{2}{*}{(\textit{$u-g$})}  & 10 d & $0.13$                          & $0.38$                           &  $\geq 0.1$            &  $\geq 0.25$  \\
                                   & 20 d & $0.01$                          & $\mathbf{7.2\,\times\,10^{-4}}$  &  $\mathbf{< 0.01}$     &  $\mathbf{< 0.01}$  \\\hline
\end{tabular}}
\tablefoot{Most significant distinctions (p$^\mathrm{KS}< 5\times10^{-3}$ and p$^\mathrm{AD}< 0.01$) are highlighted in bold.}
\label{t:pvalues_KS_AD_colour_phases}
\end{table}

\subsection{Colours at specific phases}\label{sec:colour.phases}

First we analyse the colours at 10 and 20 days, since, as mentioned in Sect.~\ref{sec:colour.curves.parameters}, the break in the colour curves typically occurs between these phases. As described in Sect.~\ref{sec:stats.methods}, K-S and A-D tests were performed for the comparison cases a and b. Figure \ref{f:CDF_10d_20d} shows the CDFs of both phases for all colours and includes the p values of both tests in each panel, which are also listed in Table \ref{t:pvalues_KS_AD_colour_phases}.

At 10 days, statistically significant differences are found only in the (\textit{$g-r$}) colour in both K-S tests. Although the A-D tests for the (\textit{$g-i$}) colour also yield p$\,<0.01$, which may indicate differences between the samples, the K-S tests fail to reject the null hypothesis. Notably, for the (\textit{$u-g$}) colour, the tests yield the highest p values. This is consistent with the observed similarity in the early-time colour curves among all classes in (\textit{$u-g$}) (see Fig.~\ref{f:colour_curves}, right panel).

On the other hand, the p$^\mathrm{KS}$ values at 20 days for (\textit{$g-r$}), (\textit{$g-i$}), and (\textit{$u-g$}) colours generally suggest differences between the parent distributions, except in (\textit{$B-V$}) for the A-D test and in (\textit{$g-i$}) for the K-S test, both for case (b). In most instances, the p values at 20 days are lower than those at 10 days, indicating a more pronounced distinction among the populations. This behaviour is particularly evident in the (\textit{$B-V$}) and (\textit{$u-g$}) colours (see Fig.~\ref{f:colour_curves}), with clear separations between the groups at 20 days. Nevertheless the distinctions in colours at 10 days and 20 days are, in general, weaker than those found in Sect.~\ref{sec:rise.distributions} for the peak absolute magnitudes.

\begin{figure}
\centering
\includegraphics[width=.85\columnwidth]{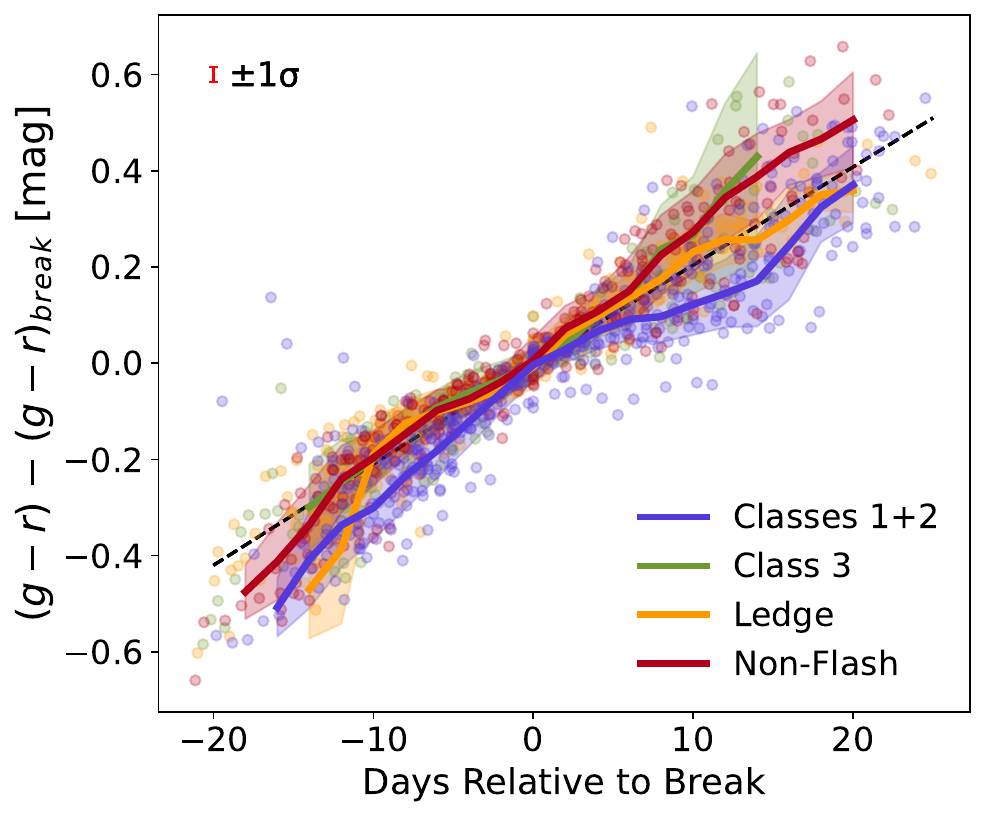}
\includegraphics[width=.825\columnwidth]{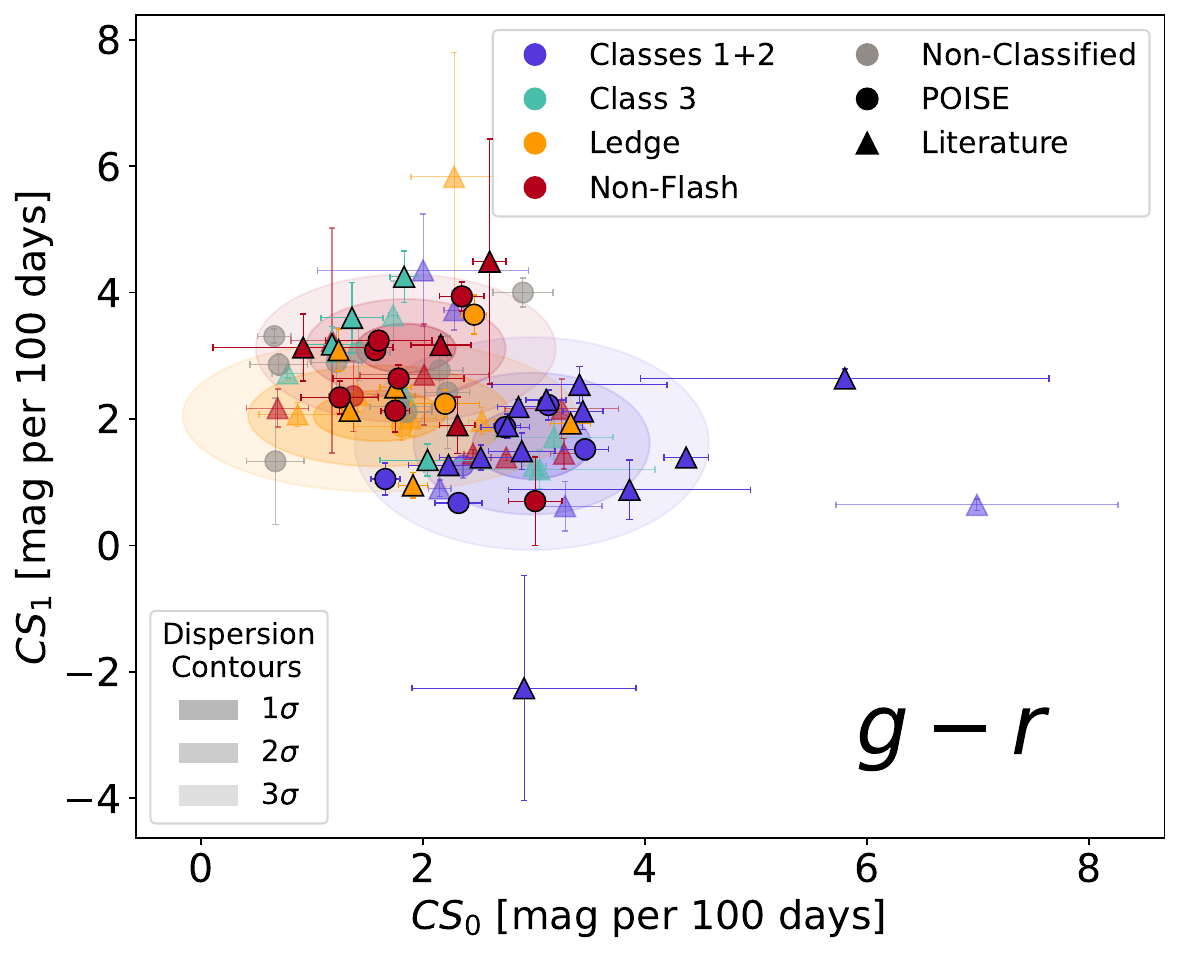}
\caption{Top: (\textit{$g-r$}) colour curves shifted to the phase and colour of the break for class 1+2 (blue), class 3 (green), ledge (orange), and non-flash SNe (red). The circles represent the data points of all SNe, while the solid lines show the weighted mean colour curve for the gold sample of each group, with the shaded regions indicating the 1$\sigma$ uncertainty, accounting for both the dispersion and measurement uncertainties. The black dashed line illustrates a slope equal to the average of $cs_0$ and $cs_1 $ of all SNe in the gold sample. The red error bar in the top-left corner indicates the typical $1\sigma$ photometric uncertainty of the individual data points. 
Bottom: $cs_{0}$ vs. $cs_{1}$ for the (\textit{$g-r$}) colour for the gold (solid markers with black edges) and complete (shaded markers) samples. Circles and triangles indicate SNe from the POISE and literature samples, respectively. The ellipses are centred on the weighted mean value of each parameter for each group, and their widths represent 1$\sigma$, 2$\sigma$, and 3$\sigma$ weighted standard deviation (from darker to lighter tones), computed using the gold sample only. Class 1+2 SNe show, on average, steeper $cs_{0}$ and shallower $cs_{1}$ than ledge, class 3, and non-flash SNe. Class 3 SNe are equally distributed around the other classes.}  
\label{f:colour.breaks_s0.s1}
\end{figure}

Figure \ref{f:colour.phases} displays the (\textit{$g-r$}) colour at 20 days vs. the peak absolute $g$-band magnitude for the full SN sample, grouped by their flash-spectroscopic classification. Class 1+2 SNe show bluer colours and higher luminosities compared to non-flash SNe, while class 3 and ledge SNe occupy an intermediate region between these two groups, sharing characteristics with both populations. A similar behaviour is observed in the other colours at 20 days. Notably, the slope of the reddening vector deviates from the trend defined by the SN distribution in this parameter space, thereby supporting the conclusion that host-galaxy extinction (at least for standard extinction laws) is not the primary driver of the observed luminosity-colour dispersion among the different classes. 

\cite{dejaeger2018} report a comparable trend between the peak absolute magnitude and colour at 15 and 30 days, finding moderate but significant correlations ($r\approx0.34-0.57$, p$\approx10^{-3}-0.1$) for a sample of 20--35 SNe~II. This is not observed at later epochs (50 and 70 days), which again suggests that redder SNe~II are intrinsically fainter at early phases. They propose that this behaviour could be attributed to differences in progenitor radii or CSM interaction rather than extinction effects. In this context, Fig.~\ref{f:colour.phases} supports the idea that the presence of flash-ionisation features, and therefore CSM interaction, may be responsible for the differences in colours at early times.

Additionally, we performed Pearson correlation tests\footnote{Via \textsc{pearsonr} from the \textsc{scipy.stats}.} between the analysed colours at 10 and 20 days and the peak absolute magnitude in their respective bands, considering the full SN sample regardless of their flash classification. In all cases, we find significant correlations, with Pearson coefficients in the range $r\,=\,0.34-0.66$ (p $\,\approx10^{-11}-10^{-3}$) at 10 days and $r\,=\,0.40-0.75$ (p $\,\approx10^{-14}-10^{-4}$) at 20 days. These results are consistent with those of \cite{dejaeger2018} and further reinforce the presence of a luminosity-colour relation at early phases, likely linked to the effects of CSI.

\begin{table*}
\centering
\caption{Mean-weighted values of colour-curve parameters.}
\resizebox{\textwidth}{!}{
\begin{tabular}{cccccccccccc}
\hline\hline
\multirow{2}{*}{Colour} & \multirow{2}{*}{\begin{tabular}[c]{@{}l@{}}Flash \\ class\end{tabular}} & \multicolumn{4}{c}{Gold sample}           & \multicolumn{4}{c}{Complete sample}       \\
   &   & \begin{tabular}[c]{@{}c@{}}$\langle t_{\mathrm{break}} \rangle$ \\ {[}days{]}\end{tabular}      &\begin{tabular}[c]{@{}c@{}}$\langle cs_{0} \rangle$ \\ {[}mag 100d$^{-1}${]}\end{tabular}      & \begin{tabular}[c]{@{}c@{}}$\langle cs_{1} \rangle$ \\ {[}mag 100d$^{-1}${]}\end{tabular}     & \begin{tabular}[c]{@{}c@{}}$\langle \Delta cs \rangle$ \\ {[}mag 100d$^{-1}${]}\end{tabular}  & \begin{tabular}[c]{@{}c@{}} Number \\ of SNe\end{tabular}  & \begin{tabular}[c]{@{}c@{}}$\langle t_{\mathrm{break}} \rangle$ \\ {[}days{]}\end{tabular}&  \begin{tabular}[c]{@{}c@{}}$\langle cs_{0} \rangle$ \\ {[}mag 100d$^{-1}${]}\end{tabular}       & \begin{tabular}[c]{@{}c@{}}$\langle cs_{1} \rangle$ \\ {[}mag 100d$^{-1}${]}\end{tabular}     & \begin{tabular}[c]{@{}c@{}}$\langle \Delta cs \rangle$ \\ {[}mag 100d$^{-1}${]}\end{tabular} & \begin{tabular}[c]{@{}c@{}} Number \\ of SNe\end{tabular}     \\ \hline
\multirow{5}{*}{\textit{$g-r$}}   & Class 1+2 & $14.90   \pm 4.10$ & $2.98 \pm 0.53$ & $1.61 \pm 0.56$ & $1.29  \pm 0.76$ & 17 & $14.80  \pm 3.55$ & $2.78 \pm 0.57$ & $1.52 \pm 0.62$ & $1.20  \pm 0.89$ & 23 \\
                                  & Class 3   & $17.40   \pm 3.24$ & $1.67 \pm 0.32$ & $2.61 \pm 1.31$ & $-1.59 \pm 1.38$ & 4  & $13.52  \pm 3.76$ & $1.80 \pm 0.47$ & $2.08 \pm 0.78$ & $-0.16 \pm 1.68$ & 10 \\
                                  & Ledge     & $16.48   \pm 3.19$ & $1.61 \pm 0.59$ & $2.05 \pm 0.40$ & $-0.34 \pm 1.01$ & 7  & $19.57  \pm 2.99$ & $1.68 \pm 0.60$ & $2.05 \pm 0.37$ & $-0.32 \pm 0.94$ & 13 \\
                                  & Non-flash & $17.20   \pm 3.30$ & $1.84 \pm 0.45$ & $3.12 \pm 0.39$ & $-1.25 \pm 0.68$ & 11 & $16.39  \pm 1.92$ & $2.11 \pm 0.63$ & $2.10 \pm 0.89$ & $0.04  \pm 1.35$ & 19 \\
                                  & POISE     & $16.16   \pm 3.24$ & $2.26 \pm 0.82$ & $2.09 \pm 0.93$ & $-0.85 \pm 1.43$ & 22 & $16.40  \pm 3.19$ & $2.24 \pm 0.79$ & $2.07 \pm 0.92$ & $-0.49 \pm 1.42$ & 27 \\
                                  & All       & $15.60   \pm 3.72$ & $2.10 \pm 0.83$ & $2.04 \pm 0.78$ & $-0.17 \pm 1.42$ & 47 & $15.94  \pm 3.63$ & $2.08 \pm 0.72$ & $1.89 \pm 0.76$ & $0.11 \pm 1.40$  & 75 \\ \hline
\multirow{5}{*}{\textit{$B-V$}}   & Class 1+2 & $13.88   \pm 3.40$ & $2.62 \pm 0.67$ & $1.13 \pm 0.79$ & $1.41  \pm 1.12$ & 11 & $15.25  \pm 3.79$ & $2.57 \pm 0.62$ & $1.16 \pm 0.75$ & $1.35  \pm 0.97$ & 17\\
                                  & Class 3   & $17.67   \pm 5.63$ & $1.20 \pm 0.24$ & $4.64 \pm 1.27$ & $-2.51 \pm 0.78$ & 2  & $13.12  \pm 3.70$ & $1.18 \pm 0.28$ & $3.00 \pm 1.16$ & $-1.31 \pm 1.64$ & 5 \\
                                  & Ledge     & $13.96   \pm 1.58$ & $0.72 \pm 0.36$ & $2.98 \pm 0.59$ & $-2.54 \pm 0.62$ & 8  & $14.02  \pm 1.63$ & $0.82 \pm 0.42$ & $2.98 \pm 0.59$ & $-2.53 \pm 0.62$ & 9 \\
                                  & Non-flash & $17.37   \pm 3.88$ & $1.76 \pm 0.25$ & $3.64 \pm 0.41$ & $-1.92 \pm 0.35$ & 8  & $16.93  \pm 3.43$ & $2.22 \pm 1.27$ & $3.38 \pm 0.69$ & $-0.66 \pm 1.96$ & 12 \\
                                  & POISE     & $15.95   \pm 3.97$ & $2.05 \pm 0.63$ & $2.33 \pm 1.50$ & $-0.71 \pm 2.03$ & 19 & $16.15  \pm 3.71$ & $2.01 \pm 0.69$ & $2.39 \pm 1.50$ & $-0.94 \pm 2.06$ & 25 \\
                                  & All       & $15.16   \pm 3.33$ & $1.68 \pm 0.81$ & $2.46 \pm 1.39$ & $-1.15 \pm 2.02$ & 34 & $15.01  \pm 3.33$ & $1.84 \pm 1.07$ & $2.49 \pm 1.29$ & $-0.83 \pm 2.05$ & 52 \\\hline
\multirow{5}{*}{\textit{$g-i$}}   & Class 1+2 & $13.63   \pm 4.47$ & $4.63 \pm 0.78$ & $1.91 \pm 0.80$ & $2.41  \pm 1.01$ & 14 & $14.36  \pm 4.16$ & $4.08 \pm 1.03$ & $1.71 \pm 0.89$ & $2.40  \pm 1.15$ & 21 \\
                                  & Class 3   & $15.51   \pm 3.03$ & $1.73 \pm 1.20$ & $2.16 \pm 1.78$ & $-1.41 \pm 4.73$ & 2  & $14.98  \pm 5.01$ & $1.24 \pm 1.13$ & $1.69 \pm 1.48$ & $0.76  \pm 3.76$ & 8 \\
                                  & Ledge     & $14.28   \pm 1.91$ & $1.19 \pm 0.58$ & $2.50 \pm 0.50$ & $-1.65 \pm 0.33$ & 7  & $16.15  \pm 3.55$ & $1.20 \pm 0.56$ & $2.50 \pm 0.50$ & $-1.64 \pm 0.32$ & 9 \\
                                  & Non-flash & $17.55   \pm 2.16$ & $2.47 \pm 0.97$ & $3.34 \pm 0.84$ & $-0.87 \pm 1.47$ & 9  & $15.67  \pm 3.08$ & $2.88 \pm 1.15$ & $2.62 \pm 0.78$ & $0.36  \pm 1.80$ & 15 \\
                                  & POISE     & $14.71   \pm 2.84$ & $3.39 \pm 1.54$ & $2.54 \pm 1.14$ & $0.01 \pm 2.25$  & 23 & $14.28  \pm 3.04$ & $3.32 \pm 1.58$ & $2.56 \pm 1.13$ & $-0.05 \pm 2.22$ & 26 \\
                                  & All       & $14.71   \pm 3.59$ & $2.35 \pm 1.64$ & $2.51 \pm 0.93$ & $-0.37 \pm 2.09$ & 40 & $15.18  \pm 3.76$ & $2.27 \pm 1.56$ & $2.33 \pm 0.98$ & $0.12  \pm 2.24$ & 63 \\\hline
\multirow{5}{*}{\textit{$u-g$}}   & Class 1+2 & $21.92   \pm 1.69$ & $3.26 \pm 0.76$ & $5.18 \pm 1.98$ & $-0.60 \pm 2.01$ & 12 & $21.85  \pm 1.74$ & $3.26 \pm 0.77$ & $5.11 \pm 1.93$ & $-0.61 \pm 2.06$ & 16 \\
                                  & Class 3   & $-$                & $-$             & $- $            & $-$              & 1  & $15.87  \pm 5.25$ & $2.85 \pm 1.60$ & $5.08 \pm 2.71$ & $-1.90 \pm 4.30$ & 5 \\
                                  & Ledge     & $13.31   \pm 2.54$ & $4.37 \pm 2.15$ & $8.87 \pm 1.15$ & $-5.07 \pm 1.38$ & 7  & $13.29  \pm 2.37$ & $4.32 \pm 1.92$ & $8.88 \pm 1.08$ & $-5.05 \pm 1.25$ & 9 \\
                                  & Non-flash & $12.56   \pm 2.42$ & $3.65 \pm 1.27$ & $6.85 \pm 1.97$ & $-5.89 \pm 2.21$ & 8  & $12.65  \pm 2.41$ & $3.59 \pm 1.70$ & $6.26 \pm 1.83$ & $-4.50 \pm 1.99$ & 11 \\
                                  & POISE     & $13.16   \pm 1.38$ & $3.63 \pm 1.66$ & $7.02 \pm 2.87$ & $-3.09 \pm 4.00$ & 16 & $13.12  \pm 1.48$ & $3.89 \pm 1.53$ & $6.94 \pm 2.83$ & $-3.16 \pm 3.76$ & 21 \\
                                  & All       & $20.93   \pm 3.28$ & $3.36 \pm 1.60$ & $7.06 \pm 2.63$ & $-3.56 \pm 3.01$ & 33 & $20.70  \pm 3.48$ & $3.50 \pm 1.59$ & $6.84 \pm 2.62$ & $-3.30 \pm 3.06$ & 47 \\\hline    
\end{tabular}}
\tablefoot{`All' also includes Non-classified SNe.}
\label{t:colour.curve.parameters}
\end{table*}

\begin{table}
\centering
\caption{K-S and A-D tests' p values and significance for the comparison cases a and b of the gold and complete sample for $cs_0$, $cs_1$, $\Delta cs$, and $t_\mathrm{break}$.}
\resizebox{\columnwidth}{!}{
\begin{tabular}{ccccccc}
\hline\hline
Colour & Parameter & Sample & p$_a^\mathrm{KS}$ & p$_b^\mathrm{KS}$ & p$_a^\mathrm{AD}$ & p$_b^\mathrm{AD}$ \\ \hline
\multirow{8}{*}{(\textit{$g-r$})}  & \multirow{2}{*}{$cs_0$}              & Gold     & $\mathbf{2\,\times\,10^{-3}}$   &  $8.3\,\times\,10^{-3}$          &  $\mathbf{<0.01}$     & $\mathbf{<0.01}$ \\
                                   &                                      & Complete & $0.02$                          &  $7.8\,\times\,10^{-3}$          &  $\mathbf{<0.01}$     & $\mathbf{<0.01}$   \\
                                   & \multirow{2}{*}{$cs_1$}              & Gold     & $8.2\,\times\,10^{-3}$          &  $ 0.02$                         &  $\mathbf{<0.01}$     & $<0.025$  \\
                                   &                                      & Complete & $0.02$                          & $8.3\,\times\,10^{-3}$           &  $\mathbf{<0.01}$     & $<0.025$  \\
                                   & \multirow{2}{*}{$\Delta cs$}         & Gold     & $\mathbf{2.6\,\times\,10^{-6}}$ & $\mathbf{6.7\,\times\,10^{-5}}$  &  $\mathbf{<0.01}$     & $\mathbf{<0.01}$ \\
                                   &                                      & Complete & $\mathbf{1.1\,\times\,10^{-4}}$ & $\mathbf{7.4\,\times\,10^{-5}}$  &  $\mathbf{<0.01}$     & $\mathbf{<0.01}$ \\
                                   & \multirow{2}{*}{$t_\mathrm{break}$}  & Gold     & $0.39$                          & $0.28$                           & $\geq 0.25$           & $\geq 0.25$  \\
                                   &                                      & Complete & $0.18$                          & $0.04$                           & $\geq 0.25$           & $\geq 0.25$  \\\hline
\multirow{8}{*}{(\textit{$B-V$})}  & \multirow{2}{*}{$cs_0$}              & Gold     & $0.01$                          &  $\mathbf{2.6\,\times\,10^{-3}}$ &  $\mathbf{<0.01}$     & $\mathbf{<0.01}$  \\
                                   &                                      & Complete & $0.11$                          &  $0.04$                          &  $\geq 0.05$          & $<0.05$ \\
                                   & \multirow{2}{*}{$cs_1$}              & Gold     & $\mathbf{2.4\,\times\,10^{-4}}$ &  $\mathbf{1.1\,\times\,10^{-4}}$ &  $\mathbf{<0.01}$     &  $\mathbf{<0.01}$ \\
                                   &                                      & Complete & $\mathbf{1\,\times\,10^{-5}}$   &  $\mathbf{3.5\,\times\,10^{-6}}$ &  $\mathbf{<0.01}$     & $\mathbf{<0.01}$  \\
                                   & \multirow{2}{*}{$\Delta cs$}         & Gold     & $\mathbf{2.6\,\times\,10^{-5}}$ &  $\mathbf{1.2\,\times\,10^{-5}}$ &  $\mathbf{<0.01}$     & $\mathbf{<0.01}$  \\
                                   &                                      & Complete & $\mathbf{5.4\,\times\,10^{-4}}$ &  $\mathbf{1.1\,\times\,10^{-4}}$ &  $\mathbf{<0.01}$     & $\mathbf{<0.01}$  \\
                                   & \multirow{2}{*}{$t_\mathrm{break}$}  & Gold     & $0.93$                          &  $0.60$                          & $\geq 0.25$           &  $\geq 0.25$ \\
                                   &                                      & Complete & $0.50$                          &  $0.17$                          & $\geq 0.25$           & $\geq 0.25$ \\\hline
\multirow{8}{*}{(\textit{$g-i$})}  & \multirow{2}{*}{$cs_0$}              & Gold     & $\mathbf{1\,\times\,10^{-3}}$   &  $\mathbf{1.2\,\times\,10^{-4}}$ &  $\mathbf{<0.01}$     & $\mathbf{<0.01}$  \\
                                   &                                      & Complete & $\mathbf{1.1\,\times\,10^{-3}}$ &  $\mathbf{4.3\,\times\,10^{-4}}$ &  $\mathbf{<0.01}$     & $\mathbf{<0.01}$  \\
                                   & \multirow{2}{*}{$cs_1$}              & Gold     & $0.02$                          &  $0.01$                          &  $<0.025$             & $\mathbf{<0.01}$  \\
                                   &                                      & Complete & $\mathbf{9.2\,\times\,10^{-4}}$ &  $\mathbf{2\,\times\,10^{-4}}$   &  $\mathbf{<0.01}$     & $\mathbf{<0.01}$  \\
                                   & \multirow{2}{*}{$\Delta cs$}         & Gold     & $\mathbf{1\,\times\,10^{-3}}$   &  $\mathbf{1.7\,\times\,10^{-5}}$ &  $\mathbf{<0.01}$     & $\mathbf{<0.01}$  \\
                                   &                                      & Complete & $\mathbf{1.1\,\times\,10^{-4}}$ &  $\mathbf{1.7\,\times\,10^{-5}}$ &  $\mathbf{<0.01}$     & $\mathbf{<0.01}$  \\
                                   & \multirow{2}{*}{$t_\mathrm{break}$}  & Gold     & $0.41$                          &  $0.21$                          &  $\geq 0.25$          &  $\geq 0.25$ \\
                                   &                                      & Complete & $0.81$                          &  $0.55$                          &  $\geq 0.25$          & $\geq 0.25$ \\\hline
\multirow{8}{*}{(\textit{$u-g$})}  & \multirow{2}{*}{$cs_0$}              & Gold     & $1.0$                           &  $0.72$                          &  $\geq 0.25$          & $\geq 0.25$  \\
                                   &                                      & Complete & $0.99$                          &  $0.97$                          &  $\geq 0.25$          & $\geq 0.25$  \\
                                   & \multirow{2}{*}{$cs_1$}              & Gold     & $\mathbf{3.2\,\times\,10^{-4}}$ &  $\mathbf{7.5\,\times\,10^{-7}}$ &  $\mathbf{<0.01}$     &  $\mathbf{<0.01}$ \\
                                   &                                      & Complete & $\mathbf{5.9\,\times\,10^{-4}}$ & $\mathbf{4.4\,\times\,10^{-6}}$  &  $\mathbf{<0.01}$     & $\mathbf{<0.01}$  \\
                                   & \multirow{2}{*}{$\Delta cs$}         & Gold     & $\mathbf{1.6\,\times\,10^{-5}}$ &  $\mathbf{1.5\,\times\,10^{-5}}$ &  $\mathbf{<0.01}$     & $\mathbf{<0.01}$  \\
                                   &                                      & Complete & $\mathbf{3.6\,\times\,10^{-5}}$ &  $\mathbf{1.6\,\times\,10^{-5}}$ &  $\mathbf{<0.01}$     & $\mathbf{<0.01}$  \\
                                   & \multirow{2}{*}{$t_\mathrm{break}$}  & Gold     & $0.02$                          &  $0.02$                          &  $<0.025$             & $<0.025$          \\
                                   &                                      & Complete & $0.03$                          &  $\mathbf{4.2\,\times\,10^{-3}}$ & $<0.025$              & $\mathbf{<0.01}$  \\\hline
\end{tabular}}
\tablefoot{Most significant differences (p$^\mathrm{KS}< 5\times10^{-3}$ and p$^\mathrm{AD}< 0.01$) are highlighted in bold.}
\label{t:pvalues_KS_AD_slopes}
\end{table}

\subsection{Colour-curve slope correlations}
\label{sec:slopes.correlations}

Figure \ref{f:colour.breaks_s0.s1} (top panel) shows the mean (\textit{$g-r$}) colour curves for the gold sample (solid lines) along with the data points of all SNe, shifted in magnitude and date to align their break colour and phase. Different behaviours in colour evolution are seen depending on the flash-spectroscopic classification, with opposite patterns between class 1+2 and non-flash SNe. Class 3 and ledge objects exhibit both kinds of trends, with ledge SNe resembling non-flash SNe more closely. Similar trends are also observed in (\textit{$B-V$}), (\textit{$g-i$}), and (\textit{$u-g$}) colours (Fig.~\ref{f:others.colours.breaks}). Table~\ref{t:colour.curve.parameters} lists the weighted mean values of the colour-curve parameters for each colour, sorted by their flash-spectroscopic class. It also reports the corresponding values for the POISE sample and for the full dataset, that is, all the classes combined, including the non-classified SNe. 

In the bottom panel of Fig.~\ref{f:colour.breaks_s0.s1}, the differences described above are presented in terms of the pre- and post-break slopes ($cs_0$ and $cs_1$, respectively). The $cs_0$ vs. $cs_1$ parameter space serves to separate the bulk of class 1+2, ledge, and non-flash SNe. Nevertheless, class 3 SNe are scattered among the other classes. Within the gold sample, in the colours, class 1+2 SNe display, on average, a steeper first slope than ledge, class 3, and non-flash SNe, whereas this trend is inverted for the second slope (see Table~\ref{t:colour.curve.parameters}). Consequently, class 1+2 SNe typically show positive values of $\Delta cs$, while the other groups exhibit negative ones (see Table~\ref{t:colour.curve.parameters}). In the (\textit{$u-g$}) colour, however, the first slope shows similar values for all classes, meaning that all of them follow a similar evolution at the beginning, as can be observed in Fig.~\ref{f:colour_curves}. Nevertheless, the $cs_1$ values in this colour exhibit significant differences between class 1+2 and the other classes. Therefore, there is a clear separation in $\Delta cs_{(u-g)}$, although the values are, on average, negative. 

K-S and A-D tests were performed for $cs_0$, $cs_1$ and $\Delta cs$ parameters considering the comparison cases a and b, as described in Sect.~\ref{sec:stats.methods}. Each case was analysed separately for the gold and the complete samples. The CDFs are displayed in Fig.~\ref{f:CDFs_slopes} with the corresponding p values for the K-S tests, which are also listed in Table~\ref{t:pvalues_KS_AD_slopes} along with the results from the A-D tests. 

In the gold sample, the majority of the comparisons yield p$\,<5\times10^{-3}$,  indicating different parent distributions between the compared populations, with different degrees of significance. The exceptions are (\textit{$g-r$}) for $cs_0$ (case b) and $cs_1$ (both cases); (\textit{$B-V$}) for $cs_0$ (case a); and (\textit{$u-g$}) for $cs_0$ (both cases). The (\textit{$u-g$}) colour yields the highest p values for $cs_0$, reinforcing that the (\textit{$u-g$}) colour evolution is similar across all classes until the break. Overall, $\Delta cs$ shows the strongest and most significant distinctions across all colours, making it the most effective colour-curve parameter for separating flash-spectroscopic classes. Moreover, including class 3 and ledge SNe (case b) tends to reinforce the differences, indicating that these groups possess colour properties analogous to class 1+2 and non-flash SNe, respectively. 

When the complete sample is considered, the distinctions between the populations remain in the majority of cases, as in the gold sample, except for $cs_0$ in (\textit{$B-V$}). Nonetheless, in most cases, the significance of the differences for the complete sample is lower than for the gold sample. This may be influenced by the inclusion of colour curves with lower data quality. Again $\Delta cs$ shows the strongest distinction among the colour-curve parameters. 

These results suggest an intrinsic diversity in the early-phase colour evolution of SNe~II, likely driven by differences in their dominant power source, either CSI or post-shock cooling. A further qualitative comparison of the colour-curve behaviours with interacting models and the implications for the power mechanism is discussed in Sect.~\ref{sec:disc.models}. 

\begin{figure*}
\centering
\includegraphics[width=.85\textwidth]{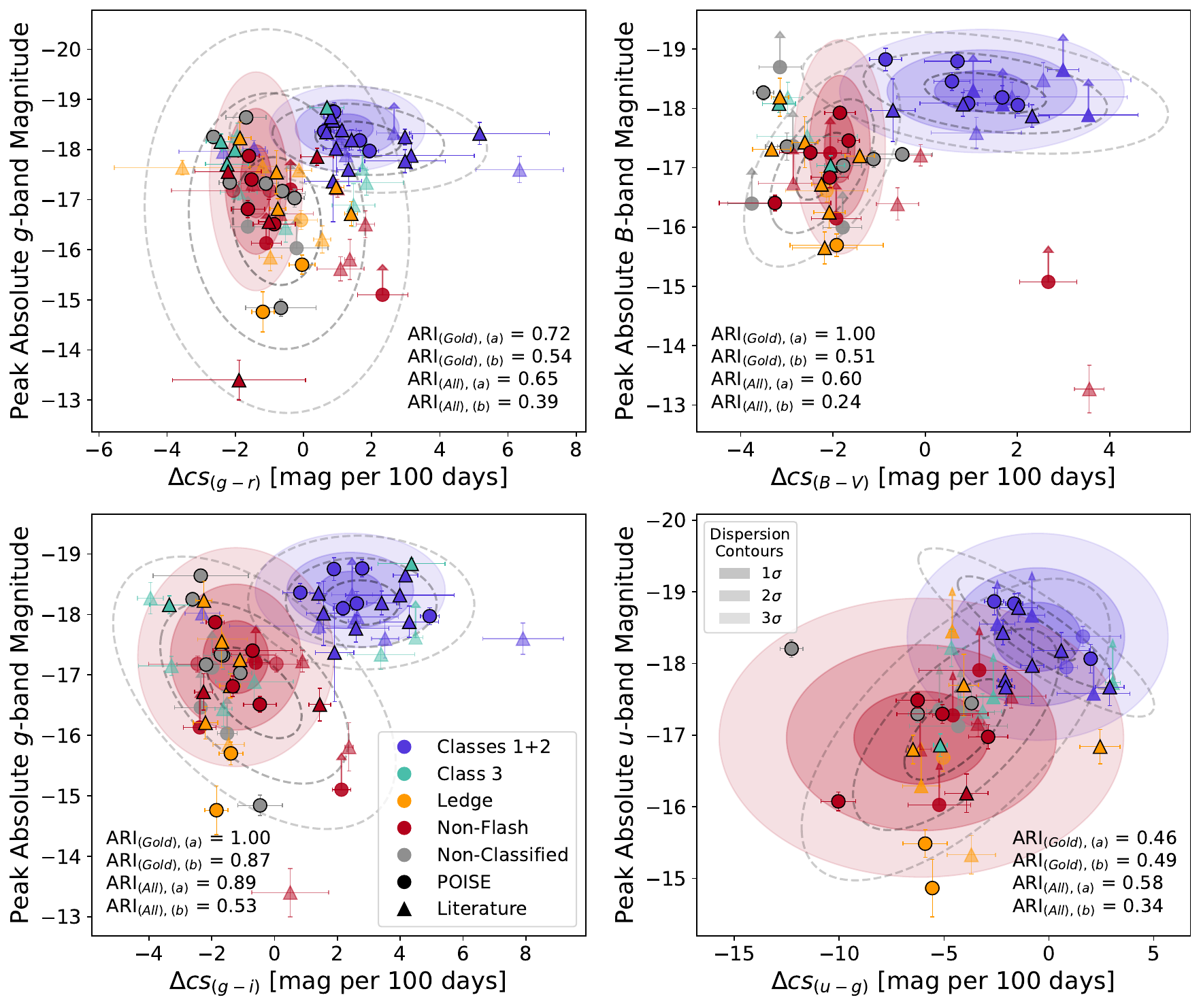}
\caption{$\Delta cs$ vs. peak absolute magnitude in the first band of each colour for (\textit{$g-r$}), (\textit{$B-V$}), (\textit{$g-i$}), and (\textit{$u-g$}) (from left to right, top to bottom). Class 1+2 (blue), class 3 (green), ledge (orange) and non-flash SNe (red) from the POISE (circles) and literature (triangles) samples are shown as solid markers with black edges for the gold sample and as shaded markers for complete samples. Non-classified SNe from the POISE sample are included for completeness. The coloured ellipses are centred on the weighted mean value of each parameter for each group, with widths corresponding to the 1$\sigma$, 2$\sigma$, and 3$\sigma$ weighted standard deviations (from darker to lighter tones), computed using only the gold sample. Dashed grey ellipses show the density contours derived from Gaussian Mixture Model clustering of the gold sample and represent the 1$\sigma$, 2$\sigma$, and 3$\sigma$ regions (from darker to lighter tones). ARI scores for comparisons cases a and b (see Sect.~\ref{sec:slopes.correlations}) are displayed in the corresponding panels.}
\label{f:diff_s_magabs}
\end{figure*}

Figure \ref{f:diff_s_magabs} illustrates the relation between $\Delta cs$ and the peak absolute magnitude in the first band of the corresponding colour. A shared trend is observed across all analysed colours, with a clear distinction between class 1+2 and the other classes. This parameter space combines the quantities that best distinguish SNe~II exhibiting clear interaction signatures in the early spectra (class 1+2 SNe) from the remaining classes.

We performed an unsupervised clustering analysis using a Gaussian Mixture Model\footnote{Via \texttt{GaussianMixture} from \texttt{scikit-learn} library.} to assess the capability of this parameter space to separate the two main populations of interest (class 1+2 and non-flash). Since our goal is to distinguish between these two groups, the model was constrained to identify only two clusters. The clustering was carried out considering only the SNe in the gold sample and excluding those with lower limits in their peak absolute magnitudes. The overdensities identified as clusters (grey dashed ellipses in Fig. \ref{f:diff_s_magabs}) closely trace the distribution of the class 1+2 and non-flash groups (blue and red ellipses, respectively).

To evaluate the reliability of the clustering results, we computed the Adjusted Rand Index\footnote{Via \texttt{adjusted$\_$rand$\_$score} from \texttt{scikit-learn} library.} (ARI, \citealp{ARI.Hubert.1985}), a clustering evaluation metric ranging from $-0.5$ (complete disagreement) to 1 (perfect agreement between true and predicted labels), where values near 0 indicate a random association. The ARI scores were calculated for the cases a and b (described in Sect.~\ref{sec:stats.methods}) and are displayed in each panel of Fig.~\ref{f:diff_s_magabs}. 

In the gold sample, case a yields two perfect agreements for the (\textit{$B-V$}) and (\textit{$g-i$}) colours, a strong consistency for (\textit{$g-r$}) colour (ARI = 0.73), and a weaker, though non-random, match for (\textit{$u-g$}) colour (ARI = 0.46). When the remaining classes are included in the analysis (case b), the ARI values decrease across all colours, except for (\textit{$u-g$}), suggesting that class 3 and ledge SNe exhibit differences relative to class 1+2 and non-flash SNe within this parameter space. Nevertheless, ARI values $\gtrsim 0.4$ still indicate that the clusters recover meaningful separations between the populations. For the complete sample, (\textit{$g-r$}) and (\textit{$g-i$}) show the highest ARI scores, although lower than those of the gold sample, indicating that these clusters continue to capture statistically significant separations. In contrast, (\textit{$B-V$}) approaches values consistent with a random association. We further verified that a K-means\footnote{Via \textsc{KMeans} from \textsc{scikit-learn} library.} clustering yields a consistent separation.  

Additionally, we note that adopting the flash classification from \cite{JG2024}, that is, reassigning the four SNe for which our classification differs from theirs (see Sect. \ref{sec:classification}) from class 3 to non-flash, leads to slightly higher ARI scores in all cases. This highlights the difficulty of differentiating between class 3 and non-flash SNe.

Overall, the early-time colour evolution exhibits a strong connection with the presence of flash features. These results provide compelling evidence that the early-time colour evolution and peak luminosity evolution can serve as an effective diagnostic to identify SNe~II with strong flash-ionisation features (i.e. class 1+2 SNe), even in the absence of early-time spectroscopy. Furthermore, this is valid for any combination of two bands producing the colour-curve slopes and peak absolute magnitudes analysed here.

\subsection{Break time, colour minimum, and flash-feature duration} \label{sec:break.time.flash}

\begin{figure}
\centering
\includegraphics[width=.95\columnwidth]{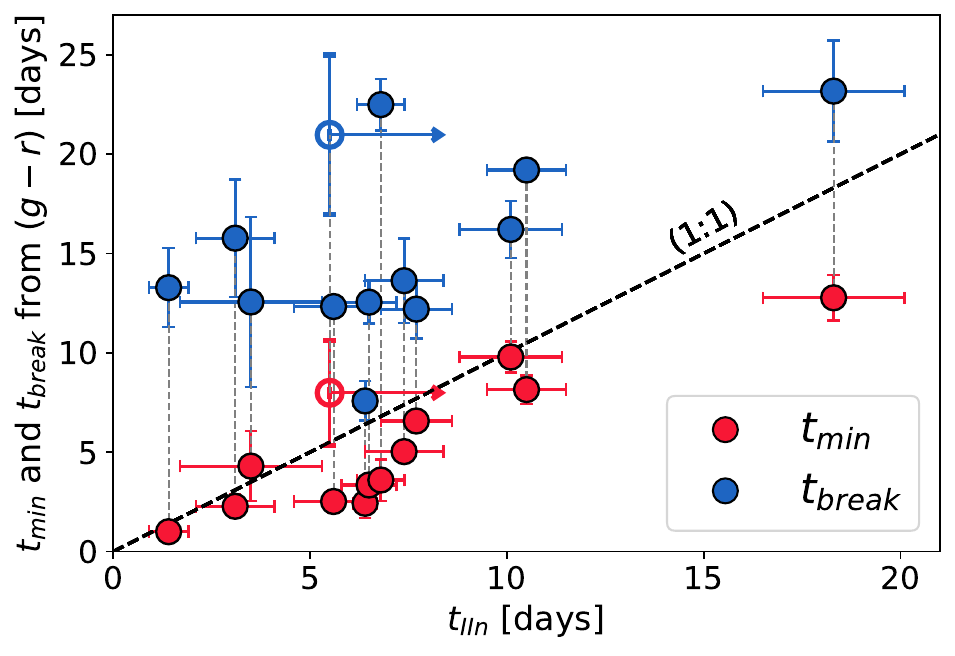}             
\caption{Duration of flash features ($t_\mathrm{IIn}$) vs. epoch of colour minimum ($t_\mathrm{min}$, red) and phase of break ($t_\mathrm{break}$, blue) derived from (\textit{$g-r$}) colour curves for 13 flash SNe. Open circles correspond to SN~2023bvh, for which $t_{\mathrm{IIn}}$ is only a lower limit. The black dashed line represents the identity (1:1) relation, while the grey dashed lines connect the two measurements corresponding of the same SN. In all cases, $t_\mathrm{IIn}$ occurs at or after $t_\mathrm{min}$ (within the uncertainties) and before $t_\mathrm{break}$, implying $t_\mathrm{min} \lesssim t_\mathrm{IIn} < t_\mathrm{break}$.}  
\label{f:times.comparisons}
\end{figure}

As indicated in Table \ref{t:colour.curve.parameters}, the moment of change in the slope of the colour curves (i.e. $t_{\mathrm{break}}$) does not show any significant difference related to the presence or absence of flash-ionisation emissions, except for the (\textit{$u-g$}) colour. 

Figure~\ref{f:CDFs_t_break} shows the break-time CDFs of all the colours. In the (\textit{$u-g$}) colour, class 1+2 SNe tend to show later break times than non-flash objects. This is not seen in any of the other colours. K-S and A-D tests performed for this parameter and considering the comparison cases a and b (see Sect.~\ref{sec:stats.methods}) yield p$\,< 5\times 10^{-3}$ for case b in the complete sample (see Table~\ref{t:pvalues_KS_AD_slopes}), indicating a statistically significant difference between the distributions. 
This finding, observed exclusively in (\textit{$u-g$}), the only colour involving the $u$ band, suggests that the same physical mechanism responsible for the enhanced $u$-band and ultraviolet luminosities in flash SNe \citep{JG2024} may also drive the delayed break times measured in this work.

In contrast, K-S and A-D tests performed for the remaining colours return p$\,\geq0.05$ in all comparison cases. Furthermore, we find no statistically significant correlations between $t_\mathrm{break}$ and any of the light-curve parameters or the remaining colour-curve parameters.   

From the spectra of flash SNe (i.e. classes 1+2+3), it is possible to determine the duration of the flash features, denoted $t_{\mathrm{IIn}}$ by \cite{JG2024}. For the flash SNe also present in the sample of \cite{JG2024}, we adopted their $t_{\mathrm{IIn}}$ values. In addition, we obtained $t_{\mathrm{IIn}}$ for SN~2021cgu (class 2) in a similar fashion, by estimating the time when the electron-scattering emission of the He \textsc{ii} $\lambda$4686 line disappears based on the available spectra. For this purpose, we compared the early spectral evolution of SN~2021cgu with those of two other well-observed events, SN~2013fs (class 3) and SN~2021qvr (class 2). In the case of SN~2023bvh (class 1), it was only possible to provide a lower limit on $t_{\mathrm{IIn}}$ given that we had a spectrum (showing flash features) at $5.5$ days and the next one was at 56 days.

Table \ref{t:tIIn.tmin} lists the duration of the flash-ionisation features. The bulk of class 1+2 SNe spans a broad range of $t_{\mathrm{IIn}}$ values, from approximately 2 to 10.5 days; only SN~2022prv lies outside this interval, with $t_\mathrm{IIn}\approx$ 18 days. Class 3 SNe, on the other hand, generally show shorter $t_{\mathrm{IIn}}$ values, between $\approx$ 0.7 and 3 days. 

As mentioned in Sect.~\ref{sec:colour.curves}, the early-time colour curves of some flash SNe~II exhibit an initial minimum. This feature is seen, for example, in SN~2022ffg (see Fig.~\ref{f:slopes_example}) and has also been reported in other well-sampled SNe~II with flash spectroscopy, such as SN~2022jox \citep{Andrews2024_22jox}, SN~2023ixf (\citealp{Hiramatsu2023_23ixf}; \citealp{Li2024.23ixf}), SN~2024bch \citep{Andrews2025}, and SN~2024ggi (\citealp{Jacobson-Galan2024.24ggi}; \citealp{Shrestha2024_24ggi}). \cite{Shrestha2024_24ggi} and \cite{Andrews2025} suggest that the end of the colour minimum in (\textit{$B-V$}) may serve as an indicator of the flash-feature duration (i.e. $t_{\mathrm{IIn}}$). However, this minimum is not always detected, mainly due to the lack of sufficiently early-time observations. In our sample, we were able to determine the epoch of the minimum ($t_\mathrm{min}$) from 13 flash SNe in the (\textit{$g-r$}) colour (see Sect.~\ref{sec:colour.curves.parameters} for measurement details and Table~\ref{t:tIIn.tmin} for the corresponding values).

In Fig.~\ref{f:times.comparisons}, we show $t_\mathrm{IIn}$ vs. $t_\mathrm{min}$ and $t_\mathrm{break}$ for the (\textit{$g-r$}) colour curves of 13 flash SNe. For all of them, $t_\mathrm{min}\leq t_\mathrm{IIn} < t_\mathrm{break}$. Consequently, $t_\mathrm{min}$ and $t_\mathrm{break}$ could be considered as lower and upper limits, respectively, on the duration of the flash-ionisation phase in SNe~II. This also implies that $t_{\mathrm{IIn}}$ lies within the range covered by $cs_0$, measured between $t_\mathrm{min}$ and $t_\mathrm{break}$. In general, the time interval between $t_\mathrm{min}$ and $t_\mathrm{IIn}$ is shorter than that between $t_\mathrm{IIn}$ and $t_\mathrm{break}$. Most minima occur within the first five days after first light, making them challenging to observe. 

Notably, SNe with longer $t_\mathrm{IIn}$ also show a delayed $t_\mathrm{min}$ in (\textit{$g-r$}). We performed Pearson correlation tests, using the \textsc{pearsonr}\footnote{\url{https://docs.scipy.org/doc/scipy-1.16.2/reference/generated/scipy.stats.pearsonr.html}} function from the \textsc{scipy.stats} library, to determine the presence of correlations between these parameters. SN~2023bvh were excluded from these test because only a lower limit on $t_\mathrm{IIn}$ is available. For the comparison between $t_\mathrm{IIn}$ and $t_\mathrm{min}$, we obtain a Pearson correlation coefficient of $r$ = 0.86 (p $ = 7\,\times\,10^{-4}$), indicating a strong and statistically significant correlation, which is also evident from Fig.~\ref{f:times.comparisons}. In contrast, the comparison between $t_\mathrm{IIn}$ and $t_\mathrm{break}$ yields a lower correlation coefficient of $r$ = 0.49 (p = 0.1), suggesting a weaker and statistically non-significant correlation. We caution, however, that these results should be considered tentative due to the limited sample size (n=12). A large sample will be required to confirm these correlations.

\section{Discussion}\label{sec:discussion}

\subsection{Ledge SNe}\label{sec:disc.ledge}

In this work, we incorporated the ledge flash-spectroscopic class for those SNe~II that showed the characteristic ledge-like feature but did not display any narrow `flash' emission. Although some flash SNe also display a ledge feature in their early-time spectra, we did not include those events in the ledge class. This differs from the scheme adopted by \cite{Bruch2023}, who considered SNe~II showing only a ledge feature as part of their flash-feature sample.  

Despite many studies discussing ledge features in SNe~II (e.g. \citealp{Hosseinzadeh2022_21yja,Bruch2023,Andrews2024_22jox,Shrestha2024_23axu}), the origin of this spectral feature remains unknown. Several physical scenarios have been proposed, based on a CSI origin: a broad, blueshifted He \textsc{ii} $\lambda$4686  emission generated by SN ejecta beneath a CSM shell (\citealp{Bullivant2018,Andrews2019-17gmr}); a blend of multiple flash-ionised lines from the CSM (\citealp{Soumagnac2020,Bruch2023}); the broadened, blueshifted interaction lines from an RSG with an extended atmosphere surrounded by a low-density CSM (\citealp{Dessart2017,Hosseinzadeh2022_21yja}); or dense fragments of outer ejecta heated by the forward shock in a tenuous progenitor wind \citep{Chugai2023_ledge}. Recently, \cite{Dessart25} reported that the ledge feature can emerge at different evolutionary phases in models spanning a wide range of CSM densities, from high mass-loss rates ($\approx 10^{-2} \,\mathrm{M}_\odot \,\mathrm{yr}^{-1}$) to low mass-loss rates ($\approx 10^{-6} \,\mathrm{M}_\odot \,\mathrm{yr}^{-1}$).

\begin{table}
\renewcommand{\arraystretch}{1.}
\centering
\caption{Previous works' mean peak absolute magnitude.}
\vspace{-0.2cm}
\resizebox{\columnwidth}{!}{
\begin{tabular}{lccc}
\hline\hline
Reference & Band & $ M_\mathrm{peak}$ [mag]& Number of SNe\\ \hline
\multirow{1}{*}{\cite{Richardson2002}}     & $B$           & $-17.03 \pm 1.05$ & 45 \\
\multirow{1}{*}{\cite{Anderson2014}}       & $V$           & $-16.74 \pm 1.01$ & 68 \\
\multirow{1}{*}{\cite{Gall2015}}           & \textit{r'/R} & $-17.11 \pm 1.03$ & 20  \\
\multirow{1}{*}{\cite{Rubin2016}}          & $R$           & $-17.16 \pm 0.98$ & 57  \\
\multirow{5}{*}{\cite{Galbany.2016}}       & $U$           & $-16.06 \pm 1.74$ & 51 \\
                                           & $B$           & $-16.43 \pm 1.19$ & 51 \\ 
                                           & $V$           & $-16.89 \pm 0.98$ & 51 \\ 
                                           & $R$           & $-16.96 \pm 1.03$ & 51 \\ 
                                           & $I$           & $-17.38 \pm 0.95$ & 51 \\ 
\multirow{1}{*}{\cite{Hinds2025}}          & $g$           & $-18.10 \pm 0.05$ & 377 \\\hline

\end{tabular}
}
\label{t:prev.works.magpeak}
\end{table}

\begin{table}
\renewcommand{\arraystretch}{1.2}
\centering
\caption{Previous works' mean time to peak.}
\vspace{-0.2cm}
\resizebox{\columnwidth}{!}{
\begin{tabular}{lccc}
\hline\hline
Reference & Band & $ t_\mathrm{peak}$ [days] & Number of SNe\\ \hline
\multirow{1}{*}{\cite{Gall2015}}           & \textit{r'/R} & $10.14 \pm 4.42$      & 20  \\
\multirow{4}{*}{\cite{GonzalezGaitan2015}} & $u$           & $5.9^{+3.0}_{-2.1}$   & 48  \\
                                           & $g$           & $6.9^{+2.8}_{-2.9}$   & 48  \\ 
                                           & $r$           & $10.6^{+7.9}_{-4.2}$  & 48  \\ 
                                           & $i$           & $12.0^{+11.2}_{-4.0}$ & 48  \\ 
\multirow{1}{*}{\cite{Rubin2016}}          & $R$           & $7.74 \pm 2.93$       & 57  \\
\multirow{3}{*}{\cite{Pessi2019}}          & $B$           & $8.3 \pm 2.0$         & 73 \\
                                           & $V$           & $12.8 \pm 2.4$        & 73 \\ 
                                           & $r$           & $16.0 \pm 3.6$        & 73 \\ \hline

\end{tabular}
}
\label{t:prev.works.tpeak}
\end{table}

Our analysis shows that ledge SNe generally exhibit photometric similarities with non-flash SNe and can be distinguished from class 1+2 SNe. Nevertheless, they display intermediate characteristics between class 1+2 and non-flash SNe in terms of the rise times in the redder bands and colour evolution, particularly in the (\textit{$g-r$}) color. \cite{Ertini.2026} report a similar behaviour for the peak absolute magnitude and post-peak decline rates, finding more significant differences when ledge and non-flash SNe were compared together against flash SNe than when ledge SNe were grouped with the flash SNe. Reproducing these comparison cases for all light- and colour-curve parameters yields consistent results, generally showing lower significance levels or even no clear distinction between the populations. The only exception is in the $i$-band time to peak, where ledge SNe follow a distribution comparable to that of flash SNe and thereby increasing the significance of the distinction relative to non-flash objects. Overall, this pattern suggests that ledge SNe share CSM properties similar to those of non-flash SNe, lending support to scenarios involving relatively low-density CSM, as mentioned above. Consequently, the sole detection of the ledge does not constitute conclusive evidence for a sufficiently dense CSM capable of producing narrow flash-ionised emission lines, in contrast to the interpretation proposed by \cite{Bruch2023}. 

\subsection{Comparison with previous works}

\subsubsection{Rise times and peak magnitudes}

Several previous studies have examined the early-time light-curve parameters of SNe~II. However, most do not incorporate a flash-spectroscopic classification as performed in this work. Tables~\ref{t:prev.works.magpeak} and \ref{t:prev.works.tpeak} list the peak absolute magnitudes and times to peak reported in previous works without this distinction, respectively. Considering both the POISE sample and the entire sample, our results for these parameters are generally consistent with those reported in the literature within the uncertainties. Furthermore, in agreement with \cite{GonzalezGaitan2015}, we also find that rise times increase with the filters' effective wavelength. 

Among the works that incorporate flash-spectroscopic classification, the adopted criteria vary. \cite{Bruch2023} classify ledge SNe as part of the group with flash features (see Sect.~\ref{sec:disc.ledge}), whereas \cite{JG2024} group ledge and non-flash SNe together in their `comparison sample' (i.e. SNe lacking flash-ionisation features). In the case of \cite{Ertini.2026}, the ledge SNe are analysed separately from the other classes, as in this work. 

These studies investigate peak-magnitude and rise-time distributions between SNe with and without flash features. \cite{JG2024} obtain distinct optical $M_{\mathrm{peak}}$ between these groups, consistent with our results, as expected given that more than half of our sample is from their dataset. Similarly, \cite{Ertini.2026} obtain comparable results for the ATLAS $o$ band (equivalent to $r+i$ bands) when comparing flash and non-flash SNe.
In contrast, \cite{Bruch2023} find no statistically significant differences in the $g$ and $r$ bands. This discrepancy cannot be explained by extinction effects, since \cite{Bruch2023}, as in our work, do not correct for host-galaxy reddening, whereas \cite{JG2024} do. Instead, we attribute the disagreement with \cite{Bruch2023} to the smaller sample they used and, most importantly, to the inclusion of ledge SNe within the flash group. 

With respect to rise times, \cite{Bruch2023} and \cite{Ertini.2026} conclude that SNe with and without flash features are drawn from the same parent distribution in the $g$, $r$, and $o$ bands. These findings contrast with our results in the $g$ band (see Sect.~\ref{sec:rise.distributions}) and arise from the same factors discussed above in the case of \cite{Bruch2023}. On the other hand, \cite{JG2024} compare ledge and non-flash SNe combined against flash SNe in the $r$ band, reporting no statistically significant differences between their parent distributions. In our framework, this corresponds to comparison case b (flash vs. non-flash + ledge) in the $r$ band (see Sect.~\ref{sec:rise.distributions}) and is consistent with the results of \cite{JG2024}.

\subsubsection{Colour-curve slopes}\label{sec:disc.colours}

The work by \cite{dejaeger2018} represents, to our knowledge, the most extensive study of colour curves for a large sample of SNe~II. However, they do not include a flash-spectroscopic classification. Since most of their colour curves lack early-time coverage, the parameters they derive describe the plateau phase, whereas our analysis focuses on the first 30 days, near the peak. Consequently, the parameters defined in both works are not directly comparable. Nonetheless, our second colour slope (i.e. $cs_1$) approximately corresponds to their first slope (denoted as $s_1$), as both describe a similar temporal phase of the colour evolution.

Considering all classes together within our gold sample (Table~\ref{t:colour.curve.parameters}), our results for $cs_1$ are in good agreement, within the uncertainties, with the values of $s_1$ reported by \cite{dejaeger2018}, $1.90\pm0.53$, $2.63\pm0.62$, and $6.66\pm2.14$ mag (100 d)$^{-1}$ in the (\textit{$g-r$}), (\textit{$B-V$}) and (\textit{$u-g$}) colours, respectively. When comparing individual classes, we find that their median $s_1$ values lie between those of our class 1+2 and non-flash SNe, suggesting that their sample likely includes a combination of events with and without flash features.

\cite{JG2024} also investigate early-time colour curves, particularly in the ultraviolet and (\textit{$g-r$}) colour, reporting bluer colours for most of the class 1 and 2 SNe. Despite not applying host-galaxy extinction corrections, we observe a similar trend in the optical range. \cite{JG2024} suggest that this behaviour may be associated with continued interaction with the CSM. 

\subsection{Clues from CSM interaction models}\label{sec:disc.models}

Theoretical models indicate that the presence of flash features in the early spectra of SNe~II arises from the interaction between the SN ejecta and a dense CSM \citep{Yaron2017,Dessart2017}. The existence of a dense CSM surrounding the progenitor star also affects the early evolution of both light and colour curves \citep{Moriya2017}. Moreover, the early-time morphology of these curves is strongly dominated by the configuration of the CSM, such as its radius, mass-loss rate ($\dot{M}$), and wind velocity, rather than by the explosion energy of the SN or the progenitor properties (such as M$_{\mathrm{ZAMS}}$ and $^{56}$Ni mass), whose influence is notable during the later evolution, in the plateau phase and beyond \citep{Moriya2018,Moriya2023,Dessart.2023}. 

A comparison between model predictions and early-time observations is crucial for constraining the progenitor evolution and its mass-loss history. Notably, the cadence and photometric precision of the POISE data enable us to detect details in the curves that may be linked to specific CSM configurations, particularly in the colour curves. 

Light-curve models predict that the optical rise times and peak absolute magnitudes are influenced by the presence of CSM. Increasing $\dot{M}$ ($\gtrsim\,10^{-3}\,\mathrm{M}_\odot\, \mathrm{yr}^{-1}$) and/or extending the CSM to larger radii yield shorter rise times and brighter peaks \citep{Dessart2017,Moriya2018,Moriya2023}. In our analysis (see Sect.~\ref{sec:light_curves}), SNe~II with strong flash features (i.e. class 1+2 SNe) display significantly brighter peak magnitudes than non-flash SNe, whereas their rise times are slightly longer, except in the $r$ and $i$ bands. At very high $\dot{M}$ ($\gtrsim 10^{-2}\,\mathrm{M}_\odot\, \mathrm{yr}^{-1}$), however, the rise times increase and the peak becomes fainter due to the longer photon diffusion times through the denser CSM \citep{Moriya2018,Forster2018}. Within our sample, three class 1+2 SNe (SN~2022ffg, SN~2022prv, and SN~2023bvh) show long rise times in the \textit{BgVri} bands ($\gtrsim \,15$ d), intermediate peak luminosities (among the class 1+2 SNe), and longer $t_{\mathrm{IIn}}$, suggesting that they may have the most massive CSM structures in our dataset. 
Figures 4 and 5 from \cite{Moriya2023} show how the rise-time distribution evolves as $\dot{M}$ increases in the \textit{g} and \textit{r} bands.
The trend observed in these figures qualitatively matches the behaviour shown in our Fig.~\ref{f:delta_t_max}, pointing to the presence of extreme values of $\dot{M}$ in our sample. These predictions assume steady or accelerated winds as the drivers of mass loss; however, other mechanisms such as binary interaction, RSG pulsations, or eruptive events could be involved. 

Regarding colour curves, Fig.~4 (bottom panel) from \cite{Moriya2018} illustrates the effect of $\dot{M}$ on the evolution of the (\textit{$g-r$}). In general, their models show a well-defined minimum in colour, which becomes broader and occurs at later times as $\dot{M}$ increases. This appears to be qualitatively consistent with our observations in Sect.~\ref{sec:break.time.flash}, with delayed minima for SNe with longer $t_\mathrm{IIn}$. In models without CSM or with $\dot{M}\, \lesssim \,10^{-4}\,\mathrm{M}_\odot\, \mathrm{yr}^{-1}$, this minimum is weak or occurs close to the explosion epoch ($\lesssim 3$ days). Therefore, the presence and timing of this feature may serve to constrain $\dot{M}$ and, by extension, the CSM properties. 

Furthermore, as also seen in Fig.~4 of \cite{Moriya2018}, a subtle break in the model colour curves appears within the same temporal range as the break we identified (see Sect.~\ref{sec:colour.curves}). The behaviour differs among models: curves with $\dot{M} \gtrsim\,10^{-3}\,\mathrm{M}_\odot\, \mathrm{yr}^{-1}$ follow trends similar to those of class 1+2 SNe, while lower $\dot{M}$ models resemble the non-flash. We note, however, that model assumptions and approximations should be taken into account before drawing any conclusions about this apparent consistency with the observational evidence. Nevertheless, this suggests that the early-time colour evolution may be a good diagnostic of progenitor mass loss. We leave a quantitative comparison for future work. 

\section{Conclusions} 
\label{sec:conclusions}

In this paper, we analysed the early-time photometric and spectroscopic properties of 86 SNe~II, including 27 unpublished SNe followed-up by the POISE collaboration, all with at least one optical spectrum. Based on the presence or absence of flash features, that is, early-time high-ionised, electron-scattered emission lines, which are signatures of CSI, we classified the sample into class 1 (strong He \textsc{ii} and N \textsc{iii} flash features), class 2 (only strong He \textsc{ii} flash features without N \textsc{iii}), class 3 (weak He \textsc{ii} flash features without N \textsc{iii}), ledge (broad, asymmetric emission feature around 4400--4800 \r{A}), and non-flash (no flash or ledge features). Our main conclusions are the following:

\begin{itemize}

    \item Class 1+2 SNe are, on average, more luminous than ledge and non-flash SNe across the \textit{uBgVri} bands. K-S and A-D tests confirm that their peak absolute magnitudes arise from different distributions in both comparison cases: class 1+2 vs. non-flash and flash (class 1+2+3) vs. non-flash + ledge (see Sect.~\ref{sec:rise.distributions}). These results are consistent with those of \cite{JG2024} and \cite{Ertini.2026}, but differs from those of \cite{Bruch2023}. 
    
    \item Rise times only show significant distinctions between flash-spectroscopic classes in the $B$ band.  
    The rise times of class 1+2 SNe are correlated across all bands, whereas for non-flash and ledge SNe the rise times in the bluer bands (\textit{BgV}) remain nearly constant while those in redder bands (\textit{ri}) increase (see Sect.~\ref{sec:rise.distributions} and Fig.~\ref{f:delta_t_max}). This suggests different powering sources for class 1+2 (CSI) and for non-flash + ledge SNe (post-shock cooling). Additionally, three class 1+2 SNe (SN~2022ffg, SN~2022prv, and SN~2023bvh) display the longest rise times in the bluer bands ($\gtrsim 15$ days). 
    According to the models of \cite{Moriya2018,Moriya2023} (see Sect.~\ref{sec:disc.models}), progenitors with the highest mass-loss rates ($\gtrsim 10^{-2}\,\mathrm{M}_\odot\, \mathrm{yr}^{-1}$) are expected to show longer rise times in the bluer bands, as photons experience longer diffusion times in a denser CSM. This suggests that these SNe are associated with high mass-loss rates.
    
    \item Class 1+2 SNe are, in general, bluer than non-flash SNe in the (\textit{$g-r$}), (\textit{$B-V$}), (\textit{$g-i$}), and (\textit{$u-g$}), consistent with the trends reported for (\textit{$g-r$}) by \cite{JG2024}. Colours at 10 and 20 days exhibit statistically significant differences between the two groups in some cases. We associate these differences with intrinsic SN properties rather than with host-galaxy extinction (see Sect.~\ref{sec:colour.phases}). These properties are potentially related to an early luminosity boost caused by CSM interaction, which can lead to bluer colours.

    \item Early-time colour curves of SNe~II (between 5 and 30 days after first light) display distinct behaviours according to the presence or absence of flash features. Representing these curves by two segments with a break, we measure the slopes ($cs_0$, $cs_1$), break time ($t_{\mathrm{break}}$), and slope change ($\Delta cs = cs_0 - cs_1$). In the gold sample (see Sect.~\ref{sec:colour.curves.parameters}), the majority of the slope parameters
    show significant differences between class 1+2 and non-flash SNe (case a; see Sect.~\ref{sec:slopes.correlations}). These distinctions persist when comparing flash SNe with non-flash + ledge SNe (case b). The inclusion of lower-quality colour curves (complete sample) generally preserves the separations in $cs_0$, $cs_1$, and $\Delta cs$ for both comparison cases. $\Delta cs$ provides the strongest and most significant distinctions for all colours. The combination of $\Delta cs$ and peak absolute magnitude defines a parameter space (see Fig.~\ref{f:diff_s_magabs}) that provides an effective diagnostic to distinguish SNe II with strong CSI, even without early-time spectroscopy. 

    \item Colour-curve break times ($t_\mathrm{break}$) show significant differences only in the (\textit{$u-g$}) colour, with flash SNe exhibiting systematically later break times than non-flash and ledge objects. In all flash SNe, this break occurs after the disappearance of the flash features ($t_{\mathrm{IIn}}$) in every colour. 
    In contrast, the colour minimum ($t_\mathrm{min}$) consistently occurs earlier than, or simultaneously with, the duration of the flash features (within uncertainties), similar to what \cite{Shrestha2024_24ggi} and \cite{Andrews2025} suggest. Therefore, the duration of the flash features is bounded by the phase of the colour minimum and the break time (see Fig.~\ref{f:times.comparisons}), providing lower and upper limits, respectively. We also find a strong and significant correlation between $t_\mathrm{IIn}$ and $t_\mathrm{min}$, whereas the correlation between $t_\mathrm{IIn}$ and $t_\mathrm{break}$ is weaker and not statistically significant. 

    \item Ledge SNe generally resemble non-flash SNe in terms of peak absolute magnitude and rise time, consistent with the results reported by \cite{Ertini.2026}, and also show colour-curve parameters similar to those of non-flash SNe. However, they tend to occupy an intermediate region between class 1+2 and non-flash SNe across the parameter spaces considered here. Class 3 SNe share several light-curve properties with class 1+2 SNe, but their colour evolution exhibits mixed behaviour, aligning with either class 1+2 or non-flash SNe depending on the parameter.

\end{itemize}

In the near future, with the operation of current surveys (ASAS-SN, ZTF, ATLAS, GOTO, Pan-STARRS) and the advent of new ones such as LSST and LS4, the number of SN~II discoveries will increase dramatically. Spectroscopic monitoring will only be achievable for a small fraction of events. Consequently, photometry will play a fundamental role in the early phases due to its easier accessibility. As we show, early-time light and colour curves of SNe~II show distinct evolutionary morphologies depending on the presence or absence of flash features. This highlights the importance of identifying photometric parameters that may serve to characterise early interaction signatures and thereby determine properties of the CSM. Notably, the trends we report are reproduced across all colours, implying that multiple bands and colour combinations can be used effectively. 

Future work involving quantitative comparisons with models using these parameters may help constrain the physical properties of the explosion and the CSM, thereby offering new insights into the mass-loss history during the final stages of the progenitor evolution. Finally, current and future surveys may lack the photometric precision and observational cadence necessary to detect the subtle differences in colour-curve parameters that we introduced here. Therefore, dedicated follow-up programmes similar to POISE can contribute substantially to future studies of SNe~II.

\section{Data availability}\label{sec:data_availability}

The POISE SN photometry tables will be available at CDS via \url{http://cdsweb.u-strasbg.fr/cgi-bin/qcat?J/A+A/} after publication. 
The spectra used to classify POISE SNe will also be available after publication on the Weizmann Interactive Supernova Data Repository (WISeREP, \citealp{wiserep}) at \url{https://www.wiserep.org}. Tables \ref{t:Lit.spectra}, \ref{t:POISE.peak.magnitude} to \ref{t:lit.rise.times}, \ref{t:POISE.slopes_gr}, and \ref{t:lit.slopes_gr}, along with additional tables, are available at \url{https://github.com/axelmendezllorca/Flashing_with_colours_tables}.

\begin{acknowledgements} 
We thank Joseph Anderson for valuable discussions. 
This research was partially funded by UNRN PI2020 40B1039, Programa Bilateral de intercambio COOPB23040.
A.M.-L. Acknowledges financial support from Consejo Nacional de Investigaciones Científicas y Técnicas (CONICET). 
C.P.G. acknowledges financial support from the Secretary of Universities and Research (Government of Catalonia) and by the Horizon 2020 Research and Innovation Programme of the European Union under the Marie Sk\l{}odowska-Curie and the Beatriu de Pin\'os 2021 BP 00168 programme. 
C.P.G. and L.G. recognise the support from the Spanish Ministerio de Ciencia e Innovaci\'on (MCIN) and the Agencia Estatal de Investigaci\'on (AEI) 10.13039/501100011033 under the PID2023-151307NB-I00 SNNEXT project, from Centro Superior de Investigaciones Cient\'ificas (CSIC) under the PIE project 20215AT016 and the program Unidad de Excelencia Mar\'ia de Maeztu CEX2020-001058-M, and from the Departament de Recerca i Universitats de la Generalitat de Catalunya through the 2021-SGR-01270 grant.
The data presented here were obtained in part with ALFOSC, which is provided by the Instituto de Astrofisica de Andalucia (IAA) under a joint agreement with the University of Copenhagen and NOT.
M.D. Stritzinger is funded by the Independent Research Fund Denmark (IRFD, grant number  10.46540/2032-00022B). J.T.H. acknowledges support from NASA through the NASA Hubble Fellowship grant HST-HF2-51577.001-A, awarded by STScI. STScI is operated by the Association of Universities for Research in Astronomy, Incorporated, under NASA contract NAS5-26555. W.B.H. acknowledges support from the National Science Foundation Graduate Research Fellowship Program under Grant No. 2236415. 

Some of the data presented herein were obtained at the W. M. Keck Observatory, which is operated as a scientific partnership among the California Institute of Technology,
the University of California, and NASA. The Observatory was made possible by the generous financial support of the W. M. Keck Foundation. The authors wish to recognise and acknowledge the very significant cultural role and reverence that the summit of Maunakea has always had within the indigenous Hawaiian community. We are most fortunate to have the opportunity to conduct observations from this mountain.

\end{acknowledgements}
%
\bibliographystyle{aa} 
\bibliography{biblio} 
%
\onecolumn

\begin{appendix}
\section{Data sample}
\FloatBarrier

\begin{table}[h!]
\centering
\resizebox{\textwidth}{!}{%
\begin{threeparttable}
\caption{SNe~II POISE sample}
\label{t:POISE.sample}
\begin{tabular}{lcccccccccc}
\hline\hline
SN   & \begin{tabular}[c]{@{}c@{}}Flash \\ classification \end{tabular} &\begin{tabular}[c]{@{}c@{}}Time of First \\ Light\\ (MJD)\end{tabular} & Host Galaxy & $z_\mathrm{H}$\tnote{$a$} & $z_\mathrm{CMB}$\tnote{$b$} & \begin{tabular}[c]{@{}c@{}}Distance\\ {[}Mpc{]}\end{tabular} & \begin{tabular}[c]{@{}c@{}}E($B - V$)$_\mathrm{MW}$\\ {[}mag{]}\end{tabular} & \begin{tabular}[c]{@{}c@{}}E($B - V$)$_\mathrm{host}$\\ {[}mag{]}\end{tabular} & \begin{tabular}[c]{@{}c@{}}Phase of First \\ Photometric \\Observation {[}days{]}\end{tabular} & \begin{tabular}[c]{@{}c@{}}Phase of First \\ Spectroscopic \\Observation {[}days{]}\end{tabular}\\ \hline
2021bmw   & Non-flash   & $59246.6\pm1.0$ & IC 4325                   &  0.0128\tablefootmark{e} & 0.0137    &  $57.8\pm5.4$\tablefootmark{$c$}          & 0.054  &  0.30 $\pm$ 0.07  & 2.7  & 3.7  \\
2021can   & Class 2     & $59252.0\pm0.3$ & SDSS J125530.88+300142.3  &  0.0207\tablefootmark{$f$} & 0.0216  &  $91.7\pm6.2$\tablefootmark{$c$}          & 0.010  &  0.03 $\pm$ 0.01  & 2.2  & 1.3  \\
2021cgu   & Class 2     & $59252.4\pm1.0$ & CGCG 038-099              &  0.0255\tablefootmark{$g$} & 0.0267  &  $113.8\pm6.7$\tablefootmark{$c$}         & 0.033  &  0.10 $\pm$ 0.03  & 1.7  & 3.8  \\
2021dbg   & Class 1     & $59259.5\pm0.9$ & MCG-01-24-014             &  0.0216\tablefootmark{$h$} & 0.0216  &  $91.7\pm6.2$\tablefootmark{$c$}          & 0.028  &  0.11 $\pm$ 0.03  & 2.7  & 1.7  \\
2021adlr  & Non-flash   & $59521.9\pm0.6$ & WISEA J031111.51-082806.6 &  0.0296\tablefootmark{$g$} & 0.0291  &  $124.0\pm6.9$\tablefootmark{$c$}         & 0.058  &  0.04 $\pm$ 0.02  & 3.0  & 2.6  \\
2021adxd  & --          & $59523.3\pm0.9$ & NGC 0198                  &  0.0175\tablefootmark{$i$} & 0.0164  &  $69.5\pm5.7$\tablefootmark{$c$}          & 0.023  &  0.08 $\pm$ 0.03  & 1.3  & 5.0  \\
2021afkk  & Class 1     & $59542.4\pm1.0$ & UGC 1971                  &  0.0152\tablefootmark{$i$} & 0.0145  &  $61.3\pm5.5$\tablefootmark{$c$}          & 0.093  &  0.12 $\pm$ 0.06  & 1.8  & 4.2  \\ 
2021agbn  & Ledge       & $59549.2\pm1.7$ & ESO 483 - G 002           &  0.0147\tablefootmark{$j$} & 0.0146  &  $61.7\pm5.5$\tablefootmark{$c$}          & 0.035  &  0.06 $\pm$ 0.02  & 1.7  & 2.8  \\
2022ewj   & --          & $59654.4\pm3.1$ & NGC 3367                  &  0.0101\tablefootmark{$k$} & 0.0113  &  $47.6\pm5.2$\tablefootmark{$c$}          & 0.025  &  0.08 $\pm$ 0.06  & 3.6  & 3.1  \\
2022eyk   & --          & $59657.2\pm3.1$ & IC 4857                   &  0.0156\tablefootmark{$j$} & 0.0153  &  $64.7\pm5.7$\tablefootmark{$c$}          & 0.072  &  0.03 $\pm$ 0.01  & 4.1  & 4.1  \\
2022fae   & Non-flash   & $59659.7\pm1.5$ & --                        &  0.04\tablefootmark{$l$}   & 0.0411  &  $177.2\pm8.1$\tablefootmark{$c$}         & 0.058  &  $<0.01$          & 3.5  & 3.6  \\
2022ffg   & Class 1     & $59663.8\pm0.5$ & CGCG 093-074              &  0.012\tablefootmark{$g$}  & 0.0131  &  $55.24\pm5.4$\tablefootmark{$c$}         & 0.035  &  0.09 $\pm$ 0.05  & 1.3  & 2.5  \\
2022frq   & Non-flash   & $59670.9\pm0.5$ & MCG-02-34-054             &  0.0226\tablefootmark{$i$} & 0.0236  &  $100.3\pm6.4$\tablefootmark{$c$}         & 0.057  &  0.14 $\pm$ 0.09  & 1.2  & 1.5  \\
2022zkc   & --          & $59884.6\pm0.6$ & GIN227                    &  0.0317\tablefootmark{$m$} & 0.0316  &  $135.4\pm7.1$\tablefootmark{$c$}         & 0.045  &  0.13 $\pm$ 0.06  & 2.5  & 13.9  \\
2022aacn  & Non-flash   & $59895.9\pm0.5$ & WISEA J004427.16+072233.0 &  0.0175\tablefootmark{$n$} & 0.0164  &  $69.4\pm5.7$\tablefootmark{$c$}          & 0.032  &  0.08 $\pm$ 0.02  & 1.2  & 2.5  \\
2022aagp  & Non-flash   & $59896.2\pm1.5$ & NGC 2777                  &  0.0049\tablefootmark{$o$} & 0.0059  &  $25.7$\tablefootmark{\textit{d,t}}       & 0.037  &  0.06 $\pm$ 0.05  & 7.2  & 1.7  \\
2022aatx  & --          & $59901.4\pm1.0$ & --                        &  0.0165\tablefootmark{$l$} & 0.0175  &  $74.1\pm5.8$\tablefootmark{$c$}          & 0.025  &  0.18 $\pm$ 0.02  & 1.9  & 6.7  \\
2022abgs  & --          & $59898.0\pm3.3$ & NGC 3508                  &  0.013\tablefootmark{$p$}  & 0.0142  &  $59.9\pm5.5$\tablefootmark{$c$}          & 0.016  &  -- \tablefootmark{w} & 8.4  & 11.5  \\
2022acko  & Ledge       & $59919.0\pm0.9$ & NGC 1300                  &  0.0053\tablefootmark{$i$} & 0.0048  &  $19.0\pm2.9$\tablefootmark{\textit{d,u}} & 0.026  &  0.05 $\pm$ 0.03  & 1.0  & 1.0  \\
2023bvh   & Class 1     & $59990.5\pm2.6$ & KUG 0843+275              &  0.0187\tablefootmark{$q$} & 0.0195  &  $82.7\pm6.0$\tablefootmark{$c$}          & 0.047  &  $<0.01$          & 6.6  & 5.6  \\
2023bvj   & Non-flash   & $59991.2\pm2.0$ & NGC 3021                  &  0.0051\tablefootmark{$i$} & 0.006   &  $28.6\pm2.3$\tablefootmark{\textit{d,v}} & 0.01   &  0.46 $\pm$ 0.15  & 5.0  & 2.1  \\
2023cmz   & --          & $59998.4\pm0.6$ & --                        &  0.025\tablefootmark{$l$}  & 0.0254  &  $108.3\pm6.5$\tablefootmark{$c$}         & 0.086  &  $<0.01$          & 1.8  & 5.8  \\
2023cnd   & --          & $59998.8\pm0.5$ & WISEA J143907.21-091832.4 &  0.0263\tablefootmark{$h$} & 0.027   &  $115.1\pm6.7$\tablefootmark{$c$}         & 0.08   &  0.06 $\pm$ 0.02  & 1.5  & 5.6  \\
2023cps   & Ledge       & $60001.7\pm0.5$ & CGCG 031-018              &  0.0136\tablefootmark{$r$} & 0.0144  &  $60.8\pm5.5$\tablefootmark{$c$}          & 0.024  &  0.33 $\pm$ 0.12  & 1.4  & 1.4  \\
2023ddh   & --          & $60015.3\pm2.9$ & ESO 340 - G 005           &  0.0216\tablefootmark{$h$} & 0.021   &  $89.7\pm6.1$\tablefootmark{$c$}          & 0.053  &  0.12 $\pm$ 0.04  & 4.1  & 5.1  \\
2023ddy   & --          & $60016.6\pm1.3$ & --                        &  0.059\tablefootmark{$l$}  & 0.0602  &  $263.4\pm9.9$\tablefootmark{$c$}         & 0.033  &  $<0.02$          & 3.6  & 4.6  \\
2023dtd   & Non-flash   & $60022.6\pm0.6$ & NGC 5260                  &  0.0217\tablefootmark{$s$} & 0.0226  &  $96.1\pm6.3$\tablefootmark{$c$}          & 0.071  &  0.34 $\pm$ 0.14  & 1.7  & 2.7  \\
\hline
\end{tabular}
\tablefoot{\tablefoottext{a}~Heliocentric redshift. \tablefoottext{b}~Redshift referred to the CMB. \tablefoottext{c}~Distance estimated from the redshift relative to the CMB (see Sect.~\ref{sec:distances}). \tablefoottext{d}~Redshift-independent distance. \tablefoottext{e}~\cite{Kaldare2003}. \tablefoottext{f}~\cite{JG2024}. \tablefoottext{g}~\cite{SDSS.DR13.Albareti2017}. \tablefoottext{h}~\cite{6dF.Jones2009}. \tablefoottext{i}~\cite{Springob2005}. \tablefoottext{j}~\cite{Mathewson1996}. \tablefoottext{k}~\cite{Lu1993}. \tablefoottext{l}~This work. \tablefoottext{m}~\cite{Rines2003}. \tablefoottext{n}~\cite{Haynes2018}. \tablefoottext{o}~\cite{SDSS.DR3.Abazajian2005}. \tablefoottext{p}~\cite{HYPERLEDA.Paturel2003}. \tablefoottext{q}~\cite{dellAntonio1996}. \tablefoottext{r}~\cite{Falco1999}. \tablefoottext{s}~\cite{Theureau1998}. \tablefoottext{t}~\cite{Karachentsev2015}. \tablefoottext{u}~\cite{Bostroem.acko.2023}. \tablefoottext{v}~Mean Cepheids distance (\citealp{Riess2009,Riess2011,Riess2016,Fiorentino2013}). \tablefoottext{w}~Only one low-resolution spectrum was available, which prevented us from measuring the Na I D line. }
\end{threeparttable}
}
\end{table}

\begin{table}[h!]
\centering
\begin{threeparttable}
\caption{List of classification spectra of the POISE sample.}
\label{t:POISE.spectra}
\begin{tabular}{lccccccc}
\hline\hline
SN   & \begin{tabular}[c]{@{}c@{}}Flash \\ classification \end{tabular} & MJD & \begin{tabular}[c]{@{}c@{}}Phase\\  (days)\end{tabular} & Telescope & Instrument & Data source \\ \hline
2021bmw   & Non-flash  & 59250.3 & 3.7  & Baade & IMACS & This work (POISE) \\
2021can   & Class 2    & 59253.0 & 1.0  & NOT & ALFOSC & \cite{JG2024}  \\
2021cgu   & Class 2    & 59256.2 & 3.8  & Baade & IMACS & This work (POISE) \\
2021dbg   & Class 1    & 59261.2 & 1.7  & NTT & EFOSC2 & ePESSTO+ (TNS, public) \\
2021adlr  & Non-flash  & 59524.5 & 2.6  & UH2.2 & SNIFS & This work (SCAT) \\
2021afkk  & Class 1    & 59546.6 & 4.2  & Lijang 2.4m & YFOSC & TNS (public) \\ 
2021agbn  & Ledge      & 59552.0 & 2.8  & NTT & EFOSC2 & ePESSTO+ (TNS, public) \\
2022fae   & Non-flash  & 59663.3 & 3.6  & UH2.2  & SNIFS & This work (SCAT) \\
2022ffg   & Class 1    & 59667.4 & 3.6  & Keck & LRIS & This work (SCAT)\\
2022frq   & Non-flash  & 59672.4 & 1.5  & FTN  & FLOYDS-N & TNS (public) \\
2022aacn  & Non-flash  & 59898.4 & 2.5  & UH2.2  & SNIFS & This work (SCAT) \\
2022aagp  & Non-flash  & 59897.8 & 1.7  & 3.8m Seimei  & KOOLS-IFU & TNS (public) \\
2022acko  & Ledge      & 59920.0 & 1.0  & SALT & RSS & \cite{Bostroem.acko.2023} \\
2023bvh   & Class 1    & 59996.1 & 5.6  & NTT & EFOSC2 & ePESSTO+ (TNS, public) \\
2023bvj   & Non-flash  & 59993.3 & 2.1  & P60  & SEDM & ZTF (TNS, public) \\ 
2023cps   & Ledge      & 60003.1 & 1.4  & NTT & EFOSC2 & ePESSTO+ (TNS, public)  \\
2023dtd   & Non-flash  & 60025.3 & 2.7  & Baade & IMACS & This work (POISE) \\
\hline
\end{tabular}
\end{threeparttable}
\end{table}

\begin{table}
\centering
\resizebox{\textwidth}{!}{
\begin{threeparttable}
\caption{SNe~II literature sample}
\label{t:Lit.sample}
\begin{tabular}{lcccccccc}
\hline\hline
SN   & \begin{tabular}[c]{@{}c@{}}Flash \\ classification \end{tabular} &  \begin{tabular}[c]{@{}c@{}}Time of First \\ Light\\ (MJD)\end{tabular} &  Host Galaxy & \begin{tabular}[c]{@{}c@{}}Heliocentric \\ Redshift\\ $z_\mathrm{H}$\end{tabular} & \begin{tabular}[c]{@{}c@{}}Redshift\\ referred to CMB \\ $z_\mathrm{CMB}$\end{tabular} & \begin{tabular}[c]{@{}c@{}}Distance\\ {[}Mpc{]}\end{tabular} & \begin{tabular}[c]{@{}c@{}}E($B - V$)$_\mathrm{MW}$\\ {[}mag{]}\end{tabular} & \begin{tabular}[c]{@{}c@{}}E($B - V$)$_\mathrm{host}$\\ {[}mag{]}\end{tabular}  \\ \hline
2012aw     & Ledge      & $56002.1 \pm 1.0$ & NGC 3351                      & 0.0026\tablefootmark{$e$}          & 0.0038  & $9.8 \pm 0.2$\tablefootmark{\textit{b,ah}}& 0.025 & 0.18 $\pm$ 0.04 \\
2013ab     & Ledge      & $56339.0 \pm 1.0$ & NGC 5669                      & 0.0046\tablefootmark{$e$}          & 0.0053  & $24.0 \pm 0.9$\tablefootmark{$b$}         & 0.024 & 0.11 $\pm$ 0.02 \\
2013fs     & Class 3    & $56571.1 \pm 0.02$\tablefootmark{$c$}& NGC 7610   & 0.0118\tablefootmark{$e$}          & 0.0106  & $44.7 \pm 5.2$\tablefootmark{$a$}         & 0.036 & 0.02 $\pm$ 0.004 \\
2013hj     & Ledge      & $56636.8 \pm 2.0$ & MCG -02-24-3                  & 0.0069\tablefootmark{$f$}          & 0.0079  & $33.2 \pm 4.9$\tablefootmark{$a$}         & 0.047 & 0.16 $\pm$ 0.07 \\
2014G      & Class 2    & $56669.6 \pm 1.7$ & NGC 3448                      & 0.0043\tablefootmark{$g$}          & 0.0051  & $23.7 \pm 8.8$\tablefootmark{$b$}         & 0.011 & 0.29 $\pm$ 0.06 \\
2016X      & Class 3    & $57406.4 \pm 1.1$ & UGC 08041                     & 0.0044\tablefootmark{$e$}          & 0.0055  & $15.1 \pm 3.0$\tablefootmark{$b$}         & 0.019 & $< 0.007$ \\
2016aqf    & Non-flash  & $57443.7 \pm 1.0$ & NGC 2101                      & 0.004\tablefootmark{$h$}           & 0.0042  & $13.3 \pm 0.8$\tablefootmark{$b$}         & 0.048 & 0.03 $\pm$ 0.006 \\
2016iyz    & Ledge      & $57741.4 \pm 1.0$ & IC 2151                       & 0.0104\tablefootmark{$i$}          & 0.0106  & $44.6 \pm 5.2$\tablefootmark{$a$}         & 0.064 & 0.17 $\pm$ 0.04\\
2017ahn    & Class 1    & $57792.3 \pm 0.5$ & NGC 3318                      & 0.009\tablefootmark{$j$}           & 0.0103  & $43.1 \pm 5.1$\tablefootmark{$a$}         & 0.069 & 0.21 $\pm$ 0.04 \\
2017eaw    & Ledge      & $57886.2 \pm 1.0$ & NGC 6946                      & 0.0013\tablefootmark{$k$}          & -0.00047& $6.7 \pm 0.1$\tablefootmark{$b$}          & 0.303 & 0.04 $\pm$ 0.01 \\
2017gmr    & Class 3    & $57999.2 \pm 1.0$ & NGC 988                       & 0.005\tablefootmark{$l$}           & 0.0043  & $19.6 \pm 1.4$\tablefootmark{$b$}         & 0.024 & 0.32 $\pm$ 0.07 \\
2018zd     & Class 1    & $58178.4 \pm 0.2$\tablefootmark{$d$} & NGC 2146   & 0.003\tablefootmark{$k$}           & 0.0029  & $18.4 \pm 4.5$\tablefootmark{$d$}         & 0.085 & 0.17 $\pm$ 0.03 \\
2018cuf    & Non-flash  & $58291.9 \pm 0.7$ & IC 5092                       & 0.011\tablefootmark{$j$}           & 0.0105  & $44.2 \pm 5.1$\tablefootmark{$a$}         & 0.028 & 0.27 $\pm$ 0.06 \\
2018dfc    & Class 2    & $58302.8 \pm 0.5$ & SDSS J164807.68+241815.4      & 0.0365\tablefootmark{$m$}          & 0.0366  & $157.2 \pm 7.6$\tablefootmark{$a$}        & 0.075 & 0.14 $\pm$ 0.03 \\
2018fif    & Class 3    & $58350.9 \pm 0.5$ & UGC 00085                     & 0.017\tablefootmark{$e$}           & 0.0163  & $69.1 \pm 5.7$\tablefootmark{$a$}         & 0.096 & 0.18 $\pm$ 0.04 \\
2019edo    & Non-flash  & $58599.8 \pm 0.5$ & NGC 4162                      & 0.0086\tablefootmark{$e$}          & 0.0096  & $40.4 \pm 5.1$\tablefootmark{$a$}         & 0.031 & $< 0.01$ \\
2019enr    & Non-flash  & $58606.3 \pm 1.0$ & NGC 2919                      & 0.0082\tablefootmark{$e$}          & 0.0092  & $38.6 \pm 5.0$\tablefootmark{$a$}         & 0.026 & 0.08 $\pm$ 0.02 \\
2019nvm    & Non-flash  & $58713.7 \pm 0.5$ & UGC 10858                     & 0.0181\tablefootmark{$m$}          & 0.018   & $76.2 \pm 5.8$\tablefootmark{$a$}         & 0.026 & 0.21 $\pm$ 0.04 \\
2019nyk    & Class 2    & $58713.4 \pm 2.0$ & HIPASS J0015-08               & 0.021\tablefootmark{$n$}           & 0.0199  & $84.5 \pm 6.0$\tablefootmark{$a$}         & 0.034 & 0.09 $\pm$ 0.05\\
2019ofc    & Class 3    & $58717.1 \pm 0.9$ & --                            & 0.021\tablefootmark{$o$}           & 0.0203  & $86.2 \pm 6.1$\tablefootmark{$a$}         & 0.013 & $<0.02$ \\
2019qch    & Class 1    & $58739.6 \pm 0.5$ & WISEA J182914.07+410228.6     & 0.024\tablefootmark{$p$}           & 0.0236  & $100.5 \pm 6.4$\tablefootmark{$a$}        & 0.103 & $<0.1$ \\
2019ust    & Class 1    & $58799.8 \pm 0.5$ & UGC 00548                     & 0.022\tablefootmark{$e$}           & 0.021   & $89.2 \pm 8.0$\tablefootmark{$a$}         & 0.054 & 0.28 $\pm$ 0.06 \\
2020ekk    & Ledge      & $58918.7 \pm 1.1$ & UGC 10528                     & 0.0142\tablefootmark{$q$}          & 0.0143  & $60.5 \pm 5.3$\tablefootmark{$a$}         & 0.042 & $<0.02$ \\
2020fqv    & Class 3    & $58937.8 \pm 1.5$ & NGC 4568                      & 0.0074\tablefootmark{$m$}          & 0.0085  & $35.7 \pm 5.0$\tablefootmark{$a$}         & 0.029 & 0.81 $\pm$ 0.18 \\ 
2020jfo    & Ledge      & $58973.2 \pm 2.0$ & M61                           & 0.0052\tablefootmark{$r$}          & 0.0064  & $16.2 \pm 1.8$\tablefootmark{$b$}         & 0.021 & 0.14 $\pm$ 0.03 \\
2020lfn    & Class 3    & $58996.8 \pm 0.5$ & SDSS J162656.94+201447.8      & 0.044\tablefootmark{$s$}           & 0.0442  & $190.3 \pm 8.3$\tablefootmark{$a$}        & 0.063 & 0.08 $\pm$ 0.02 \\
2020mjm    & Non-flash  & $59010.7 \pm 1.5$ & UGC 09299                     & 0.0051\tablefootmark{$e$}          & 0.0059  & $29.4 \pm 6.1$\tablefootmark{$b$}         & 0.039 & $<0.008$ \\
2020nif    & Class 3    & $59022.7 \pm 1.5$ & NGC 4939                      & 0.010\tablefootmark{$t$}           & 0.0116  & $49.0 \pm 5.2$\tablefootmark{$a$}         & 0.035 & 0.19 $\pm$ 0.04 \\
2020pni    & Class 1    & $59045.7 \pm 0.5$ & UGC 09684                     & 0.0168\tablefootmark{$m$}          & 0.0172  & $73.0 \pm 5.8$\tablefootmark{$a$}         & 0.031 & 0.18 $\pm$ 0.04 \\
2020tlf    & Class 1    & $59108.7 \pm 1.5$ & NGC 5731                      & 0.0084\tablefootmark{$u$}          & 0.0088  & $37.2 \pm 5.0$\tablefootmark{$a$}         & 0.015 & 0.02 $\pm$ 0.01 \\
2020xua    & Class 3    & $59141.6 \pm 1.7$ & UGC 12218                     & 0.021\tablefootmark{$v$}           & 0.0203  & $89.2 \pm 5.8$\tablefootmark{$a$}         & 0.077 & 0.08 $\pm$ 0.02 \\
2020abjq   & Class 1    & $59184.9 \pm 0.5$ & UGC 00678                     & 0.018\tablefootmark{$e$}           & 0.0173  & $73.3 \pm 5.8$\tablefootmark{$a$}         & 0.023 & 0.11 $\pm$ 0.02 \\
2020acbm   & Ledge      & $59192.7 \pm 0.5$ & LSBC F831-08                  & 0.0216\tablefootmark{$e$}          & 0.0208  & $88.6 \pm 6.1$\tablefootmark{$a$}         & 0.03  & 0.08 $\pm$ 0.02 \\
2021zj     & Class 1    & $59222.4 \pm 1.0$ & SDSS J111632.91+290546.5      & 0.046\tablefootmark{$m$}           & 0.0469  & $198.3 \pm 17.1$\tablefootmark{$a$}       & 0.012 & $<0.032$ \\
2021aek    & Class 1    & $59226.5 \pm 1.0$ & I SZ 091                      & 0.0223\tablefootmark{$w$}          & 0.0235  & $99.5 \pm 6.3$\tablefootmark{$a$}         & 0.042 & 0.36 $\pm$ 0.10 \\
2021ass    & Non-flash  & $59230.6 \pm 0.5$ & NGC 0684                      & 0.0118\tablefootmark{$e$}          & 0.0109  & $45.9 \pm 5.0$\tablefootmark{$a$}         & 0.074 & 0.12 $\pm$ 0.02 \\
2021gmj    & Non-flash  & $59292.2 \pm 1.0$ & NGC 3310                      & 0.0033\tablefootmark{$k$}          & 0.0039  & $18.7 \pm 3.4$\tablefootmark{$b$}         & 0.019 & 0.11 $\pm$ 0.02 \\
2021gvm    & Class 3    & $59292.8 \pm 1.5$ & NGC 5185                      & 0.0247\tablefootmark{$m$}          & 0.0256  & $109.4 \pm 6.6$\tablefootmark{$a$}        & 0.021 & 0.017 $\pm$ 0.003 \\
2021jtt    & Class 3    & $59317.7 \pm 1.5$ & NGC 2955                      & 0.0234\tablefootmark{$e$}          & 0.0242  & $103.0 \pm 6.4$\tablefootmark{$a$}        & 0.009 & 0.39 $\pm$ 0.08 \\
2021mqh    & Class 1    & $59351.0 \pm 1.5$ & MCG-01-30-021                 & 0.021\tablefootmark{$x$}           & 0.0227  & $96.3 \pm 8.2$\tablefootmark{$a$}         & 0.04  & $<0.01$ \\
2021ont    & Class 2    & $59365.8 \pm 0.5$ & WISE J162642.49+390842.7      & 0.0283\tablefootmark{$m$}          & 0.0284  & $121.4 \pm 6.8$\tablefootmark{$a$}        & 0.008 & 0.14 $\pm$ 0.03 \\
2021qvr    & Class 2    & $59385.6 \pm 2.0$ & NGC 7678                      & 0.0116\tablefootmark{$e$}          & 0.0104  & $44.1 \pm 5.1$\tablefootmark{$a$}         & 0.042 & 0.15 $\pm$ 0.03 \\
2021rhk    & Non-flash  & $59395.3 \pm 1.0$ & ECO 01022                     & 0.0225\tablefootmark{$y$}          & 0.0234  & $97.7 \pm 6.3$\tablefootmark{$a$}         & 0.022 & $<0.06$ \\
2021tyw    & Class 1    & $59416.9 \pm 0.5$ & UGC 12354                     & 0.013\tablefootmark{$z$}           & 0.0118  & $49.4 \pm 5.3$\tablefootmark{$a$}         & 0.204 & $<0.03$ \\
2021ucg    & Non-flash  & $59419.6 \pm 1.0$ & UGC 12188                     & 0.0172\tablefootmark{\textit{aa}}  & 0.0161  & $68.2 \pm 5.9$\tablefootmark{$a$}         & 0.1   & $<0.006$ \\
2021uoy    & Non-flash  & $59427.4 \pm 1.0$ & CGCG 453-026                  & 0.0335\tablefootmark{$e$}          & 0.0323  & $138.6 \pm 7.2$\tablefootmark{$a$}        & 0.053 & 0.11 $\pm$ 0.02 \\
2021vaz    & Non-flash  & $59432.2 \pm 0.4$ & NGC 1961                      & 0.0131\tablefootmark{\textit{ab}}  & 0.013   & $55.0 \pm 5.4$\tablefootmark{$a$}         & 0.011 & 0.18 $\pm$ 0.04 \\
2021yja    & Ledge      & $59464.4 \pm 0.5$ & NGC 1325                      & 0.0053\tablefootmark{$e$}          & 0.0048  & $23.7 \pm 4.7$\tablefootmark{$b$}         & 0.02  & 0.18 $\pm$ 0.04 \\
2022dml    & Class 2    & $59634.0 \pm 1.5$ & WISEA J161729.19+142507.4     & 0.03\tablefootmark{$p$}            & 0.0302  & $129.2 \pm 7.0$\tablefootmark{$a$}        & 0.045 & 0.07 $\pm$ 0.01 \\
2022fuc    & Ledge      & $59671.8 \pm 0.4$ & NGC 4545                      & 0.0091\tablefootmark{$e$}          & 0.0095  & $40.0 \pm 5.0$\tablefootmark{$a$}         & 0.01  & $<0.008$\\
2022ibv    & Class 3    & $59689.8 \pm 0.6$ & ESO 437-G064                  & 0.0144\tablefootmark{$w$}          & 0.0155  & $60.6 \pm 5.5$\tablefootmark{$a$}         & 0.051 & 0.07 $\pm$ 0.02 \\
2022jox    & Class 2    & $59707.1 \pm 1.1$ & ESO 435-G014                  & 0.0089\tablefootmark{\textit{ac}}  & 0.01    & $42.0 \pm 5.1$\tablefootmark{$a$}         & 0.082 & 0.09 $\pm$ 0.02 \\
2022jzc    & Non-flash  & $59714.4 \pm 1.0$ & NGC 4088                      & 0.0249\tablefootmark{$e$}          & 0.0031  & $16.0 \pm 1.5$\tablefootmark{$b$}         & 0.017 & 0.36 $\pm$ 0.08 \\
2022ovb    & Non-flash  & $59773.5 \pm 0.9$ & UGC 12005                     & 0.0183\tablefootmark{\textit{ad}}  & 0.0172  & $72.9 \pm 5.8$\tablefootmark{$a$}         & 0.12  & $<0.014$\\
2022pgf    & Class 1    & $59781.8 \pm 0.5$ & NGC 5894                      & 0.0082\tablefootmark{$u$}          & 0.0084  & $35.2 \pm 4.9$\tablefootmark{$a$}         & 0.01  & 0.49 $\pm$ 0.10 \\
2022prv    & Class 1    & $59783.3 \pm 1.0$ & IC 1132                       & 0.0151\tablefootmark{\textit{ae}}  & 0.0155  & $65.5 \pm 5.6$\tablefootmark{$a$}         & 0.054 & $<0.009$ \\
2023ixf    & Class 1    & $60082.7 \pm 0.08$& M101                          & 0.0008\tablefootmark{\textit{af}}  & 0.0012  & $6.9 \pm 0.1$\tablefootmark{$b$}          & 0.008 & 0.04 $\pm$ 0.01 \\
2024ggi    & Class 1    & $60410.8 \pm 0.4$ & NGC 3621                      & 0.0022\tablefootmark{\textit{ag}}  & 0.0035  & $6.7 \pm 0.2$\tablefootmark{$b$}          & 0.071 & 0.13 $\pm$ 0.02 \\\hline
\end{tabular}
\tablefoot{\tablefoottext{a}~Distance estimated from the redshift relative to the CMB (see Sect.~\ref{sec:distances}). \tablefoottext{b}~Redshift-independent distance. \tablefoottext{c}~\cite{Yaron2017}. \tablefoottext{d}~\cite{Zhang2020_18zd}. \tablefoottext{e}~\cite{Springob2005}. \tablefoottext{f}~\cite{Garcia1994}. \tablefoottext{g}~\cite{deVaucouleurs1991}. \tablefoottext{h}~\cite{Lauberts1989}. \tablefoottext{i}~\cite{Mathewson1996}. \tablefoottext{j}~\cite{Wong2006}. \tablefoottext{k}~\cite{Bekiaris2016}. \tablefoottext{l}~\cite{Koribalski2004}. \tablefoottext{m}~\cite{SDSS.DR13.Albareti2017}. \tablefoottext{n}~\cite{Garcia1992}. \tablefoottext{o}~\cite{Neumann2023}. \tablefoottext{p}~\cite{JG2024}. \tablefoottext{q}~\cite{Jones2018}. \tablefoottext{r}~\cite{Bear-Way2024}. \tablefoottext{s}~\cite{Bruch2023}. \tablefoottext{t}~\cite{Koss2022}. \tablefoottext{u}~\cite{VanDriel2016}. \tablefoottext{v}~\cite{Schneider1990}. \tablefoottext{w}~\cite{6dF.Jones2009}. \tablefoottext{x}~\cite{Bilicki2014}. \tablefoottext{y}~\cite{Moffett2015}. \tablefoottext{z}~\cite{Taylor2014}. \tablefoottext{aa}~\cite{Huchra1999}. \tablefoottext{ab}~\cite{Strauss1992}. \tablefoottext{ac}~\cite{Kregeel2004}. \tablefoottext{ad}~\cite{Siljeg2024}. \tablefoottext{ae}~\cite{Martinsson2013}. \tablefoottext{af}~\cite{Hiramatsu2023_23ixf}. \tablefoottext{ag}~\cite{Kacharov2018}. \tablefoottext{ah}~\cite{DallOra2014}. }
\end{threeparttable}
}
\end{table}

\twocolumn

\begin{figure}
\centering
\includegraphics[width=0.93\columnwidth]{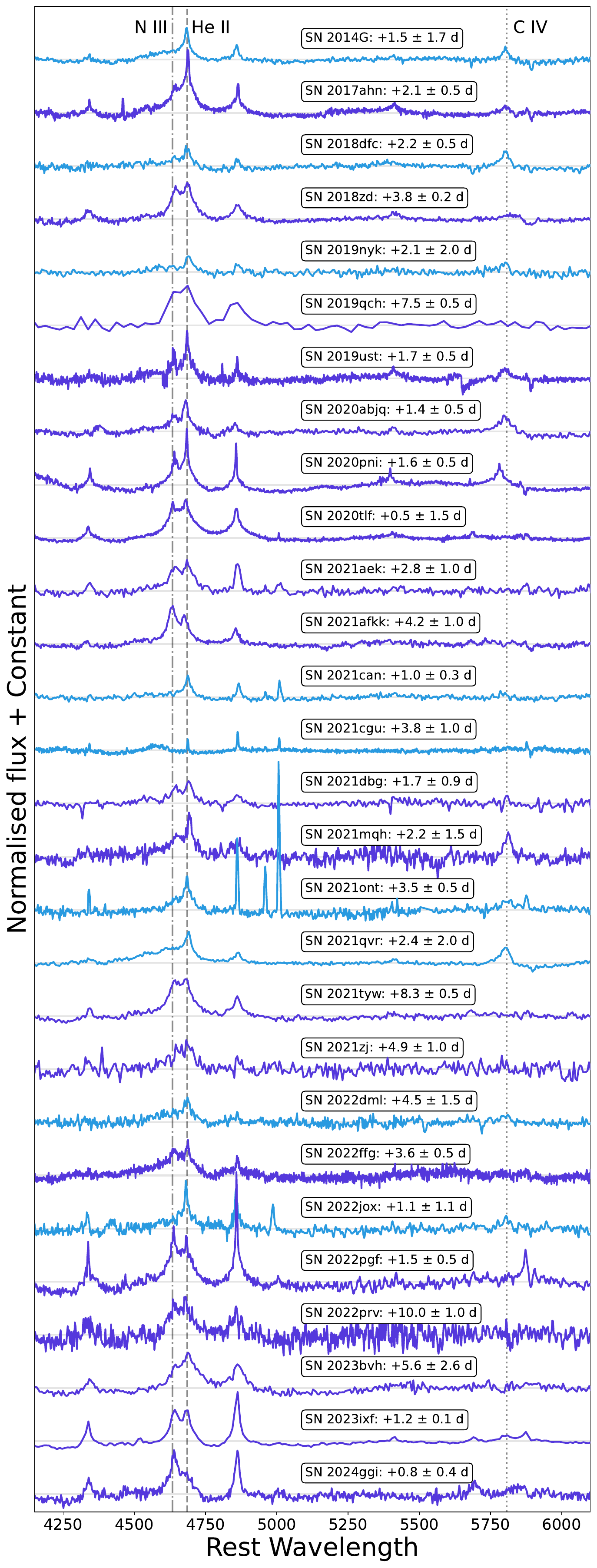}
\caption{Sample spectra of class 1+2 SNe. Some spectra were obtained more than four days after first light but were still used to determine the flash class. The main flash features are indicated with vertical dashed lines. These SNe exhibit intense He~\textsc{ii} and N~\textsc{iii} emission (class 1, blue) or intense He~\textsc{ii} emission but no N~\textsc{iii} (class 2, light blue). The SN name and the phase relative to the first-light epoch are labelled above each spectrum.}
\label{f:c1+2_spectra}
\end{figure}

\begin{figure}
\centering
\includegraphics[width=0.93\columnwidth]{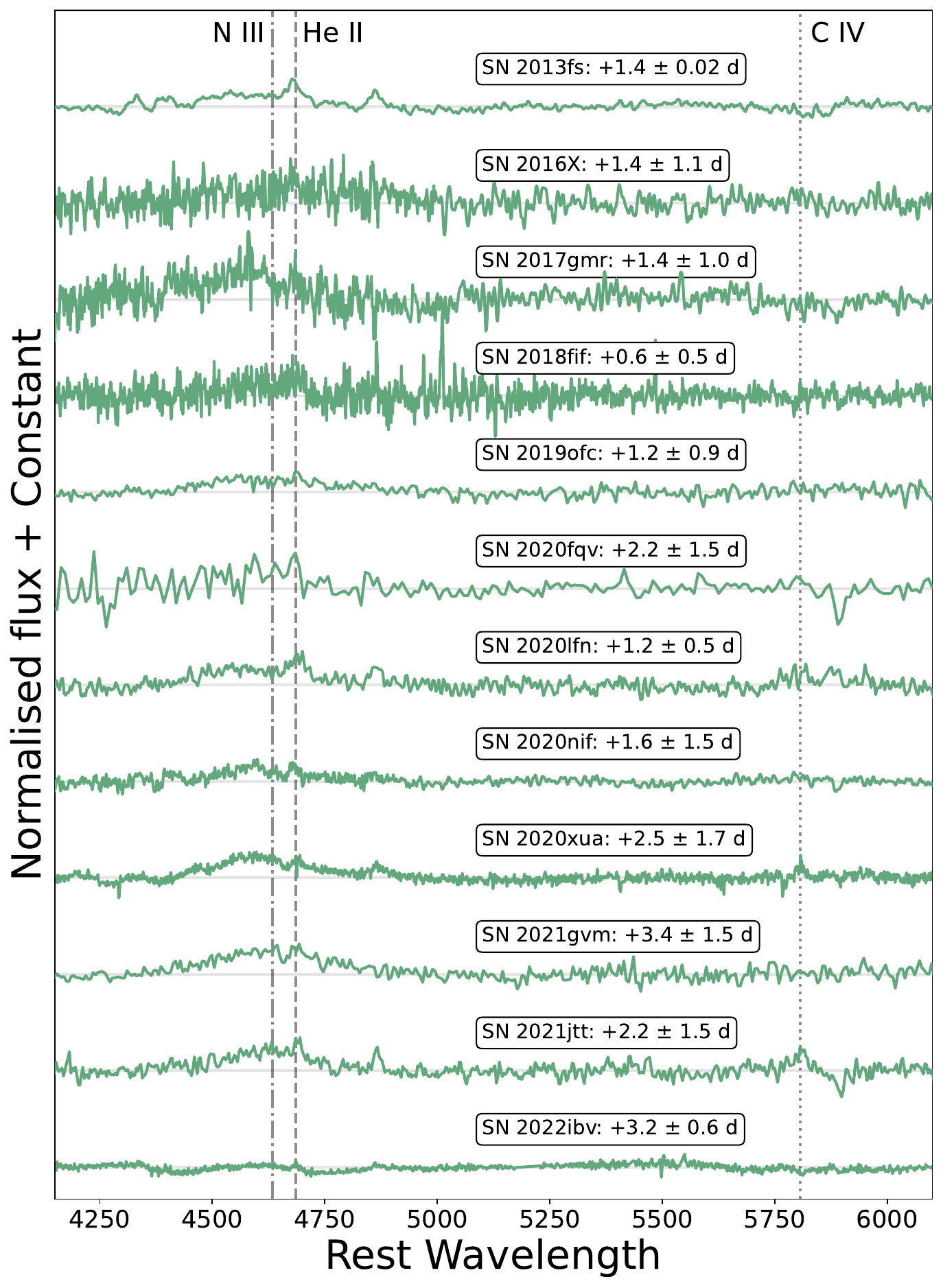}
\caption{Sample spectra of class 3 SNe. All spectra were obtained within four days of first light. The main flash features are indicated with vertical dashed lines. These SNe exhibit weak He~\textsc{ii} emission. The SN name and the phase relative to the first-light epoch are labelled above each spectrum.}
\label{f:c3_spectra}
\end{figure}

\clearpage

\begin{table}[h!]
\centering
\resizebox{\columnwidth}{!}{%
\begin{threeparttable}
\caption{List of classification spectra of the literature sample.}
\label{t:Lit.spectra}
\begin{tabular}{lccccc}
\hline\hline
SN   & \begin{tabular}[c]{@{}c@{}}Flash \\ classification \end{tabular} & MJD & \begin{tabular}[c]{@{}c@{}}Phase\\  (days)\end{tabular}  & Data source \\ \hline
2012aw     & Ledge     & 56003.0 & 0.9 &      \cite{12aw.spectrum.2014} \\
2013ab     & Ledge     & 56341.7  & 2.2 &       \cite{13ab.spectrum.2015}   \\
2013fs     & Class 3   &  56572.5 & 1.4 &       \cite{Yaron2017}      \\
2013hj     & Ledge     &  56639.1 & 2.3 &   WISeREP (public)      \\
2014G      & Class 2 &  56671.1 & 1.5 &         \cite{14G_spectrum.2016} \\
2016X      & Class 3   & 57407.7  & 1.4 &          \cite{Huang2018.2016X}       \\
... \\
\hline
\end{tabular}

\tablefoot{The complete table is available in machine-readable form.}
\end{threeparttable}
}
\end{table}

\begin{figure}
\centering
\includegraphics[width=0.93\columnwidth]{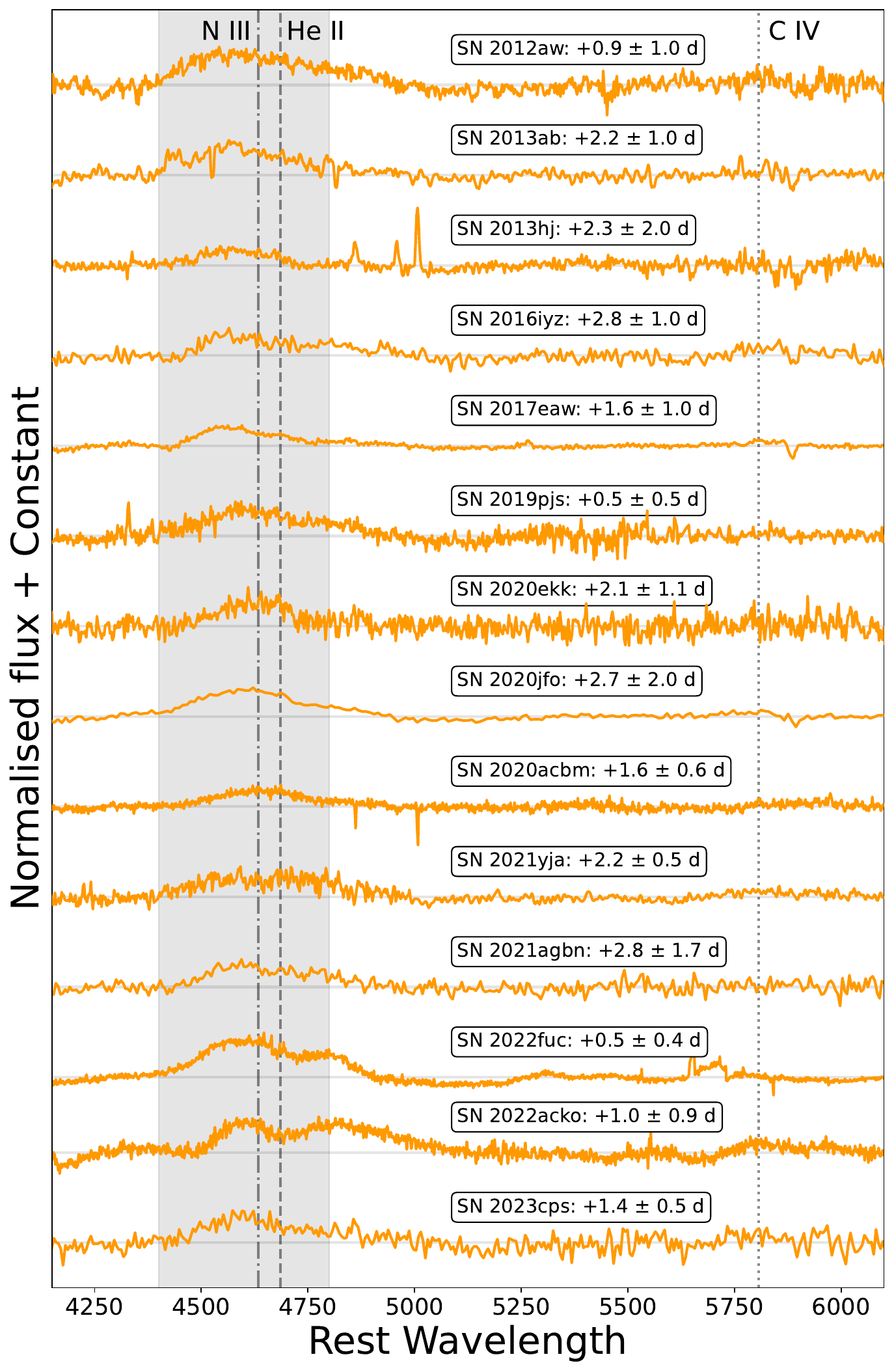}
\caption{Sample spectra of ledge SNe. All spectra were obtained within four days of first light. These SNe show a broad, asymmetric emission feature typically around 4400 -- 4800 \r{A} (shaded region). The main flash features are indicated with vertical dashed lines. The SN name and the phase relative to the first-light epoch are labelled above each spectrum.}
\label{f:ledge_spectra}
\end{figure}

\begin{figure}
\centering
\includegraphics[width=0.9\columnwidth]{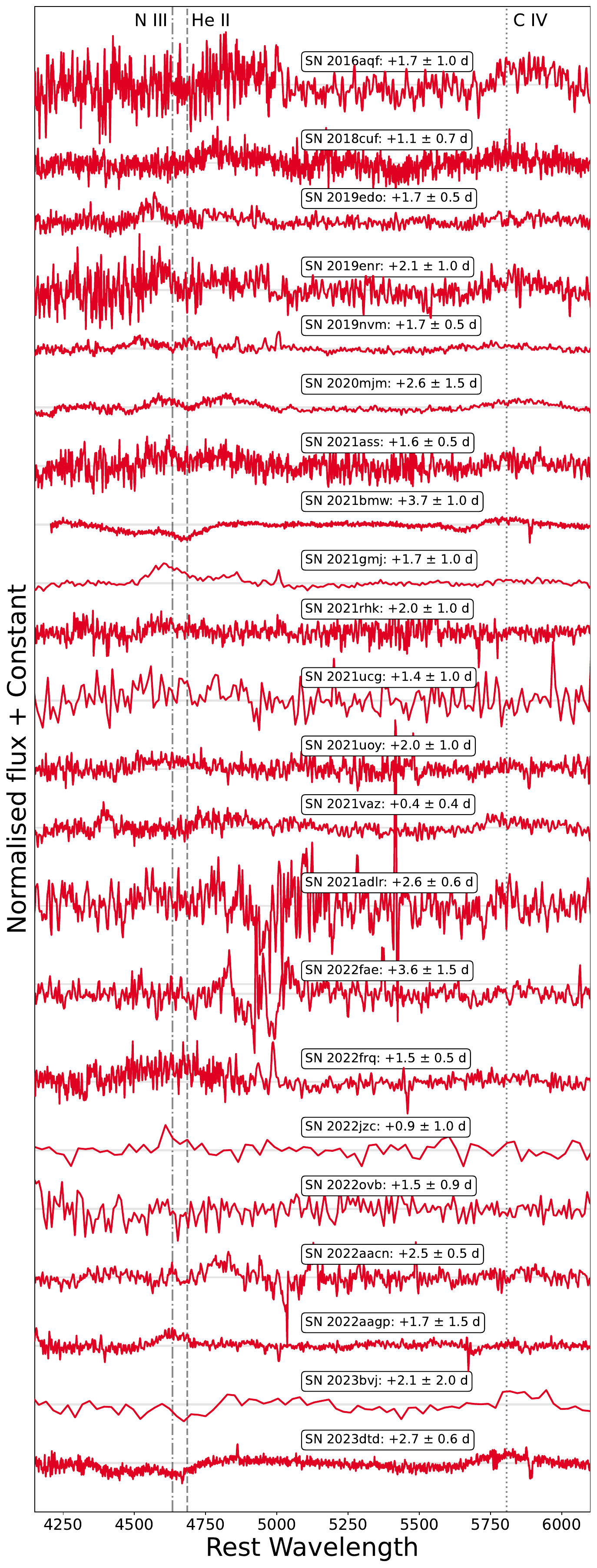}
\caption{Sample spectra of non-flash SNe. All the spectra were obtained within four days of first light. These SNe do not show any flash-ionisation emission features. Some events (SN~2019edo, SN~2020mjm, SN~2020gmj, and SN~2022aagp) exhibit a broad, symmetric emission feature blueshifted relative to the He~\textsc{ii} line. This emission is attributed to photospheric He \textsc{ii} and is not associated with CSI. The main flash features are indicated with vertical dashed lines. The SN name and the phase relative to the first-light epoch are labelled above each spectrum.}
\label{f:nf_spectra}
\end{figure}

\onecolumn
\section{Light-curves parameters}

\begin{table}[h!]
\centering
\tiny
\caption{POISE sample peak absolute magnitude (AB magnitudes)}
\label{t:POISE.peak.magnitude}
\vspace{-0.2cm}
\begin{tabular}{lcccccc}
\hline\hline
SN   & \begin{tabular}[c]{@{}c@{}}$M_{\mathrm{peak},u}$ \\ {[}mag{]}\end{tabular} &  \begin{tabular}[c]{@{}c@{}}$M_{\mathrm{peak},B}$ \\ {[}mag{]}\end{tabular} &\begin{tabular}[c]{@{}c@{}}$M_{\mathrm{peak},g}$ \\ {[}mag{]}\end{tabular} & \begin{tabular}[c]{@{}c@{}}$M_{\mathrm{peak},V}$ \\ {[}mag{]}\end{tabular} & \begin{tabular}[c]{@{}c@{}}$M_{\mathrm{peak},r}$ \\ {[}mag{]}\end{tabular} & \begin{tabular}[c]{@{}c@{}}$M_{\mathrm{peak},i}$ \\ {[}mag{]}\end{tabular} \\ \hline
2021bmw     &  $< -16.03$         &  $<-16.15$             &  $<-16.13$            &   $-16.22 \pm 0.19$    &  $-16.34 \pm 0.19$     &   $-16.34 \pm 0.19$           \\
2021can     &  $-18.38 \pm 0.17$  &  $-18.06 \pm 0.14$     &  $-17.97 \pm 0.14$    &   $-17.91 \pm 0.14$    &  $-17.73 \pm 0.14$     &   $-17.54 \pm 0.14$        \\
2021cgu     &  $-17.95 \pm 0.12$  &  $-18.09 \pm 0.12$     &  $-18.10 \pm 0.12$    &   $-18.20 \pm 0.12$    &  $-18.25 \pm 0.12$     &   $-18.31 \pm 0.12$  \\
2021dbg     &  $-18.84 \pm 0.14$  &  $-18.81 \pm 0.14$     &  $-18.76 \pm 0.14$    &   $-18.75 \pm 0.14$    &  $-18.65 \pm 0.14$     &   $-18.55 \pm 0.14$   \\
2021adlr    &  $-17.48 \pm 0.12$  &  $-17.47 \pm 0.13$     &  $-17.40 \pm 0.11$    &   $-17.36 \pm 0.11$    &  $-17.44 \pm 0.11$     &   $-17.38 \pm 0.11$  \\
2021adxd    &  $-14.22 \pm 0.19$  &  $-14.66 \pm 0.18$     &  $-14.84 \pm 0.22$    &   $-15.21 \pm 0.17$    &  $-15.58 \pm 0.17$     &   $-15.99 \pm 0.21$      \\
... \\
\hline
\end{tabular}
\vspace{-0.2cm}
\tablefoot{The complete table is available in machine-readable form.}
\end{table}

\begin{table}[h!]
\centering
\tiny
\caption{Literature sample peak absolute magnitude (AB magnitudes).}
\label{t:lit.peak.magnitude}
\vspace{-0.2cm}
\begin{tabular}{lccccccc}
\hline\hline
SN   & \begin{tabular}[c]{@{}c@{}}$M_{\mathrm{peak},u}$ \\ {[}mag{]}\end{tabular} &  \begin{tabular}[c]{@{}c@{}}$M_{\mathrm{peak},B}$ \\ {[}mag{]}\end{tabular} &\begin{tabular}[c]{@{}c@{}}$M_{\mathrm{peak},g}$ \\ {[}mag{]}\end{tabular} & \begin{tabular}[c]{@{}c@{}}$M_{\mathrm{peak},V}$ \\ {[}mag{]}\end{tabular} & \begin{tabular}[c]{@{}c@{}}$M_{\mathrm{peak},r}$ \\ {[}mag{]}\end{tabular} & \begin{tabular}[c]{@{}c@{}}$M_{\mathrm{peak},i}$ \\ {[}mag{]}\end{tabular} \\ \hline
2012aw      &  $-16.80 \pm 0.20$  &  $-16.72 \pm 0.20$     &  $-16.82 \pm 0.20$    &   $-16.79 \pm 0.20$    &  $-16.88 \pm 0.20$     &   --           \\
2013ab      &  $-17.04 \pm 0.08$  &  $-17.20 \pm 0.08$     &  $-17.25 \pm 0.08$    &   $-17.08 \pm 0.08$    &  $-17.26 \pm 0.08$     &   $-17.20 \pm 0.08$        \\
2013fs      &  $-17.33 \pm 0.25$  &  $-17.53 \pm 0.25$     &  $-17.34 \pm 0.25$    &   $-17.37 \pm 0.26$    &  $-17.32 \pm 0.25$     &   $-17.27 \pm 0.25$           \\
2013hj      &  $< -18.45$         &  $-18.19 \pm 0.32$     &  $-18.23 \pm 0.32$    &   $-18.18 \pm 0.32$    &  $-18.15 \pm 0.32$     &   $-18.07 \pm 0.32$        \\
2014G       &  --                 &  --                    &  $-17.37 \pm 0.81$    &   --                   &  $-17.54 \pm 0.81$     &   $-17.60 \pm 0.81$           \\
2016X       &  --                 &  $-16.84 \pm 0.43$     &  $-16.89 \pm 0.43$    &   $-16.86 \pm 0.43$    &  $-16.90 \pm 0.43$     &   $-16.80 \pm 0.43$        \\
...     \\
\hline
\end{tabular}
\vspace{-0.2cm}
\tablefoot{The complete table is available in machine-readable form.}
\end{table}

\begin{table}[h!]
\centering
\tiny
\caption{POISE sample time to peak}
\label{t:POISE.rise.times}
\vspace{-0.2cm}
\begin{tabular}{lccccccc}
\hline\hline
SN   & \begin{tabular}[c]{@{}c@{}}$t_{\mathrm{peak},u}$ \\ {[}days{]}\end{tabular} &  \begin{tabular}[c]{@{}c@{}}$t_{\mathrm{peak},B}$ \\ {[}days{]}\end{tabular} &\begin{tabular}[c]{@{}c@{}}$t_{\mathrm{peak},g}$ \\ {[}days{]}\end{tabular} & \begin{tabular}[c]{@{}c@{}}$t_{\mathrm{peak},V}$ \\ {[}days{]}\end{tabular} & \begin{tabular}[c]{@{}c@{}}$t_{\mathrm{peak},r}$ \\ {[}days{]}\end{tabular} & \begin{tabular}[c]{@{}c@{}}$t_{\mathrm{peak},i}$ \\ {[}days{]}\end{tabular} \\ \hline
2021bmw     &  $<2.69$          &  $<2.70$           &  $<2.68$            &   $4.83 \pm 1.07$    &   $5.93 \pm 1.20$     &  $5.91 \pm 1.0$           \\
2021can     &  $4.05 \pm 0.36$  &  $4.65 \pm 0.36$   &  $4.86 \pm 0.36$    &   $5.46 \pm 0.39$    &   $6.47 \pm 0.53$     &  $8.89 \pm 0.45$        \\
2021cgu     &  $6.84 \pm 1.10$  &  $10.53 \pm 1.03$  &  $10.98 \pm 1.08$   &   $12.72 \pm 1.37$   &   $14.74 \pm 1.64$    &  $18.80 \pm 1.37$  \\
2021dbg     &  $7.18 \pm 0.96$  &  $8.77 \pm 0.96$   &  $8.99 \pm 0.96$    &   $9.84 \pm 0.97$    &   $10.36 \pm 0.96$    &  $11.47 \pm 1.0$   \\
2021adlr    &  $7.08 \pm 0.98$  &  $9.38 \pm 0.64$   &  $12.13 \pm 0.86$   &   $16.16 \pm 1.91$   &   $22.39 \pm 0.62$    &  $38.64 \pm 1.18$  \\
2021adxd    &  $8.40 \pm 2.07$  &  $12.92 \pm 1.26$  &  $11.88 \pm 1.36$   &   $12.01 \pm 1.14$   &   $13.67 \pm 1.85$    &  $14.23 \pm 1.11$      \\
... \\ \hline
\end{tabular}
\vspace{-0.2cm}
\tablefoot{The complete table is available in machine-readable form.}
\end{table}

\begin{table}[h!]
\centering
\tiny
\caption{Literature sample time to peak}
\label{t:lit.rise.times}
\vspace{-0.2cm}
\begin{tabular}{lcccccc}
\hline\hline
SN   & \begin{tabular}[c]{@{}c@{}}$t_{\mathrm{peak},u}$ \\ {[}days{]}\end{tabular} &  \begin{tabular}[c]{@{}c@{}}$t_{\mathrm{peak},B}$ \\ {[}days{]}\end{tabular} &\begin{tabular}[c]{@{}c@{}}$t_{\mathrm{peak},g}$ \\ {[}days{]}\end{tabular} & \begin{tabular}[c]{@{}c@{}}$t_{\mathrm{peak},V}$ \\ {[}days{]}\end{tabular} & \begin{tabular}[c]{@{}c@{}}$t_{\mathrm{peak},r}$ \\ {[}days{]}\end{tabular} & \begin{tabular}[c]{@{}c@{}}$t_{\mathrm{peak},i}$ \\ {[}days{]}\end{tabular} \\ \hline
2012aw     &  $6.68 \pm 1.02$  &  $8.68 \pm 1.05$   &  $9.75 \pm 1.19$    &   $10.49 \pm 1.54$   &  $23.14 \pm 1.67$     &   --           \\
2013ab     &  $6.22 \pm 1.20$  &  $6.49 \pm 1.11$   &  $8.40 \pm 1.11$    &   $8.03 \pm 1.03$    &  $9.03 \pm 1.08$      &   $8.48 \pm 1.03$        \\
2013fs     &  $4.31 \pm 0.85$  &  $4.39 \pm 0.74$   &  $6.82 \pm 0.55$    &   $7.59 \pm 0.49$    &  $8.88 \pm 0.59$      &   $9.92 \pm 0.53$           \\
2013hj     &  $< 3.49$         &  $4.47 \pm 2.15$   &  $6.33 \pm 2.03$    &   $8.28 \pm 2.08$    &  $10.55 \pm 2.02$     &   $11.25 \pm 2.04$        \\
2014G      &  --               &  --                &  $10.84 \pm 1.73$   &   --                 &  $14.64 \pm 1.84$     &   $15.84 \pm 1.83$           \\
2016X      &  --               &  $7.99 \pm 1.10$   &  $10.70 \pm 1.36$   &   $9.52 \pm 1.10$    &  $14.24 \pm 1.55$     &   $13.16 \pm 1.78$        \\
...     \\
\hline
\end{tabular}
\vspace{-0.2cm}
\tablefoot{The complete table is available in machine-readable form.}
\end{table}

\begin{figure*}[h!]
\centering
\includegraphics[width=0.85\textwidth]{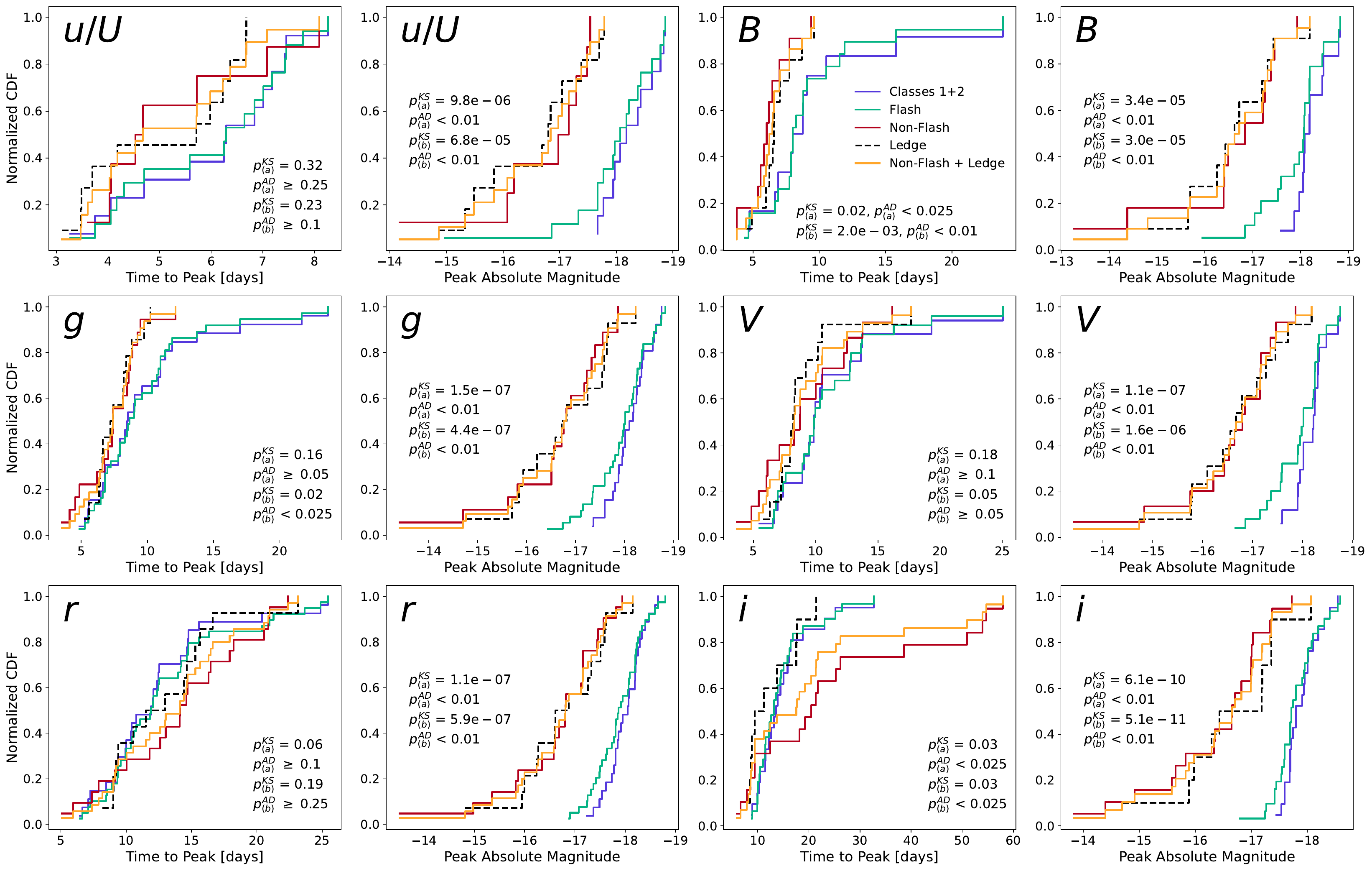}
\caption{Cumulative distribution functions of time to peak (first and third columns) and peak absolute magnitude (second and fourth columns) in \textit{u/U}, $B$, $g$, $V$, $r$, and $i$ bands (as indicated in each panel). Two comparison cases are presented: (a) class 1+2 (blue) vs. non-flash (red); and (b) flash (green) vs. non-flash + ledge (orange). The corresponding p values of the K-S and A-D tests for each comparison are indicated in the respective panels and in Table~\ref{t:KS_AD_peak_m_t}. The CDF of ledge SNe is plotted separately with dashed black lines.}
\label{f:CDFs.LC.all.bands}
\end{figure*}

\FloatBarrier

\section{Colour-curves parameters}

\begin{table}[h!]
\centering
\tiny
\caption{(\textit{$g-r$}) colour-curve parameters of POISE sample}
\label{t:POISE.slopes_gr}
\vspace{-0.2cm}
\begin{tabular}{lccccccc}
\hline\hline
SN  & \begin{tabular}[c]{@{}c@{}}Colour-curve\\ Sample\end{tabular}   & \begin{tabular}[c]{@{}c@{}}$(g-r)_{\mathrm{\mathrm{10d}}}$ \\ {[}mag{]}\end{tabular} & \begin{tabular}[c]{@{}c@{}}$(g-r)_{\mathrm{\mathrm{20d}}}$ \\ {[}mag{]}\end{tabular} & \begin{tabular}[c]{@{}c@{}}$cs_0$ \\ {[}mag per 100d{]}\end{tabular} &  \begin{tabular}[c]{@{}c@{}}$cs_1$ \\ {[}mag per 100d{]}\end{tabular} &\begin{tabular}[c]{@{}c@{}}$\Delta cs$ \\ {[}mag per 100d{]}\end{tabular} & \begin{tabular}[c]{@{}c@{}}$t_{\mathrm{\mathrm{break}}}$ \\ {[}days{]}\end{tabular} \\
\hline
2021bmw   & Gold     & $0.28 \pm 0.006$  & $0.48 \pm 0.01$   & $1.25 \pm 0.35$     & $2.34 \pm 0.26$  &  $-1.09 \pm 0.44$    &   $13.16 \pm 4.12$  \\
2021can   & Gold     & $-0.05 \pm 0.004$ & $0.18 \pm 0.01$   & $3.46 \pm 0.21$     & $1.52 \pm 0.17$  &  $1.94 \pm 0.27$     &   $13.70 \pm 1.23$  \\
2021cgu   & Complete & $0.07 \pm 0.004$  & $0.26 \pm 0.02$   & $2.36 \pm 0.12$     & $1.26 \pm 0.19$  &  $1.10 \pm 0.23$     &   $15.71 \pm 1.95$  \\
2021dbg   & Gold     & $-0.10 \pm 0.005$ & $0.11 \pm 0.005$  & $2.75 \pm 0.28$     & $1.88 \pm 0.05$  &  $0.87 \pm 0.29$     &   $12.62 \pm 1.80$  \\
2021adlr  & Gold     & $-0.11 \pm 0.005$ & $0.07 \pm 0.007$  & $1.57 \pm 0.06$     & $3.09 \pm 0.11$  &  $-1.52 \pm 0.12$    &   $18.71 \pm 0.65$  \\
2021adxd  & Gold     & $0.73 \pm 0.01$   & $0.82 \pm 0.02$   & $0.68 \pm 0.33$     & $1.08 \pm 0.82$  &  $-0.40 \pm 0.88$    &   $21.32 \pm 4.75$  \\
... \\ \hline
\end{tabular}
\vspace{-0.2cm}
\tablefoot{The complete table is available in machine-readable form.}
\end{table}

\begin{table}[h!]
\centering
\tiny
\caption{(\textit{$g-r$}) colour-curve parameters of literature sample}
\label{t:lit.slopes_gr}
\vspace{-0.2cm}
\begin{tabular}{lcccccccc}
\hline\hline
SN  & \begin{tabular}[c]{@{}c@{}}Colour-curve\\ Sample\end{tabular} & \begin{tabular}[c]{@{}c@{}}$(g-r)_{\mathrm{\mathrm{10d}}}$ \\ {[}mag{]}\end{tabular} & \begin{tabular}[c]{@{}c@{}}$(g-r)_{\mathrm{\mathrm{20d}}}$ \\ {[}mag{]}\end{tabular} & \begin{tabular}[c]{@{}c@{}}$cs_0$ \\ {[}mag per 100d{]}\end{tabular} &  \begin{tabular}[c]{@{}c@{}}$cs_1$ \\ {[}mag per 100d{]}\end{tabular} &\begin{tabular}[c]{@{}c@{}}$\Delta cs$ \\ {[}mag per 100d{]}\end{tabular} & \begin{tabular}[c]{@{}c@{}}$t_{\mathrm{\mathrm{break}}}$ \\ {[}days{]}\end{tabular} \\
\hline
2012aw  & Gold     & $-0.01 \pm 0.008$  & $0.19 \pm 0.007$  & $1.75  \pm 0.14$ &	$2.50 \pm 0.18$	& $-0.75 \pm 0.23$	& $17.93 \pm 1.85$  \\
2013ab  & Gold     & $0.03 \pm 0.01$    & $0.22 \pm 0.02$   & $1.91  \pm 0.13$ &	$0.95 \pm 0.20$ & $0.96 \pm 0.23$	& $18.24 \pm 1.00 $ \\
2013fs  & Complete & $0.04 \pm 0.03$    & $0.16 \pm 0.03$   & $3.05  \pm 1.04$ &	$1.20 \pm 0.31$	& $1.85 \pm 1.09$	& $13.28 \pm 1.99$  \\
2013hj  & Gold     & $-0.05 \pm 0.007$  & $0.05 \pm 0.007$  & $1.24  \pm 0.06$ &	$3.09 \pm 0.33$	& $-1.86 \pm 0.33$	& $23.20 \pm 2.27 $ \\
2014G   & Gold     & $0.12 \pm 0.01$    & $0.36 \pm 0.02$   & $3.41  \pm 0.79$ &	$2.54 \pm 0.29$	& $0.87 \pm 0.84$	& $12.56 \pm 4.27$  \\
2016X   & Complete & $-0.03 \pm 0.02$   & $0.17 \pm 0.03$   & $3.18  \pm 0.53$ &	$1.71 \pm 0.35$	& $1.47 \pm 0.63$	& $13.72 \pm 1.47$  \\
...     \\
\hline
\end{tabular}
\vspace{-0.2cm}
\tablefoot{The complete table is available in machine-readable form.}
\end{table}

\begin{figure}[h!]
\centering
\includegraphics[width=.95\textwidth]{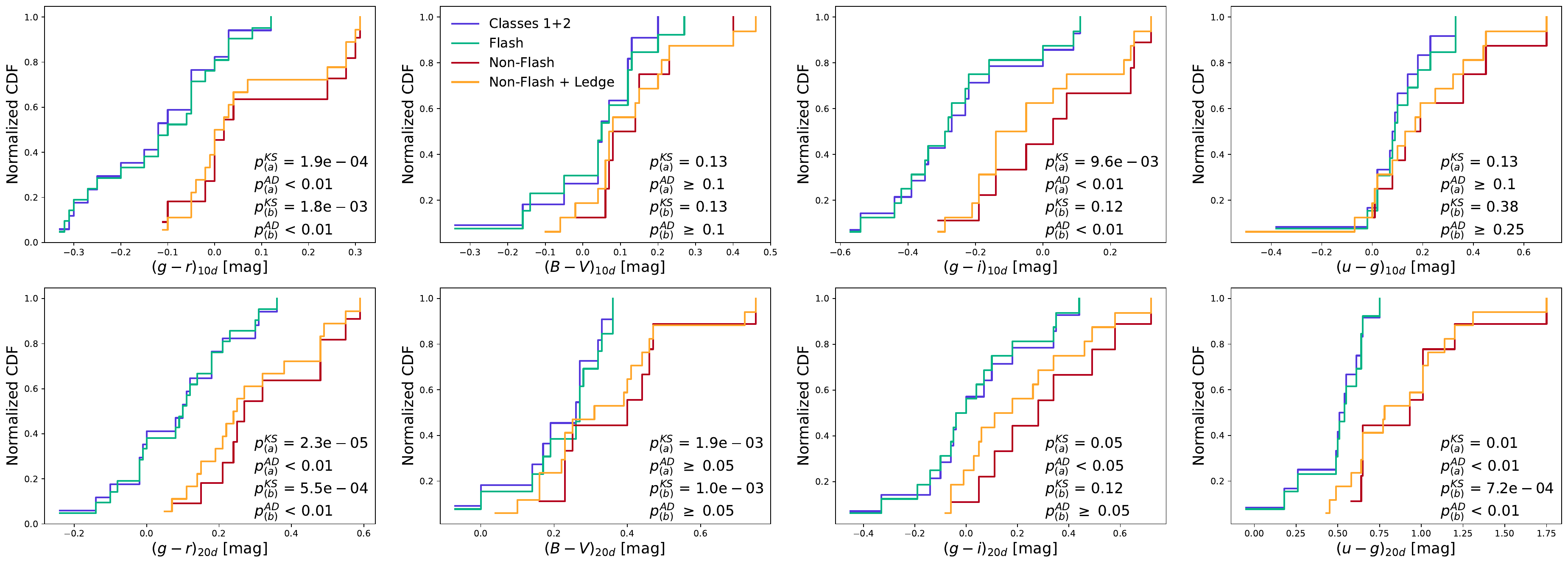}             
\caption{Cumulative distribution functions of colour at 10 days (\textit{top row}) and 20 days (\textit{bottom row}) in (\textit{$g-r$}), (\textit{$B-V$}), (\textit{$g-i$}), and (\textit{$u-g$}) colours (left to right). Class 1+2 (blue), flash (class 1+2+3, green), non-flash (red), and non-flash + ledge SNe (orange) are shown. The p values of the K-S and A-D tests for the comparison cases (a) class 1+2 vs. non-flash and (b) flash vs. non-flash + ledge SNe are given in each panel and also listed in Table~\ref{t:pvalues_KS_AD_colour_phases}.}  
\label{f:CDF_10d_20d}
\end{figure}

\begin{table}
\centering
\tiny
\begin{threeparttable}
\caption{Phase of the colour-curve minimum in (\textit{$g-r$}) ($t_\mathrm{min}$), duration of the flash features ($t_\mathrm{IIn}$), and phase of the break ($t_\mathrm{break}$).}
\label{t:tIIn.tmin}
\vspace{-0.2cm}
\begin{tabular}{lcccc}
\hline\hline
SN        & \begin{tabular}[c]{@{}c@{}} Flash\\ classification \end{tabular} & $t_\mathrm{min}$ [days] & $t_\mathrm{IIn}$ [days] & $t_\mathrm{break}$ [days] \\ \hline
2013fs    & Class 3 & $1.0 \pm 0.4$  & $1.4 \pm 0.5$  & $13.3 \pm 2.0$  \\
2014G     & Class 2 & $4.3 \pm 1.8$  & $3.5 \pm 1.8$  & $12.6 \pm 4.3$  \\
2016X     & Class 3 & --             & $1.5 \pm 0.2$  & $13.7 \pm 1.5$  \\
2017ahn   & Class 1 & $2.4 \pm 0.7$  & $6.4 \pm 0.3$  & $7.6 \pm 1.0$  \\
2017gmr   & Class 3 & --             & $1.9 \pm 0.5$  & $21.9 \pm 1.0$  \\
2018zd    & Class 1 & $6.6 \pm 0.3$  & $7.7 \pm 0.9$  & $12.2 \pm 1.5$ \\
2018dfc   & Class 2 & $2.3 \pm 0.6$  & $3.1 \pm 1.0$  & $15.8 \pm 2.9$   \\
2019nyk   & Class 2 & --             & $4.6 \pm 0.5$  & $17.8 \pm 2.0$  \\
2019ust   & Class 1 & --             & $3.5 \pm 0.6$  & $14.92 \pm 0.5$  \\
2020fqv   & Class 3 & --             & $3.0 \pm 0.1$  & $17.3 \pm 1.8$  \\
2020lfn   & Class 3 & --             & $0.7 \pm 0.6$  & $13.9 \pm 1.8$  \\
2020nif   & Class 3 & --             & $1.4 \pm 1.2$  & $12.9 \pm 1.5$  \\
2020pni   & Class 1 & $2.5 \pm 0.6$  & $5.5 \pm 1.0$  & $12.3 \pm 0.5$  \\
2020abjq  & Class 1 & --             & $3.2 \pm 1.0$  & $13.6 \pm 1.8$  \\
2021can   & Class 2 & --             & $2.5 \pm 0.6$  & $13.7 \pm 1.2$  \\
2021cgu   & Class 2 & --             & $5.0 \pm 1.6$  & $15.7 \pm 1.9$  \\
2021dbg   & Class 1 & --             & $1.9 \pm 0.8$  & $12.2 \pm 1.5$  \\
2021jtt   & Class 3 & --             & $2.5 \pm 0.3$  & $21.6 \pm 2.3$  \\
2021mqh   & Class 1 & --             & $3.8 \pm 0.9$  & $14.9 \pm 1.5$  \\
2021ont   & Class 2 & --             & $4.2 \pm 1.1$  & $19.9 \pm 3.0$  \\
2021qvr   & Class 2 & --             & $3.2 \pm 2.1$  & $17.7 \pm 3.1$  \\
2021tyw   & Class 1 & $9.8 \pm 0.8$  & $13.6 \pm 1.3$ & $16.2 \pm 1.4$   \\
2021afkk  & Class 1 & $3.6 \pm 1.0$  & $8.3 \pm 0.6$  & $22.5 \pm 1.3$   \\
2022dml   & Class 2 & --             & $5.2 \pm 1.6$  & $12.7 \pm 1.6$ \\
2022ffg   & Class 1 & $3.3 \pm 0.5$  & $6.5 \pm 0.7$  & $12.5 \pm 1.1$ \\
2022ibv   & Class 3 & --             & $1.5 \pm 0.6$  & $11.4 \pm 0.4$ \\ 
2022pgf   & Class 1 & $8.1 \pm 0.7$  & $10.3 \pm 1.0$ & $19.2 \pm 0.5$    \\
2022prv   & Class 1 & $12.8 \pm 1.1$ & $18.3 \pm 1.8$ & $23.1 \pm 2.5$   \\
2023bvh   & Class 1 & $8.0 \pm 2.6$  & { $> 5.5$}     & $21.0 \pm 4.0$   \\
2023ixf   & Class 1 & $5.0 \pm 0.4$  & $7.3 \pm 1.0$  & $13.6 \pm 2.1$   \\
2024ggi   & Class 1 & --             & $4.1 \pm 1.6$  & $11.0 \pm 0.4$ \\ 
\hline
\end{tabular}
\end{threeparttable}
\end{table}

\begin{figure}
\centering
\includegraphics[width=.85\textwidth]{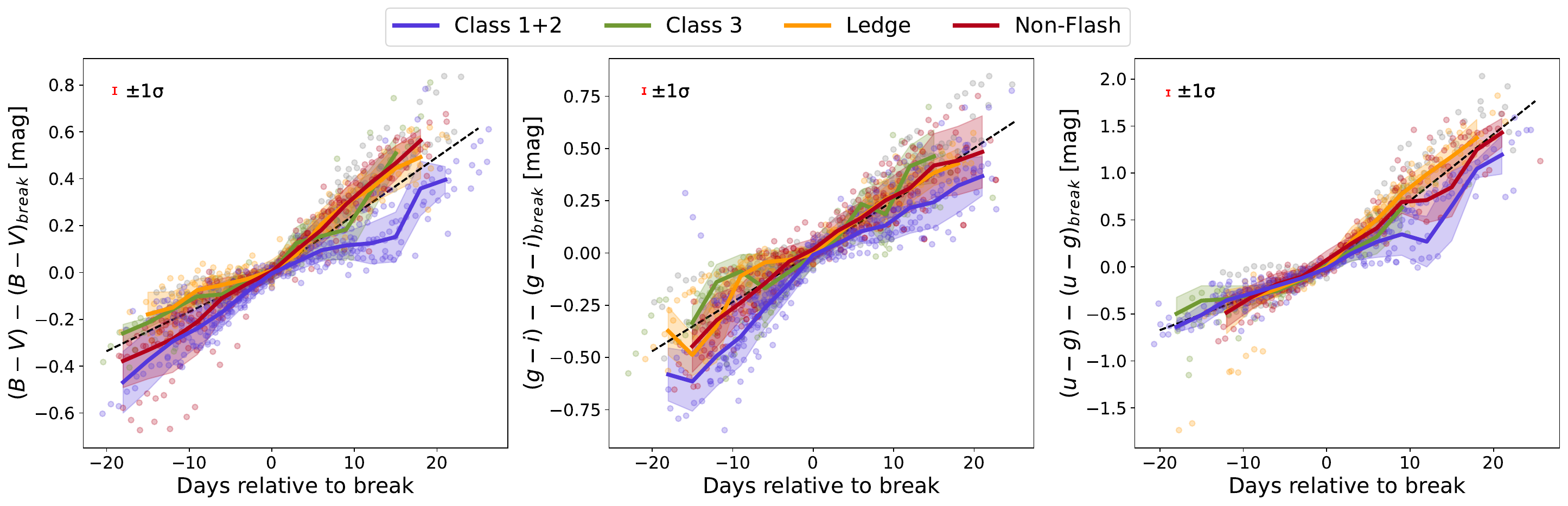}
\caption{Left to right: (\textit{$B-V$}), (\textit{$g-i$}), and (\textit{$u-g$}) colour curves shifted to the phase and colour of the break. Class 1+2 (blue), class 3 (green), ledge (orange), and non-flash SNe (red) are shown. The circles represent the data points of all SNe, while the solid lines show the weighted mean colour curve for each group in the gold sample, with the shaded regions indicating the 1$\sigma$ uncertainty, accounting for both the dispersion and measurement uncertainties. The black dashed lines show the weighted mean slopes of $cs_0$ (negative phases) and $cs_1$ (positive phases), calculated separately for all SNe. The red error bar in the top-left corner of each panel indicates the typical $1\sigma$ photometric uncertainty of the individual data points. }
\label{f:others.colours.breaks}
\end{figure}

\begin{figure}
\centering
\includegraphics[width=.85\textwidth]{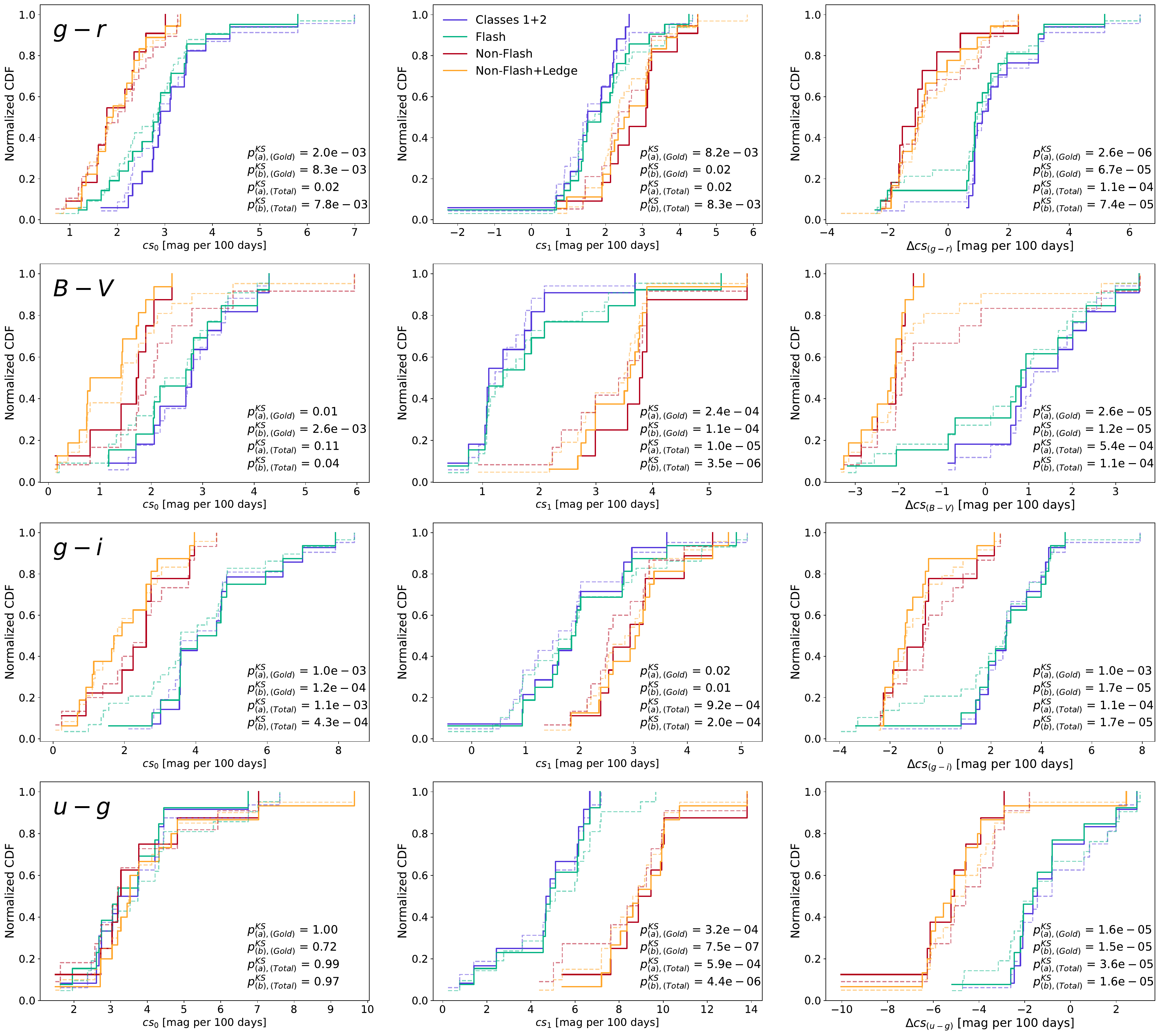}             
\caption{Cumulative distribution functions of $cs_0$ (left column), $cs_1$ (middle column), and $\Delta cs$ (right column) for (\textit{$g-r$}), (\textit{$B-V$}), (\textit{$g-i$}), and (\textit{$u-g$}) colours (as indicated in each row). Coloured lines correspond to class 1+2 (blue), flash (class 1+2+3, green), non-flash (red), and non-flash + ledge SNe (orange) from the gold (solid lines) and complete samples (dashed lines). The corresponding p values of the K-S test are shown in each panel. Table~\ref{t:pvalues_KS_AD_slopes} lists the p values from the K-S and A-D tests for comparison cases a and b.}  
\label{f:CDFs_slopes}
\end{figure}

\begin{figure}
\centering
\includegraphics[width=.95\textwidth]{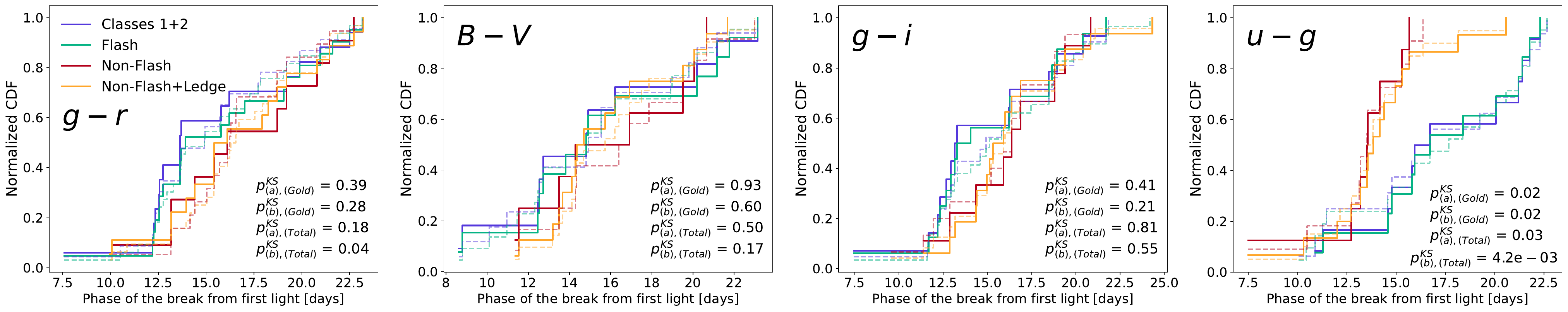}             
\caption{Cumulative distribution functions of break phase, $t_\mathrm{break}$, for (\textit{$g-r$}), (\textit{$B-V$}), (\textit{$g-i$}), and (\textit{$u-g$}) (as indicated in each panel). Coloured lines correspond to class 1+2 (blue), flash (class 1+2+3, green), non-flash (red), and non-flash + ledge SNe (orange) from the gold (solid lines) and complete (dashed lines) samples. The corresponding p values of the K-S test are shown in each panel. Table~\ref{t:pvalues_KS_AD_slopes} lists the p values from the K-S and A-D tests for comparison cases a and b.}  
\label{f:CDFs_t_break}
\end{figure}

\end{appendix}
\end{document}